\documentclass{article}
\pdfoutput=1
\usepackage[left=3cm, right=3cm, bottom=3cm]{geometry}
\usepackage{amsthm,amssymb,amsmath}
\usepackage{authblk}
\usepackage{bm}       
\usepackage{matlab-prettifier}
\usepackage{xcolor}
\usepackage{comment}
\usepackage{epsf}
\usepackage{graphicx} 
\usepackage{subcaption}
\usepackage[normalem]{ulem}
\usepackage{url}

\newcommand{\nameref}[1]{\ref{#1}}
\begin{document}

\title{Random Evolutionary Dynamics in Predator-Prey Systems Yields Large, Clustered Ecosystems}

\author[1,2,3]{Christian H.S. Hamster}
\affil[1]{Dutch Institute for Emergent Phenomena, University of Amsterdam, Amsterdam, The Netherlands}
\affil[2]{Korteweg-De Vries Institute for Mathematics, University of Amsterdam, Amsterdam, The Netherlands}
\affil[3]{Biometris, Wageningen University \& Research, Wageningen, The Netherlands}

\author[4,5]{Jorik Schaap}
\affil[4]{PhotoCatalytic Synthesis Group, University of Twente, Enschede, The Netherlands}
\affil[5]{Physical Chemistry and Soft Matter, Wageningen University \& Research, Wageningen, The Netherlands}

\author[3]{Peter van Heijster}

\author[5,6]{Joshua A. Dijksman}
\affil[6]{Van der Waals-Zeeman Institute, Institute of Physics, University of Amsterdam, Amsterdam, The Netherlands.}


\maketitle
\begin{abstract}
We study the effect of introducing new species through evolution into communities. We use the setting of predator-prey systems. Predator-prey dynamics is classically well modeled by Lotka-Volterra (LV) equations, also when multiple predator and prey species co-exist. We use a stochastic method to introduce new species in a two-trophic LV system. We find that introducing random evolving species leads to robust ecosystems in which large numbers of species coexist. Crucially, in these large ecosystems an emergent clustering of species is observed, tying functional differences to phylogenetic history.

\end{abstract}

\section{Introduction}
In natural ecosystems, many different species populate a finite environment. The species interact with each other, consume resources, and their composition changes in time by producing offspring and evolving, or by having their ecosystem get invaded. The dynamics of such ecosystems is extremely complex and challenging to unravel but connected to deep and important questions about ecosystem stability, adaptability, diversity (or richness) and community structures that emerge in such systems~\cite{borrelli2015selection,shtilerman2015emergence,tokita2003emergence}. For instance, the unexplained high diversity of plankton species in marine ecosystems has been dubbed ``the paradox of the plankton''~\cite{hutchinson1961paradox}. 

A century of quantitative work on ecosystem dynamics has produced important milestones such as the Lotka-Volterra (LV) predator-prey model~\cite{lotka1920analytical,volterra1926fluctuations}, MacArthur's resource-consumer model~\cite{macarthur1958},
and May's stability criterion~\cite{may1972will}. The central approach in much of the quantitative modeling is to capture the time dependence of species population size via the species' interactions with other species and the resources it consumes~\cite{vandenberg2022ecological}. This often results in a system of Ordinary Differential Equations (ODEs) that describes the interaction among many species and leads to interaction or community matrices that define a community structure. 
A large part of modeling work has focused on the coexistence of a fixed number of species~\cite{may1972will}; perhaps considering the role of extinction~\cite{galla2018dynamically}; or the invasion of new species~\cite{tilman2004niche}, often from fixed regional pools~\cite{macarthur2016theory}. 
The focus in these models often lies on the creation of communities with stable~\cite{allesina2012stability,allesina2015stability,Biroli2018marginally,fedeli2021nonlinearity,may1972will} and feasible equilibria~\cite{dougoud2018feasibility,grilli2017feasibility,song2018guideline}. Furthermore, they consider short timescales (in the eco-evolutionary sense), disregarding transient and evolutionary phenomena~\cite{hastings2010timescales,hastings2018transient,mittelbach2015ecological}. For a discussion of equilibrium and non-equilibrium hypotheses of biological diversity see~\cite{oliveira2016species}. 

Despite decades of research, many questions remain unanswered~\cite{allesina2015stability}. In recent years the role of introducing new species to the community through evolution has gained traction~\cite{bellavere2023speciation,maynard2018network,posfai2017metabolic,shtilerman2015emergence,tokita2003emergence}. For clarity of presentation, we will always call such an introduction of a new species a \emph{speciation} event.  In particular, we do not consider genetic variation within a species and also identify each step in adaptive diversification\cite{champagnat2006unifying} as a speciation event in this article.
On biological grounds, it is very natural to include speciation in ecosystem dynamics as it is common to perceive that new species come and go, yet many of the consequences of adding such dynamics remain unexplored or counterintuitive. An early attempt for a specific plankton model even showed that speciation may reduce the biodiversity in the system~\cite{shoresh2008evolution}. Note that this speciation approach is a different approach than introducing a fixed number of species with random coefficients such as~\cite{feng2024emergent}.
 
Here, we show the promise of the speciation approach by showing that speciation can lead to a complex ecosystem that settles in a dynamic equilibrium, in which many species cluster into separate families of species. We explore the dynamics of speciating ecosystems that are being built up from basic principles by evolving ecosystems from just two ancestor species (predator and prey). We observe that clusters evolve out of these two ancestors and develop a phylogenetic tree of functionally different species.  In our two trophic layer model system, we observe such clustering for both the predator and the prey dynamics under the condition that the variability in the speciation step is large enough. Note that clustering has been observed  before 
in competitive LV models where the competition is defined as the relative position along a niche axis \cite[e.g.]{pigolotti2010gaussian,saltini2023complex, scheffer2006self}.

Our specific setup, as outlined in sec.~\nameref{sec:setup}, leads to new concerns about the conditions under which such type of speciation generates dynamic equilibria. In particular, we show that the classic LV modeling framework can be extended with speciation dynamics while retaining a finite biomass when two trophic levels are imposed while simultaneously creating a large number of coexisting species. The addition of trophic levels and speciation to LV has been explored successfully before. In~\cite{haerter2016food}, stable food webs were built using the generalized LV equations, by applying strict trophic levels and additional constraints within these levels. We show that stable systems can also be achieved with fewer constraints. However, completely removing the trophic structure is not feasible, as discussed in the appendix.  

The relative importance of equilibrium and non-equilibrium dynamics has been debated for decades, see e.g.~\cite{oliveira2016species}. Our model shows that while its ecological and evolutionary timescales are initially well-separated in the sense that the exponential decay towards the stable state is much faster than the average time between speciation events, these timescales eventually become intertwined. As the increasing number of species leads to eigenvalues of the equilibrium with a small negative real part, we expect slow decay towards the equilibrium. Hence, speciation does not necessarily happen near a steady state. The resulting overlap in timescales is an important feature that has a grounding in natural systems, where the ecological timescale and evolutionary timescales overlap~\cite{sax2007ecological,stockwell2003contemporary}. 

The structure of the paper is as follows. In sec.~\nameref{sec:setup}, we discuss the setup of our speciating LV model. In sec.~\nameref{sec:2layers}, we study the general dynamical properties of the resulting systems, e.g. number of species, biomass and the influence of the variability. In sec.~\nameref{sec:2layers:drive} we focus on the formation of genealogical and phenomenological clusters and study the interactions between these clusters.

\section{Model Setup}
\label{sec:setup}

\begin{figure}[t]
  \centering
   \includegraphics[width=\linewidth]{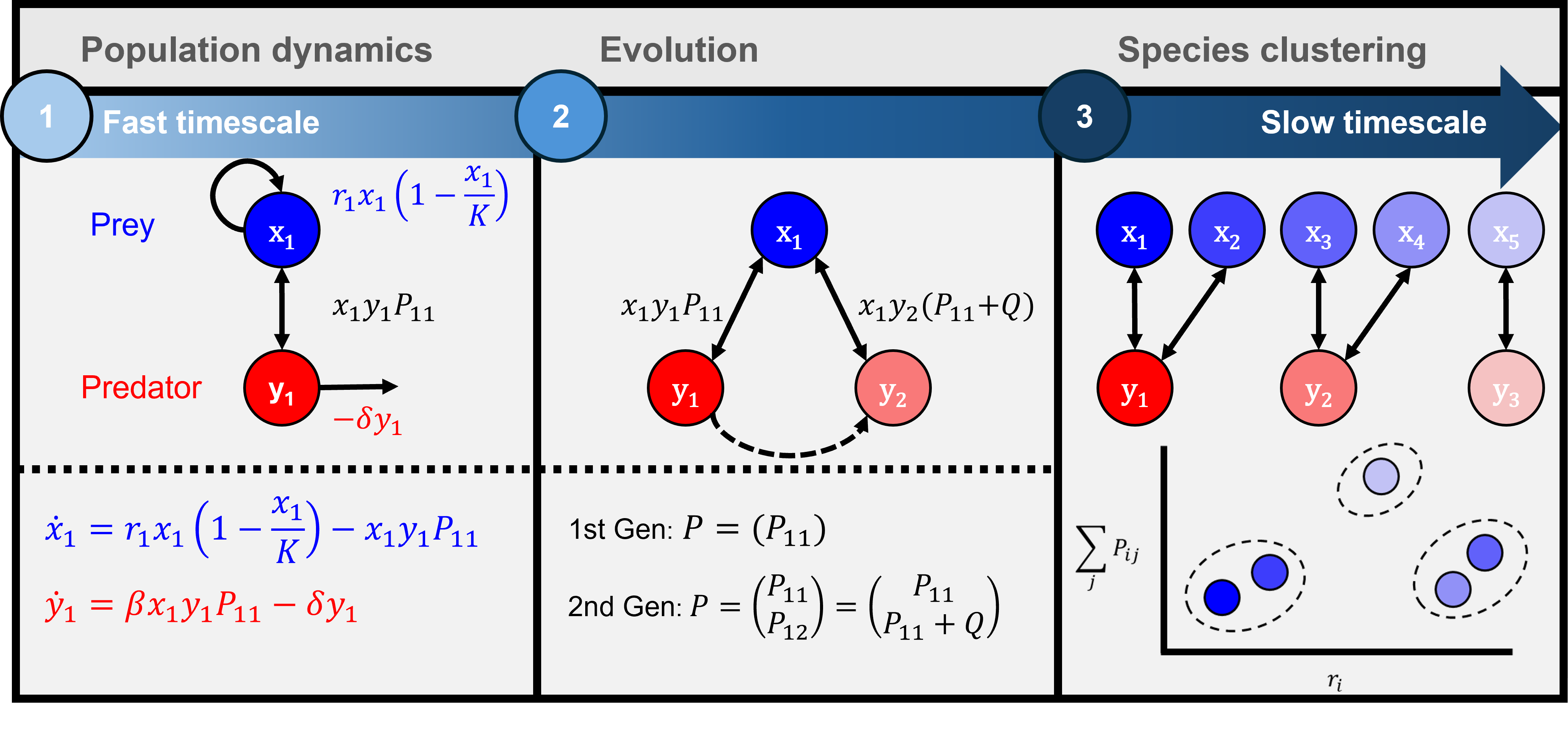}
\caption{Scheme summarizing the model setup and main outcome. 1) \textit{Population dynamics:} The population dynamics are modeled with a two-trophic LV model. The prey species ($x$) has logistic growth and is consumed by predator species ($y$), which is regulated through predation rate $P_{11}$ and a conversion factor $\beta$. The predator population decreases with a linear death rate $\delta$. The equations underlying the model are further explained in sec. \ref{sec:setup}. 2) \textit{Evolution:} We evolve the ecosystem by introducing new species stochastically. New species are generated by perturbing the interaction (shown) and growth rate (for prey, not shown) coefficients of a randomly selected ancestor species. The added perturbation $Q$ is drawn from a normal distribution with mean $0$ and variance $\eta^2$. Further details on how we add evolution to the model can be found in sec.~\ref{sec:2layers:spec}. 3) \textit{Species clustering:} After many cycles of evolution and extinction, we find functional (shown) and genealogical clustering of species. The formation of clusters is discussed in sec.~\ref{sec:2layers:drive}.}
\label{fig:vis_abstract}
\end{figure}

We discuss a general approach to implementing speciation in generalizations of the following classic LV predator-prey model:
\begin{equation}
\label{eq:int:pp}
\begin{split}
    \dot x=\,&rx\left(1-\frac{x}{K}\right)-xPy.\\
    \dot y=\,&\beta yPx-\delta y.
\end{split}
\end{equation}
In particular, we consider two trophic levels: prey $x_i$ and predators $y_j$. For the prey species, we use logistic growth combined with the negative interaction coming from the predators: 
\begin{equation}
\label{eq:2layers:prey}
    \dot x_i=r_i x_i\left(1-\frac{x_i}{K}\right)-x_i\sum_j P_{ij}y_j.
\end{equation}
Here, $r_i$ is the growth rate of prey $i$ and $K$ the carrying capacity for prey $i$, which we assume to be constant for all species. A discussion on the inclusion of the carrying capacity and its value can be found in Appendix~\ref{sec:AppB}. The interaction between prey $i$ and predator $j$ is given by the value $P_{ij}$ which is assumed to be nonnegative. 

For the predator species $y_j$, we assume a mortality rate $\delta_j$ and growth only comes via consuming prey:
\begin{equation}
\label{eq:2layers:pred}
    \dot y_j=\beta y_j\sum_i P^T_{ji}x_i-\delta_jy_j.
\end{equation}
Here, $P^T$ is the transpose of the matrix $P$ and $\beta$ is the conversion factor between the consumption of prey and the growth of the predator, so typically $0<\beta<1$, and it is assumed to be constant. We assume that there is no direct competition among the prey species and between the predators. Therefore, the interaction matrix $A$ of the total ecosystem has the following block structure:
\begin{equation}
    A=\begin{pmatrix}
        0 & -P\\
        \beta P^T&0
    \end{pmatrix}.
    \label{eq:DefA}
\end{equation}
The assumption that the prey species do not compete implies that they all occupy their own niche with carrying capacity $K$, which allows us to focus on the interaction between the predators and prey. For a graphical representation of the entire model, see Fig.~\ref{fig:vis_abstract}.

\subsection{Speciation}
\label{sec:2layers:spec}
We start our simulations on a classic predator-prey configuration \eqref{eq:int:pp} with one prey species and one predator species and parameters $r_1=r=1.1, K=10, P_{11}=P=0.4, \beta_1=\beta=0.25$ and $\delta_1=\delta=0.4$. For this setting, the classical system \eqref{eq:int:pp} has a stable fixed point 
\begin{equation}
\label{eq:CLV}
(x^*,y^*)=\left(\frac{\delta}{\beta P},\frac{r}{P}\left(1-\frac{\delta}{\beta K P}\right)\right) = (4,1.65).
\end{equation}
To model the evolutionary process, we introduce a new species after a random waiting time by assuming that this new species is a perturbation of a randomly chosen ancestor species. To spawn new species from an ancestral species, choices need to be made concerning the strength of self-interaction, growth rate, species' interaction and the sign of the species interaction. We emphasize that these model choices are crucial in determining the dynamics observed, but within a given regime of the phase space, we see that they generate similar dynamics.
Notably, in~\cite{shtilerman2015emergence}, it is assumed that the interaction between all species is always competitive, and predator-prey and mutualism-type interactions are excluded. In other words, the interaction matrix contains only negative numbers. This is an important and restrictive model choice. However, it is also a constraint that is not easily relaxed. For instance, we find that under a broad set of conditions, speciation eventually leads to blowup of the biomass in the ecosystem, if one does not constrain the sign of the interactions. We emphasize that we have no formal proof that the nonphysical biomass dynamics always occurs, but observed it for all parameter choices we could reasonably explore. Using the most unconstrained Lotka-Volterra model choice regarding interaction coefficients would be preferred, but as this most unconstrained choice only produces biologically unrealistic scenarios, we discuss it only in the Appendix~\nameref{sec:AppA}. In the main text, we limit ourselves to two trophic levels and hypothesize that imposing such a structure is a minimum requirement for maintaining a finite biomass in LV systems. By extension, we expect that multi-trophic levels retain the basic phenomenology of a two-trophic level system, but leave this for future work to verify.

When a new species is introduced as a daughter species of prey $i$, we add a new row to $P$ which is a perturbation of the $i$-th row of $P$ and replace $P^T$ with the corresponding new $P^T$. Hence, after the first speciation event, we get that $P_{11}=P=0.4$ becomes
\begin{equation}
    P=\begin{pmatrix}
        P_{11}\\
        P_{21}
    \end{pmatrix}
    =\begin{pmatrix}
        0.4\\
        0.4+\text{perturbation}
    \end{pmatrix}\,.
\end{equation}
The full interaction matrix A therefore updates as
\begin{equation}
    \begin{pmatrix}
        0 & -0.4\\
        0.1&0
    \end{pmatrix} \implies \begin{pmatrix}
        0& 0 & -0.4\\
        0 &0 & -0.4-\text{perturbation}\\
        0.1 &0.1+ 0.25 \times \text{perturbation}&0
    \end{pmatrix}.
\end{equation}
Similarly, when a new predator is introduced, we add a new row to $\beta P^T$, resulting in a corresponding new column for $P$. We assume that the added perturbation is normally distributed with zero average and standard deviation $\eta$. We will refer to $\eta$ as the size of the variability in the speciation step. If a newly added row to $P$ contains a negative element, it is set to zero to keep the structure of the trophic levels. As the rows of $P^T$ describe how much the predators can consume, it would seem that increasing these values in a speciation step is favorable, leading to a run-off in the $P$-values. However, as $P$ and $P^T$ are by definition linked, a speciation event where a predator increases its positive interaction with a prey automatically increases the predation on the prey, thus counteracting the run-off. 

\begin{table}
    \centering
    \begin{tabular}{c|c|c}
         Parameter& Meaning & Value at $t=0$   \\ \hline
$x_i(t)$ & Biomass prey species $i$ at time $t$& 4 \\
$y_j(t)$ & Biomass predator species $j$ at time $t$ & 2  \\
       $r_i$  & Growth rate prey species $i$ & 1.1 \\
$K$ & Carrying capacity & 10 \\
$\beta$ & Biomass conversion & 0.25 \\
$\delta_j$ & Mortality predator species $j$ & 0.4 \\
$P_{11}$ & Predation of the initial predator on the initial prey & 0.4\\
$P_{ij}$ & Predation of predator $j$ on prey $i$ & t.b.d.\\
$\eta$ & Standard deviation of the noise/variability in speciation &0.02\\
$p_m(0)$ & Parameter for the exponential waiting times &$10^{-3}$\\
$f$&Fraction of biomass transferred to daughter species&0.05\\
 & Extinction threshold &$10^{-3}$\\\hline
    \end{tabular}
    \caption{Table summarizing the parameters used in the speciating LV predator-prey model. t.b.d. means to be determined.  }
    \label{tab:2layers}
\end{table}

We assume that the time intervals between speciation events, which sets the mutation rate, are exponentially distributed with time-dependent parameter~$p_m$~\cite{venditti2010phylogenies}. We let~$p_m$ depend on the total biomass in the system. This means that $p_m$ is indexed with respect to the biomass at $t=0$:
\begin{equation}p_m(t)=p_m(0)\frac{\sum_ix_i(t)+\sum_jy_j(t)}{x_1(0)+y_1(0)}\,.
\end{equation}
We typically take $p_m(0)=10^{-3}$, meaning that the average waiting time for a speciation event at the start of the simulation is~$10^3$. Again, we emphasize that the model performance is largely insensitive to these model choices. We chose $x_1(0)=4$ and $y_1(0)=2$, hence $p_m(0)$ is defined with respect to a biomass of $6$, which is the order of magnitude we typically observe in the simulations as Fig.~\ref{fig:2layers} in the next section will show. Note that the initial condition sets $p_m(0)$, but does not essentially influence the simulation in any other way as the dynamics converges exponentially fast to the coexistence stable state \eqref{eq:CLV}.  When a speciation event happens, the probability that species $k$ is chosen to speciate equals the fraction of its biomass of the total biomass. 

Similar to the approaches in~\cite{caetano2021evolution,shoresh2008evolution}, we assume that at a speciation event, a fraction $f$ of $5\%$ of the biomass of the ancestor species $k$ transforms to the new daughter species, while the biomass of the other species does not change. We note that in the assumption that $5\%$ of the biomass of the ancestor species evolves, we implicitly assume that unsuccessful speciation events happen on a shorter timescale. The outcome of the simulations is largely insensitive to the specific value of $f$ that is chosen, see Appendix~\ref{sec:AppB}. 

In the model, a species is declared extinct when its abundance has dropped below the extinction threshold at a new speciation event. Upon extinction of a species, its corresponding columns and rows in $P$ and $P^T$ are deleted (reducing the dimension of the interaction matrix $A$ from $n\times n$ to $(n-1)\times(n-1)$). Note that this step happens before the new speciation step is done. That is, an extinct species cannot speciate anymore. We typically set the extinction threshold to $10^{-3}$, see Appendix \ref{sec:AppB} for a discussion of this value. A complete overview of all model settings used in our simulations (unless stated otherwise) is provided in Table~\ref{tab:2layers}.

\subsection{State variables}
\label{subsec:statevar}
Having defined our model, we can now identify several important quantities of interest, which for convenience we will call \emph{state variables}. 
The main focus here is on the species richness $S(t)$ and biomass $B(t)$, which we split out over the predators and prey. 
Hence, $S_\mathrm{pred}(t)$ stands for the number of predator species alive, while $S_\mathrm{prey}(t)$ stands for the number of prey species alive. 
We define the predator and prey biomass as 
\begin{equation}
    B_\mathrm{prey}(t)= \sum_{i=1}^{S_\mathrm{prey}(t)} x_i(t)\,, \qquad
    B_\mathrm{pred}(t)= \sum_{j=1}^{S_\mathrm{pred}(t)} y_j(t).
\end{equation}
The total species richness and biomass are now given by $S(t)=S_\mathrm{prey}(t)+S_\mathrm{pred}(t)$ and $B(t)=B_\mathrm{prey}(t)+B_\mathrm{pred}(t)$. Note that the simulations will be intrinsically stochastic due to the speciation events, meaning that these state variables will change from simulation to simulation. However, the ratio of biomass in predators and prey is a realistic biological observable~\cite{borrelli2015selection}, and, as we shall see, this ratio is also a benchmark for our simulations which shows that the model performance is stable under the addition of speciation.

We also need to be able to quantify the genealogical distance between species. As all the species descended from a small number of ancestors, i.e. we always start with one prey and one predator, we can define a genealogical distance between species with the same ancestor. In particular, we define the genealogical distance between two species (still present at the end of the simulation) as the difference between the present time and the time their lineages split.  
\begin{figure}[t]
\begin{subfigure}{.49\textwidth}
  \centering
 		\def\svgwidth{\columnwidth}
    		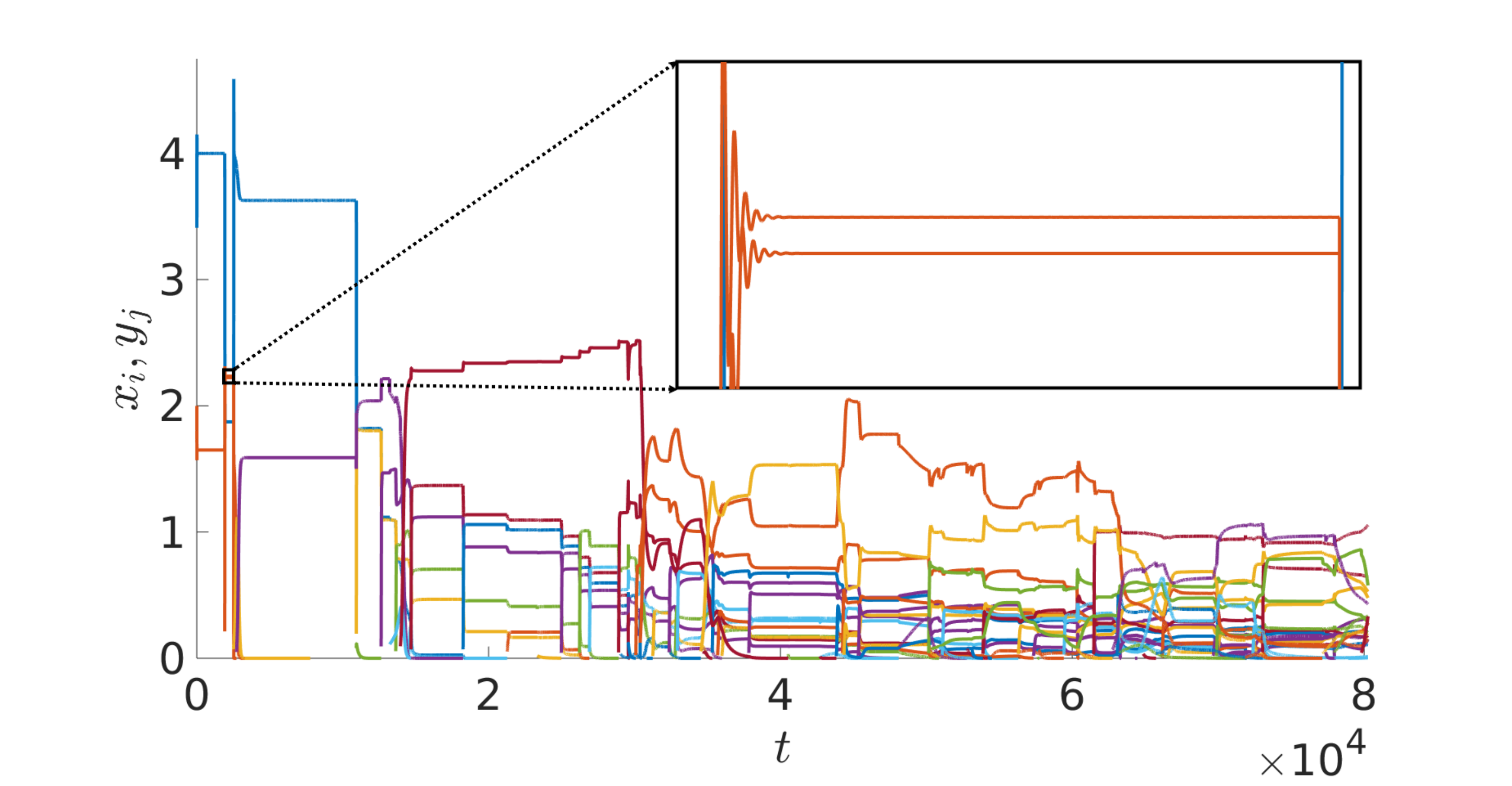
  \caption{}
  \label{fig:2layers:ExRun}
\end{subfigure}
\begin{subfigure}{.49\textwidth}
  \centering
 		\def\svgwidth{\columnwidth}
    		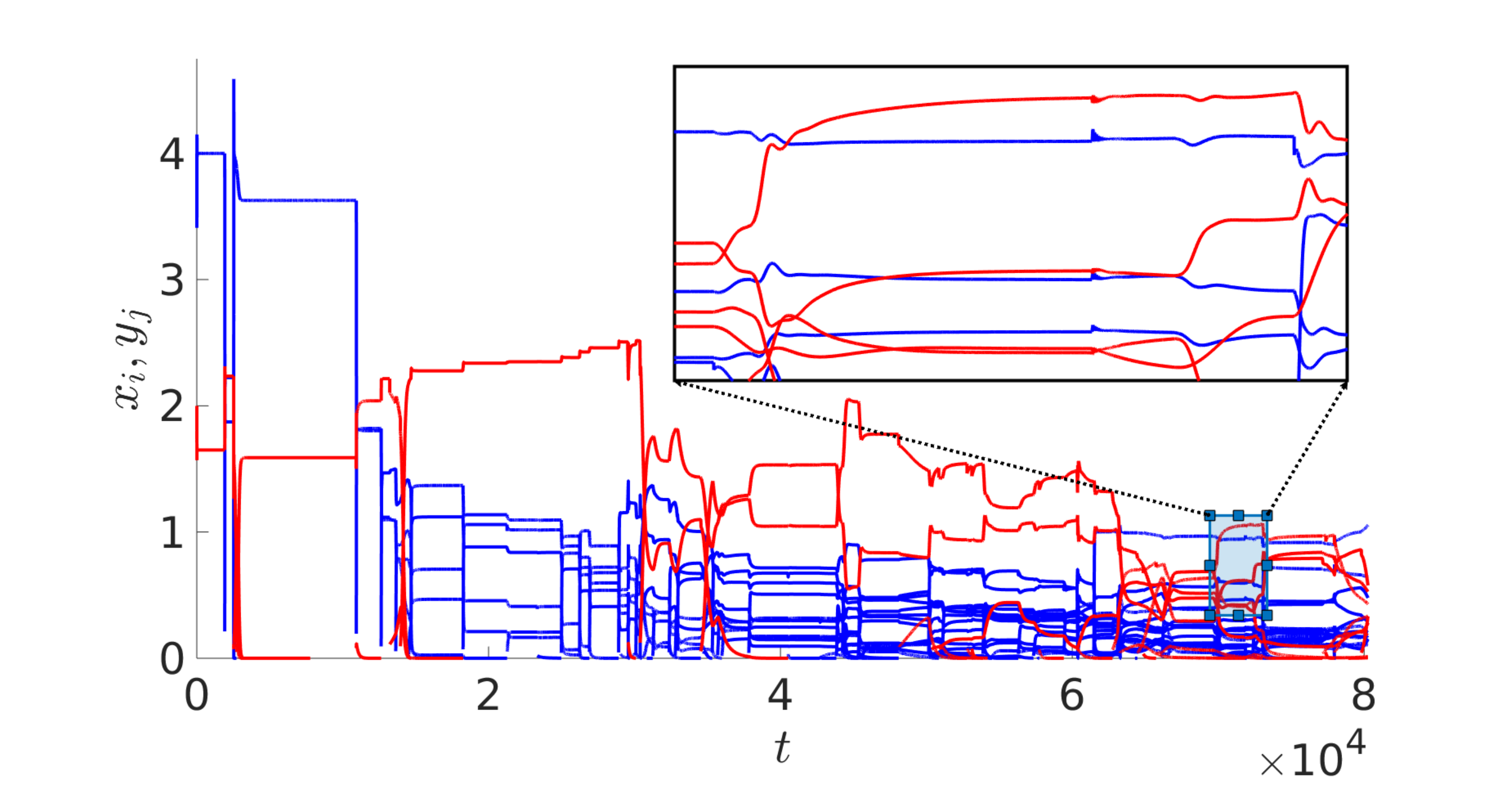
  \caption{}
  \label{fig:2layers:ExRunAnc}
\end{subfigure}
\caption{This figure shows a short-time example run of the model in Eqs.~(\ref{eq:2layers:prey}) and (\ref{eq:2layers:pred}), with an initial condition as specified in Table~\ref{tab:2layers}. In the left figure, each species is colored individually, while in the right figure, all prey species are blue and all predator species are red. The two insets highlight the difference in timescales. In the inset in (a), the convergence to a steady state after a speciation event is very fast, while the inset in (b) shows that the dynamics has not yet stabilized before a new speciation event happens. }
\label{fig:2layers:ExampleRun}
\end{figure}
\section{Large, Stable and Robust Ecosystems}
\label{sec:2layers}

To explicitly show the dynamics of all species involved, we first start with a short-time simulation; see Fig.~\ref{fig:2layers:ExampleRun} for an overview of the resultant stochastic dynamics. At the beginning of the simulation, the timescale on which the predator-prey oscillations die out is much faster than the timescale of speciation, resulting in dynamics that has many abrupt speciation events followed by long periods of stationary dynamics; see the inset. As time progresses, the timescales of the dynamics of the speciation and the species size fluctuations become more intertwined, as the inset in Fig.\ref{fig:2layers:ExRunAnc} shows. 
In this figure, the species are colored according to their trophic layer, i.e. red for prey and blue for predators. Note that the evolutionary timescale is set by $1/p_m(t)$, while the timescale of the ecosystem dynamics around a fixed point is dominated by $1/\text{Re}(\lambda_\mathrm{max})$, where $\lambda_\mathrm{max}$ is the eigenvalue of the Jacobian around the fixed point with the largest real part. Indeed, we find that at the start of the simulation, this ratio between the two timescales is 220, i.e. the evolutionary timescale is significantly longer than the ecosystem timescale. However, this ratio quickly drops to approximately one (results not shown).

\begin{figure}[t]
\begin{subfigure}{.49\textwidth}
  \centering
 		\def\svgwidth{\columnwidth}
    		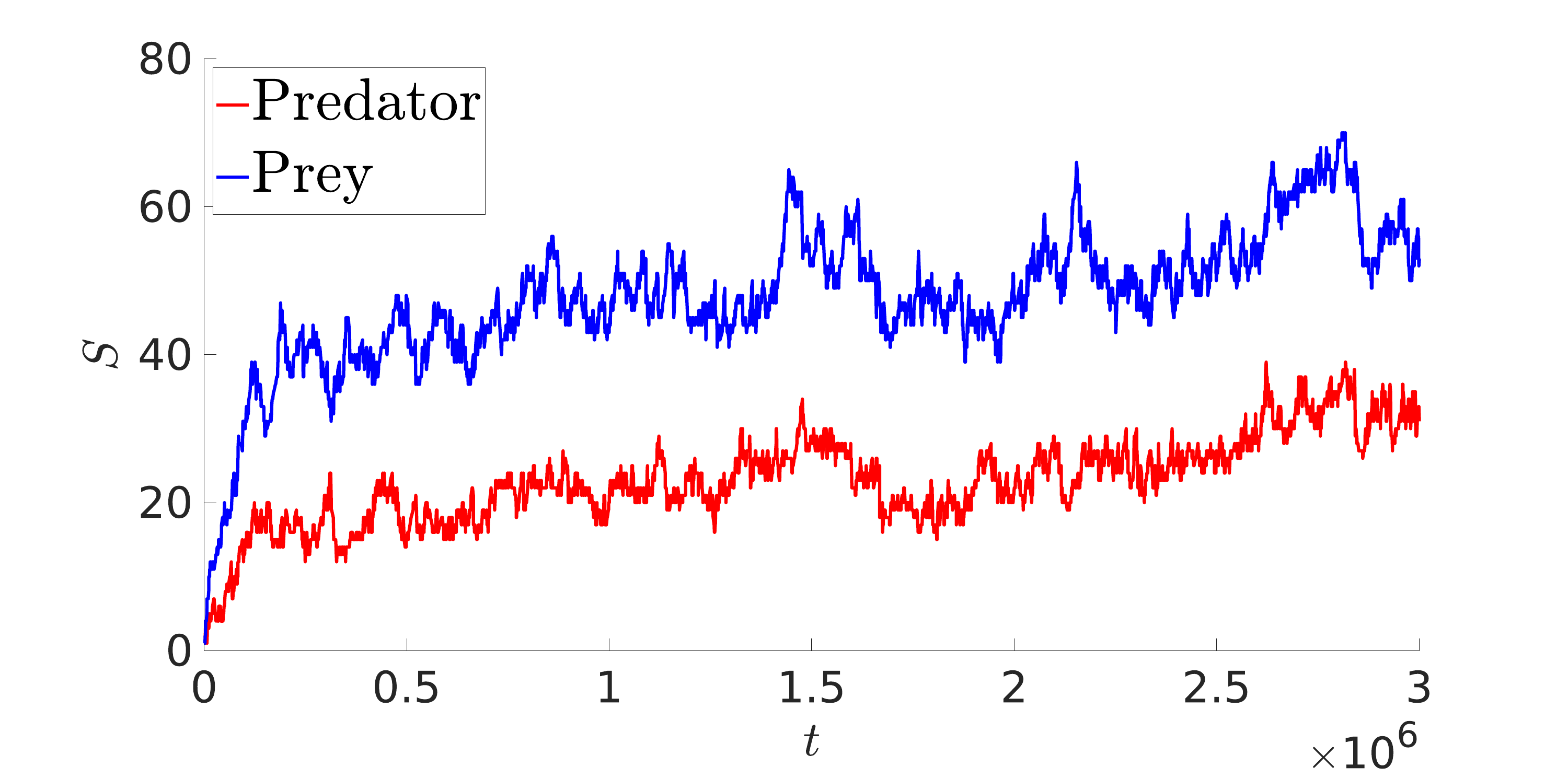
  \caption{}
  \label{fig:2layers:S}
\end{subfigure}
\begin{subfigure}{.49\textwidth}
  \centering
 		\def\svgwidth{\columnwidth}
    		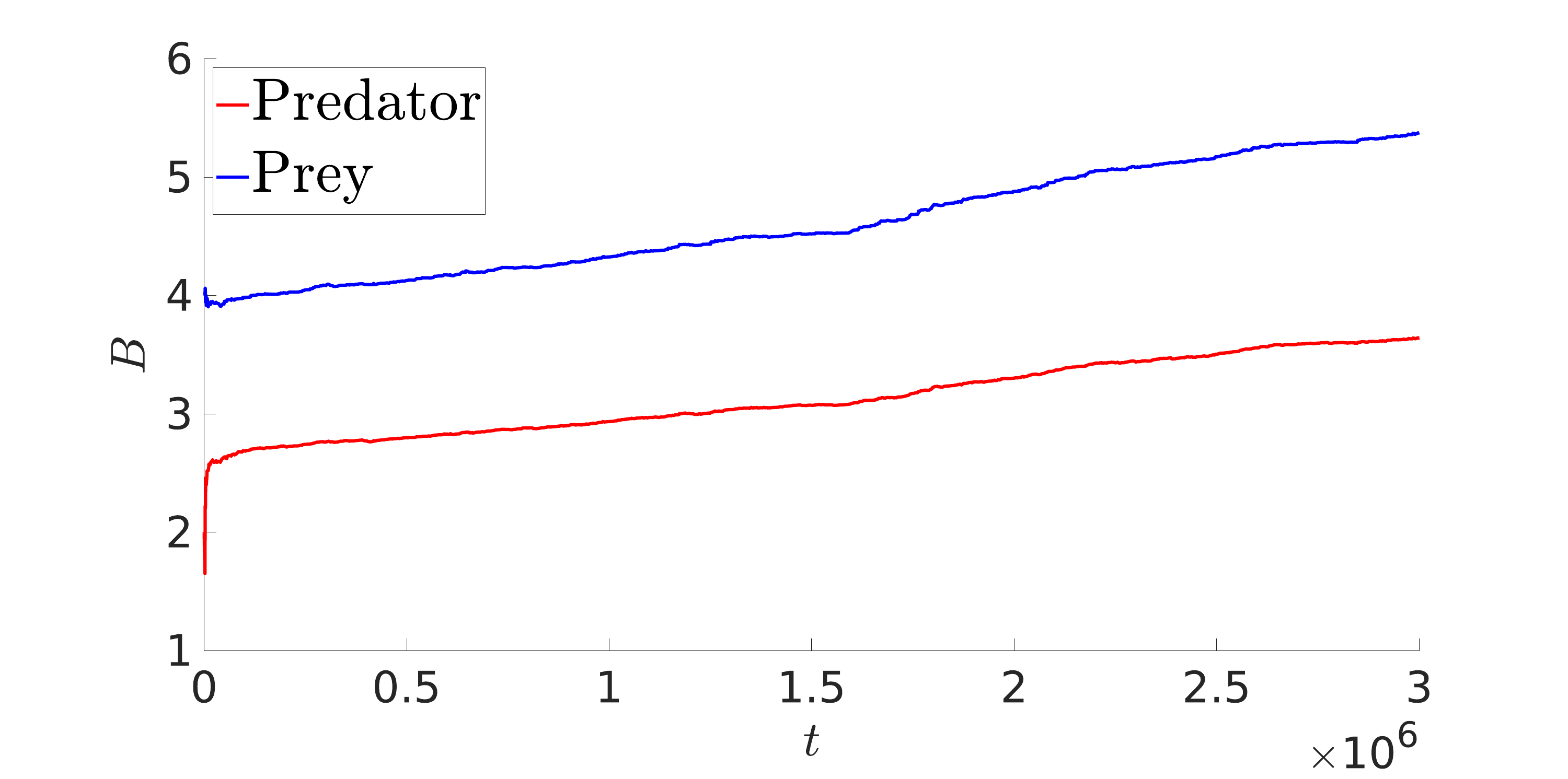
  \caption{}
  \label{fig:2layers:B}
\end{subfigure}
\caption{Long-time simulation of the model from sec.~\nameref{sec:setup}. We plot the number of species (left) and biomass (right). The total biomass of both predator and prey species appear strongly correlated and also stabilize quickly to a small slope after an initial steep growth.}
\label{fig:2layers}
\end{figure}
\begin{figure}
\begin{subfigure}{.49\textwidth}
  \centering
 		\def\svgwidth{\columnwidth}
    		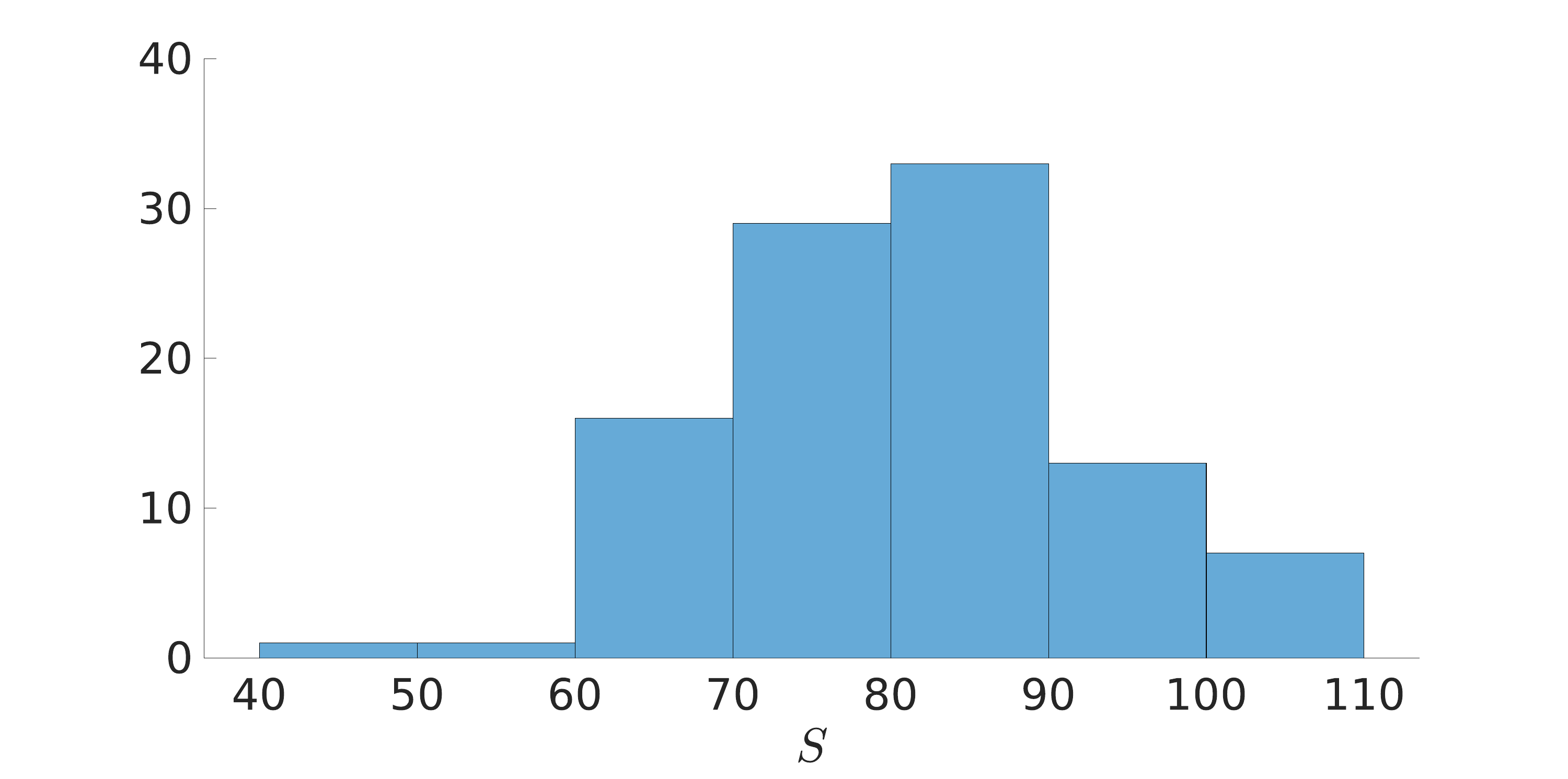
  \caption{}
  \label{fig:2layers:Histogram}
\end{subfigure}
\begin{subfigure}{.49\textwidth}
  \centering
 		\def\svgwidth{\columnwidth}
    		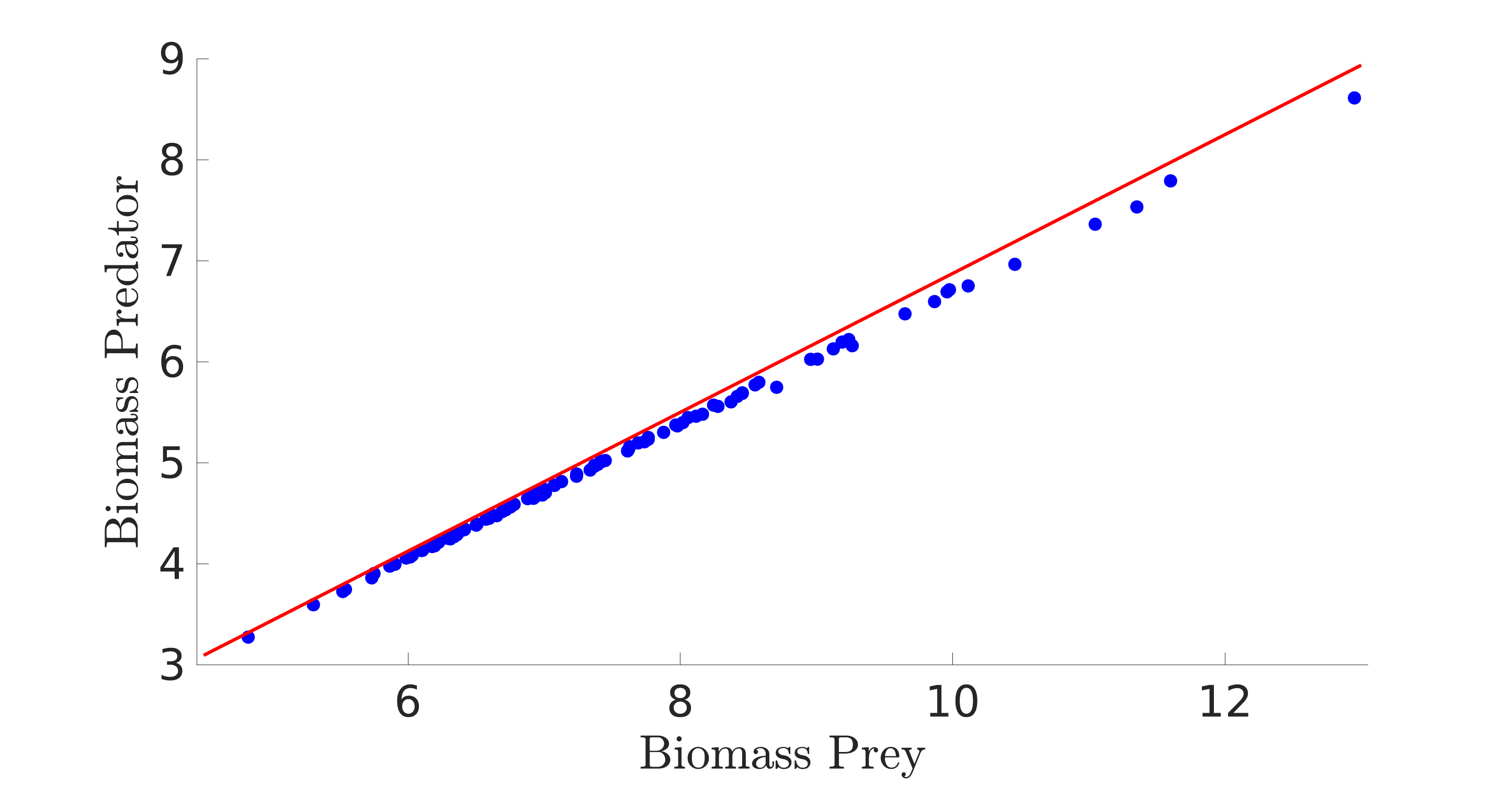
  \caption{}
  \label{fig:2layers:Bpvsr}
\end{subfigure}
\caption{Similar simulation to Fig.~\ref{fig:2layers}, but repeated 100 times. Figure (a) shows the spread in the total number of species at the end of the simulation, while Figure (b) shows for all simulations the relation between the predator and prey biomass. Clearly, there is a strong correlation between the two biomasses as dictated by the model, see Eq.~(\ref{eq:2layers:biomassratio}). The red line is the linear approximation of this equation.}
\label{fig:2layers:HB}
\end{figure}

For longer integration times, we only plot the state variables $S_\mathrm{pred}(t),S_\mathrm{prey}(t),B_\mathrm{pred}(t)$ and $B_\mathrm{prey}(t)$, as defined in the previous section, because figures like~Fig.~\ref{fig:2layers:ExampleRun} become too cluttered at longer timescales. 
Fig.~\ref{fig:2layers} shows the evolution of these state variables over a longer integration time of $T=3\cdot10^6$. The number of predator and prey species quickly stabilizes to a ratio of approximately 1:2, but the value of the biomass keeps increasing slowly. When we repeat the simulation 100 times, with the same initial condition, we find $\langle S\rangle=81$ (rounded to the nearest integer) at $T=3\cdot10^6$, but there is a significant spread as the histogram in Fig.~\ref{fig:2layers:Histogram} shows. Despite this spread, the histogram shows that the model consistently evolves towards an ecosystem with a high number of species. This observation and even the ratio of 1:2 is in line with real ecosystems~\cite{borrelli2015selection}. As Fig.~\ref{fig:2layers:B} indicates, there is a strong correlation between the biomasses of the predators and prey. This correlation holds not just in time, but also between different runs, as becomes clear from Fig.~\ref{fig:2layers:Bpvsr}. 

\begin{figure}
\begin{subfigure}{.49\textwidth}
  \centering
 		\def\svgwidth{\columnwidth}
    		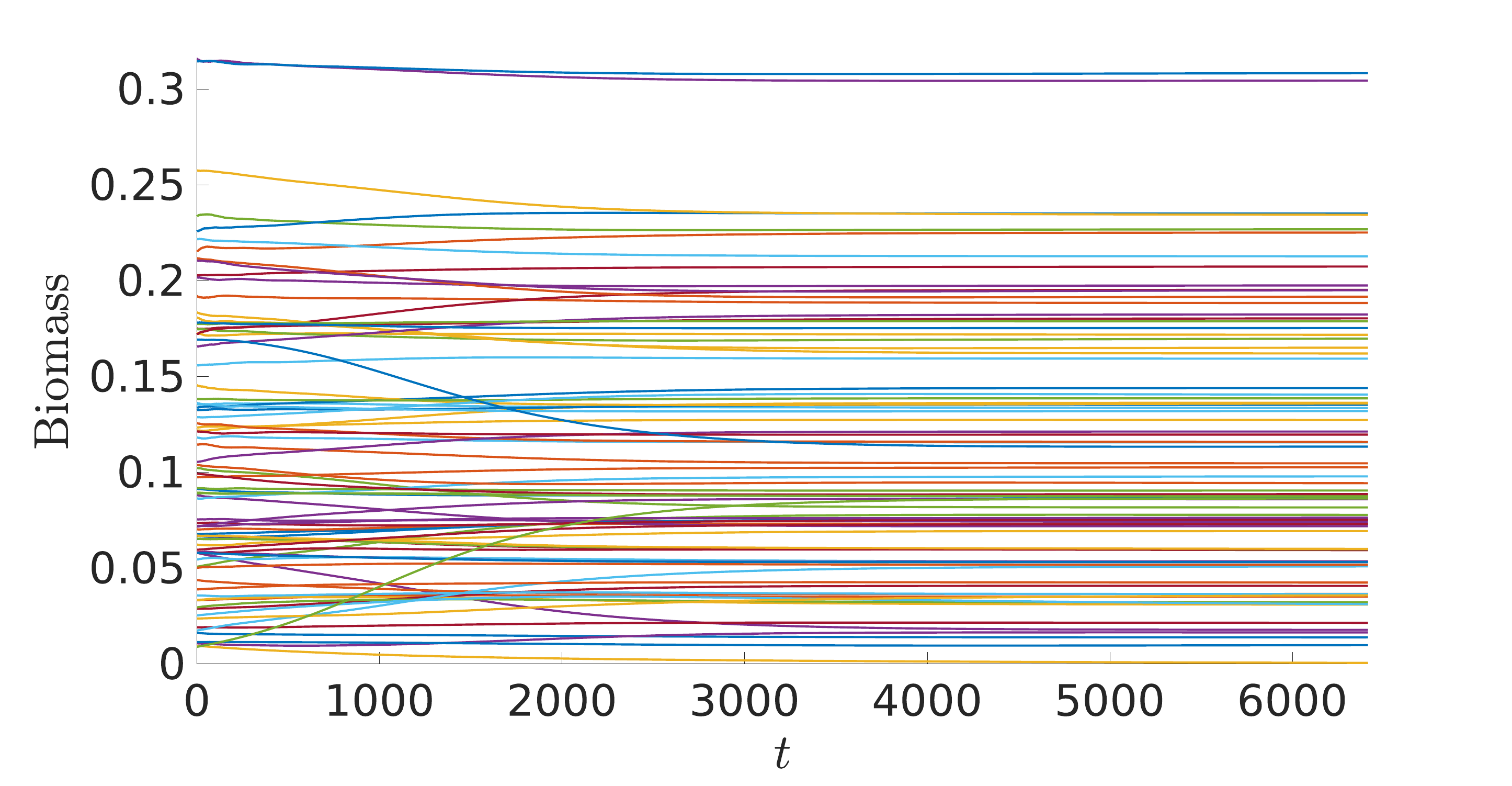
  \caption{}
  \label{fig:2layers:FinalRun}
\end{subfigure}
\begin{subfigure}{.49\textwidth}
  \centering
 		\def\svgwidth{\columnwidth}
    		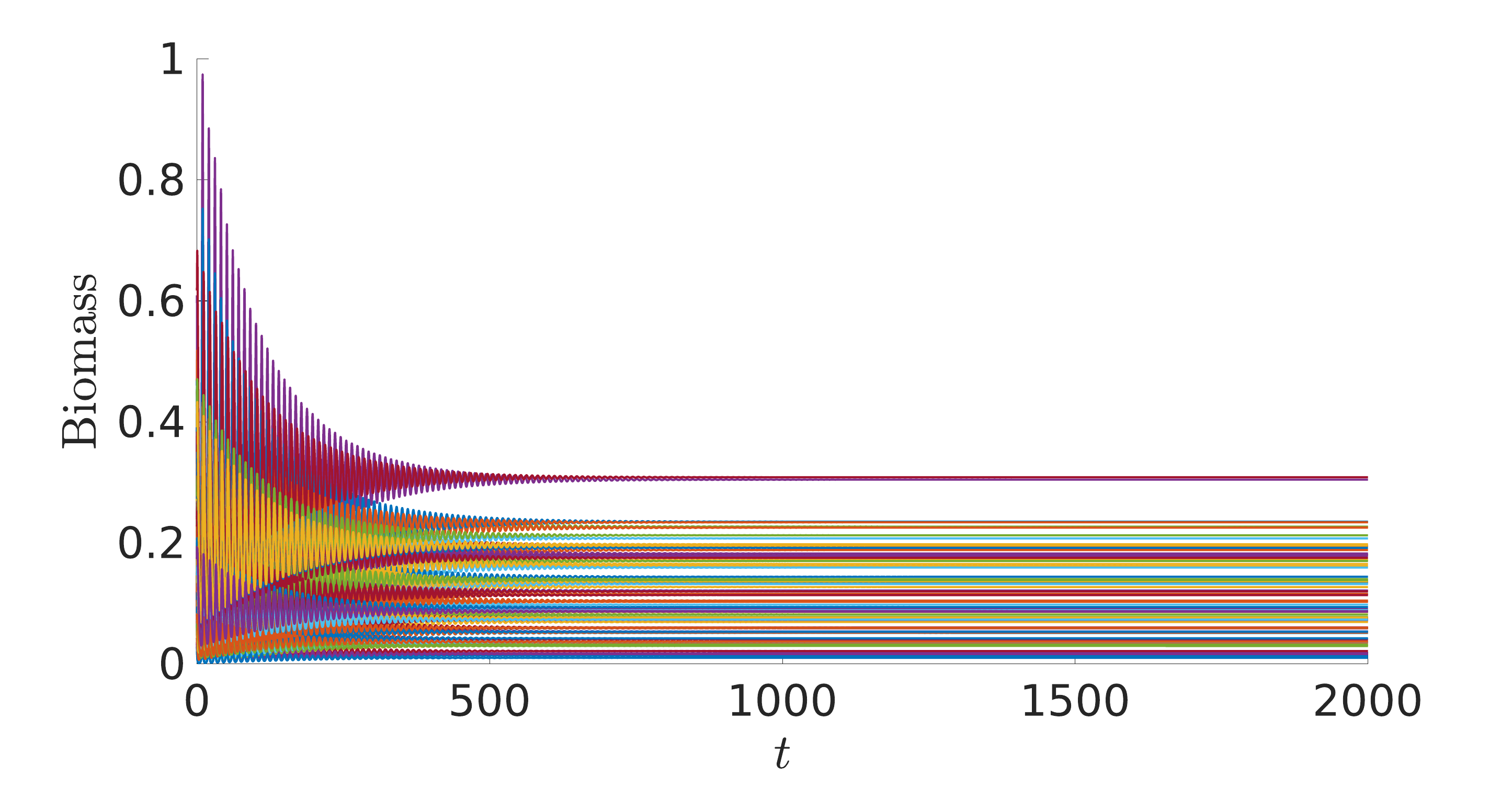
  \caption{}
  \label{fig:2layers:periodic}
\end{subfigure}
\caption{Figure~(a) shows the ecosystem dynamics after the final speciation step in Fig.~\ref{fig:2layers}. The solution is simulated until the dynamics stabilizes and here 81 out of 82 species survive (the actual integration time $10^5$ is significantly longer than shown in the figure). In Fig.~(b), we take the stationary state from Fig.~(a) and multiply it by two to create a large perturbation. This creates oscillations initially, but none of the 81 species go extinct.  }
\label{fig:2layers:final}
\end{figure}

To test the stability of the resulting ecosystem, we perform several tests. The first test is to stop speciation at the end of the simulation and let the system equilibrate while removing species that drop below the extinction threshold. As Fig.~\ref{fig:2layers:FinalRun} shows, the ecosystem from Fig.~\ref{fig:2layers} indeed equilibrates to a fixed point where 81 out of 82 species survive. Numerical evaluation of the Jacobian at this fixed point also shows that all eigenvalues are negative. Furthermore, when we add a large perturbation to the stable state, for example, when we take as initial condition two times the stable state, we see that the system converges back to the stable state after a series of predator-prey cycles and no species go extinct, see Fig.~\ref{fig:2layers:periodic}. This indicates that the system is robust and is able to support large stable ecosystems.  
Note that it takes at least up to $t=4000$ before the system becomes close to the stable state, which is significantly longer than the speciation timescale of $1/p_m=666$ for the biomass at the start of the simulation in Fig.~\ref{fig:2layers:FinalRun}. Averaged over 100 simulations, 80.32 out of 80.61 species survive. This shows that the ecosystem is not dominated by species that are bound to go extinct.  

\subsection{Stationary State of the Biomass}
Due to the nonlinearity in the logistic model, we cannot analytically solve for the stationary values of $B_{\mathrm{pred}}$ and $B_{\mathrm{prey}}$, but in the stationary state we find
\begin{equation}
\label{eq:2layers:biomassratio}
    \frac{B_{\mathrm{pred}}}{B_{\mathrm{prey}}}=\frac{r\beta}{\delta}-\frac{r\beta}{\delta K}\frac{\sum_ix_i^2}{\sum_ix_i}=\frac{r\beta}{\delta}\left(1-\frac{1}{K}\frac{\sum_ix_i^2}{B_\mathrm{prey}}\right)\approx \frac{r\beta}{\delta} + \mathcal{O}(1/K).
\end{equation}
We get an explicit, leading order (in $1/K$), expression for the biomass ratio, which is the straight line plotted in Fig.~\ref{fig:2layers:Bpvsr}. Heuristically, one can think of this limit as an infinite carrying capacity, i.e. $K\to\infty$, or, in other words, the absence of the logistic term. The ratio $B_{\mathrm{pred}}/B_{\mathrm{prey}}$ is identical to the ratio for the initial two-species ecosystem equilibrium, i.e. $y^*/x^*$ from Eq.~(\ref{eq:CLV}). This approximation is justified as $1/K$ is small and $\sum_ix_i^2/B_{\mathrm{prey}}$ is also small because most prey species have a biomass smaller than one. 

\begin{figure}[t]
\begin{subfigure}{.49\textwidth}
  \centering
 		\def\svgwidth{\columnwidth}
    		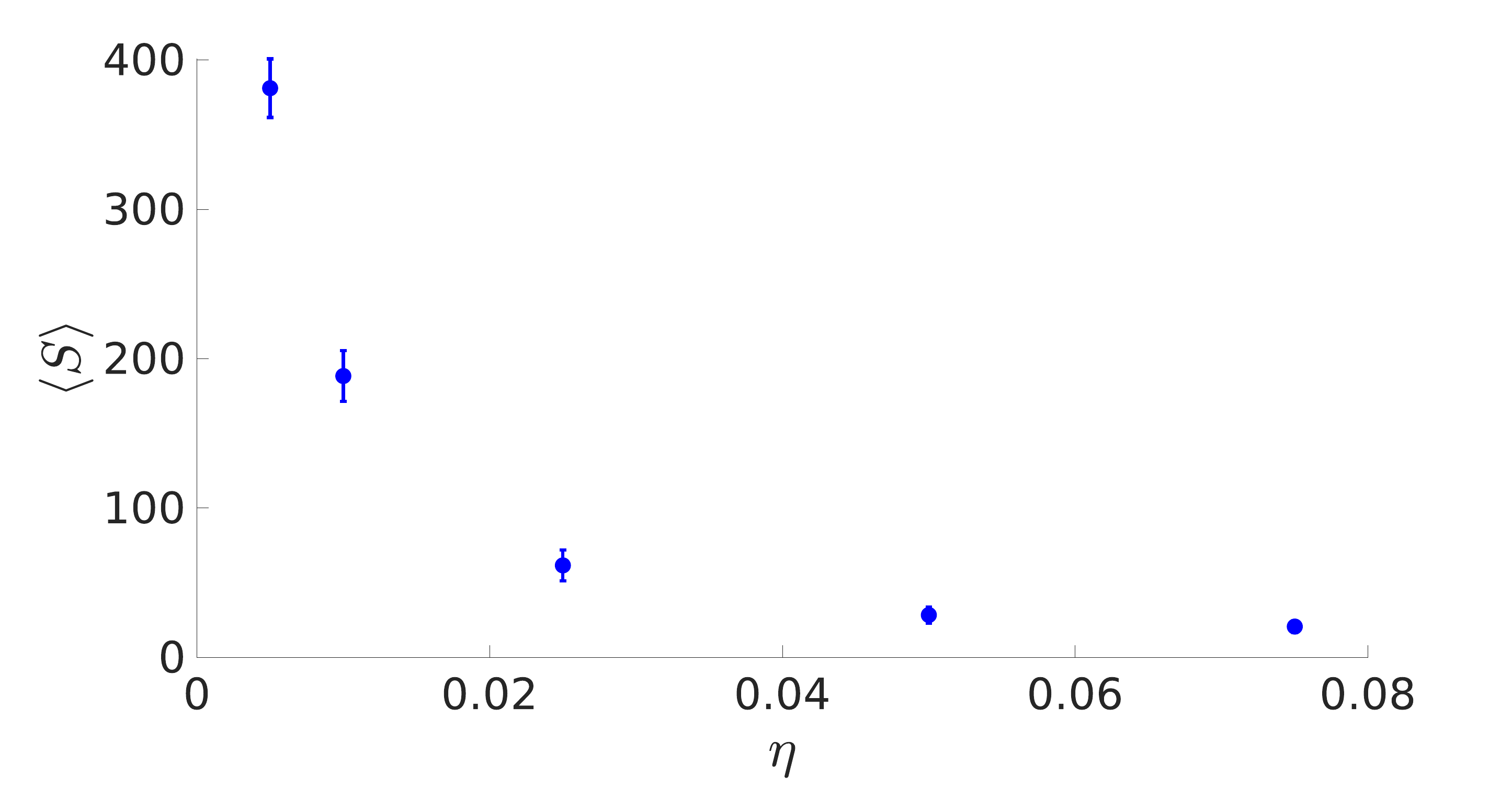
  \caption{}
  \label{fig:2layers:SvsEtaError}
\end{subfigure}
\begin{subfigure}{.49\textwidth}
  \centering
 		\def\svgwidth{\columnwidth}
    		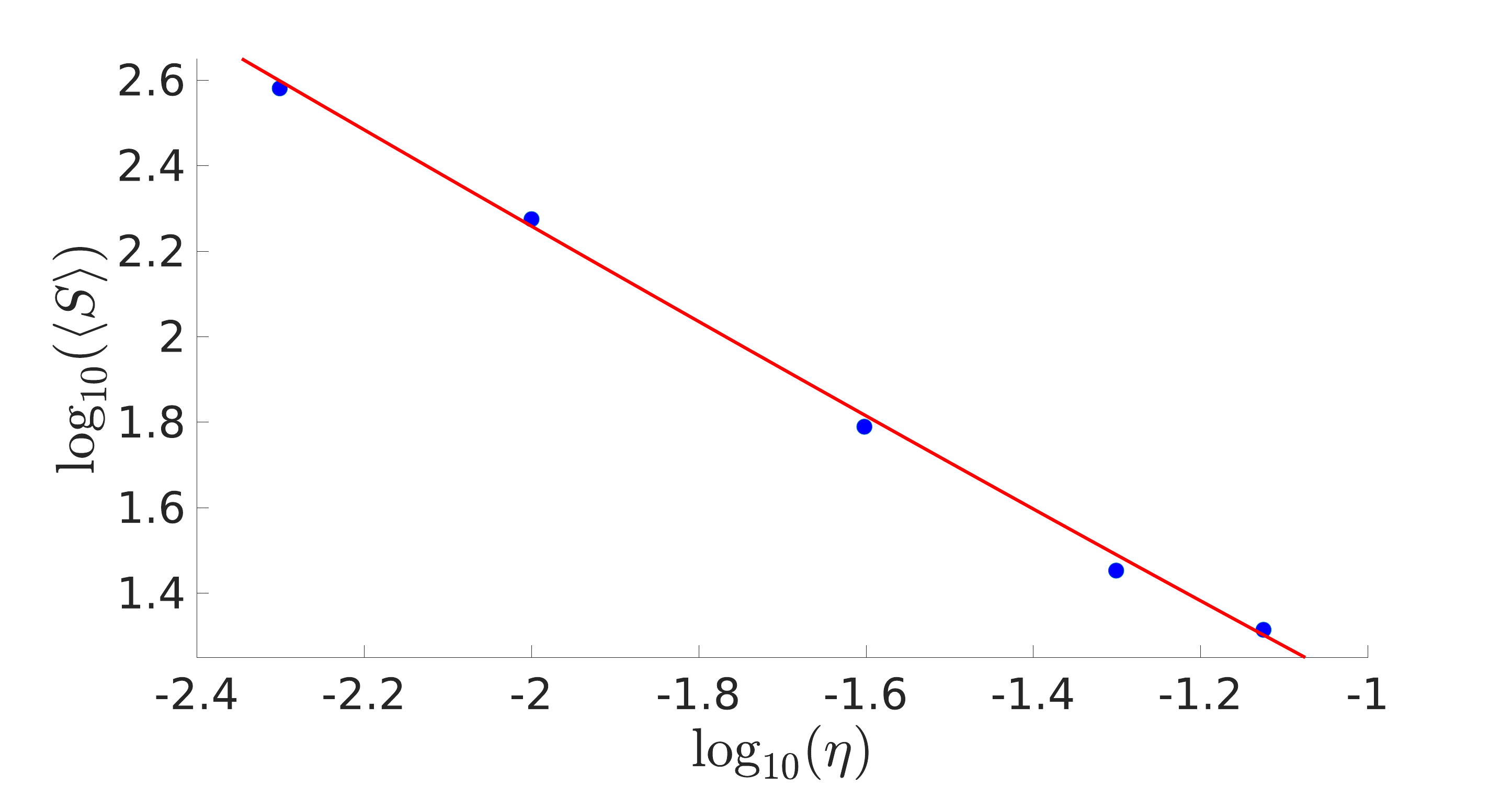
  \caption{}
  \label{fig:2layers:SvsEtaLog}
\end{subfigure}
\caption{Figure (a) shows the average number of alive species at the end of the simulation plus the standard deviation as a function of $\eta$. The average is taken over 60 iterations and the other parameters are equal to the previous figures. In figure (b) we show the same data on a log-log plot, and a linear fit shows that the data are well approximated by a power law $\langle S\rangle\sim \eta^{-1.1}$, as indicated by the red line with slope $-1.1$. }
\label{fig:2layers:SvsEta}
\end{figure}

\subsection{The role of variability}
In ecological modeling, it is a well-known fact that variability in the ecosystem is related to niche width. In a classic work by May and MacArthur~\cite{may1972niche}, it was shown that in a generalized LV model with parameters stochastically varying with variance $\sigma^2$, the niche width is inversely related to $\sigma^2$. Similar results have been found in~\cite{ackermann2004evolution}. In our setup, the parameters do not vary stochastically within a single species but vary stochastically with deviation $\eta$ between mother and daughter species. Furthermore, as a species is defined by the rows of $A$, there is no one-dimensional niche width. However, we still expect that smaller variation, i.e. smaller $\eta$, leads to more species in the system. To test this prediction, we compute for a range of $\eta$ values, the average number of predator and prey species $\langle S\rangle$. Indeed, as Fig.~\ref{fig:2layers:SvsEta} shows, an increase in the variability $\eta$ leads to a decrease in the number of coexisting species. Especially, the log-log plot in Fig.~\ref{fig:2layers:SvsEtaLog} shows that roughly $\langle S\rangle\sim\eta^{-1.1}$. Variability of speciation also affects the clustering ability; we will discuss this in \nameref{sec:2layers:eta}.

\subsection{Evolution in the growth rate $r$}
\label{sec:evor}
Until now, only the values of $P$ were affected during speciation, which resulted in the construction of large stable ecosystems. We can also allow for mutation in other parameters, such as the growth rate $r_i$ of the prey, to simulate resource competition between the prey species. In this case, we define the growth rate of a new prey species spawning from prey species $i$ as $r_i+\xi$ where $\xi$ is normally distributed with zero mean and standard deviation $\eta$ (i.e. $\xi \sim \mathcal{N}(0,\eta^2)$). Note that the same $\eta$ is used for the perturbation of $P$. 

\begin{figure}
\begin{subfigure}{.49\textwidth}
  \centering
 		\def\svgwidth{\columnwidth}
    		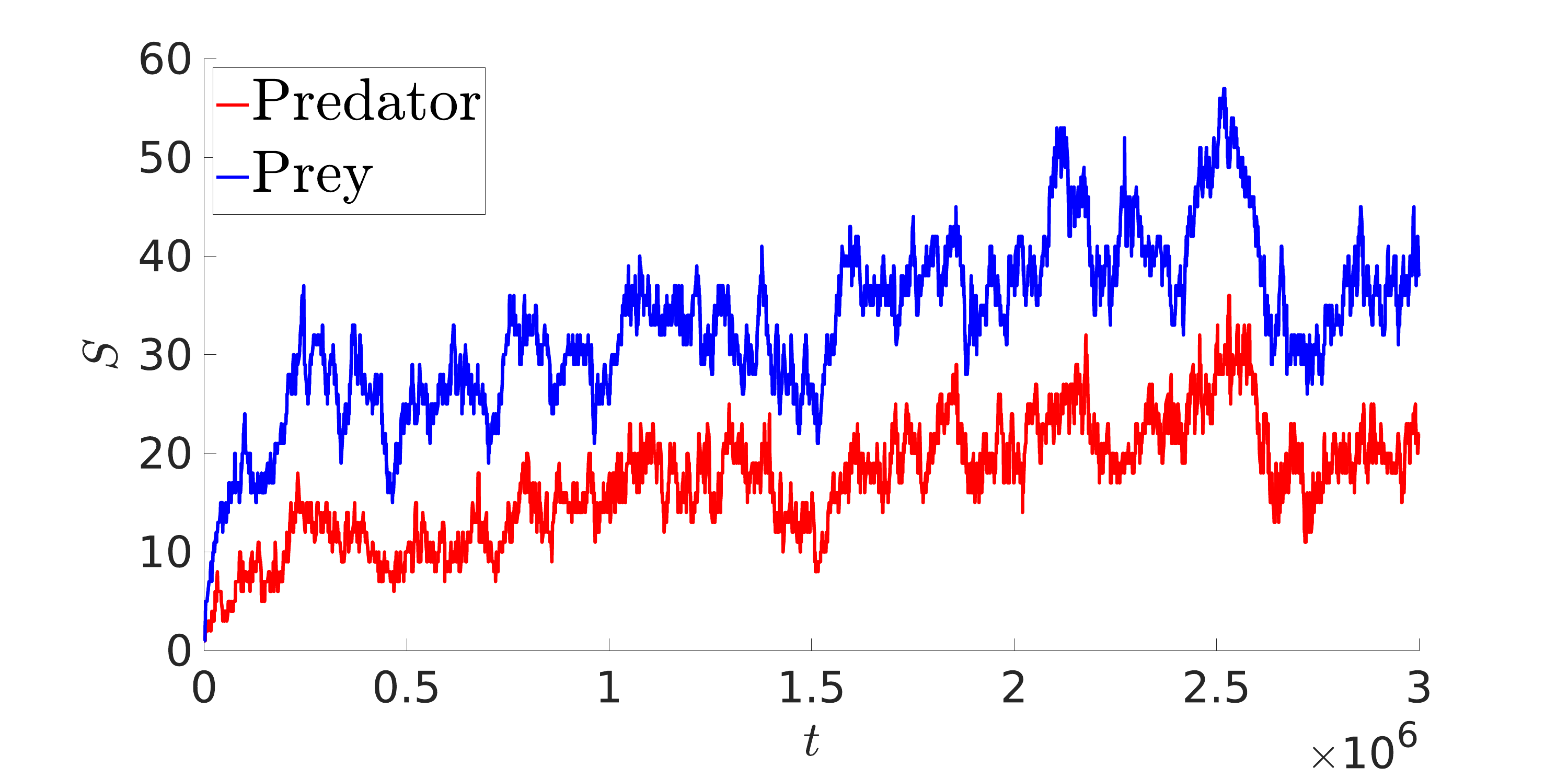
  \caption{}
  \label{fig:2layers:NewModelSvsTR}
\end{subfigure}
\begin{subfigure}{.49\textwidth}
  \centering
 		\def\svgwidth{\columnwidth}
    		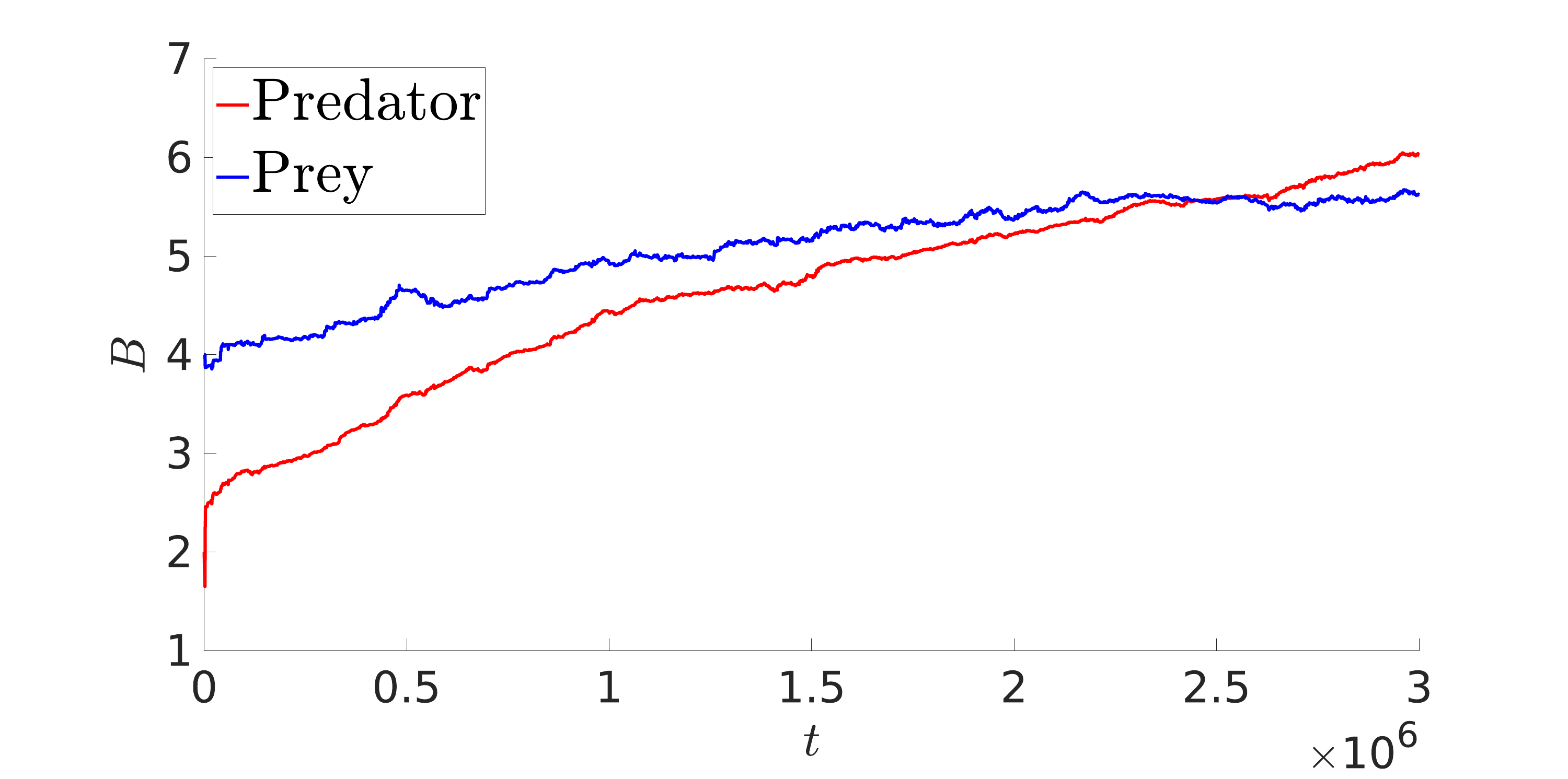
  \caption{}
  \label{fig:2layers:NewModelBvsTdelta}
\end{subfigure}
\caption{Simulation of the predator-prey LV model, with the adaptation that each newly spawned prey species has a slightly modified growth rate $r_i$. Figure~(a) shows the number of species versus time, while in Figure~(b) the biomass is shown. The biomass of the predators eventually overtakes the biomass of the prey. This is caused by the evolutionary growth of the growth rate~$r_i$, which allows predators to consume more. This apparently counterintuitive result is already present in the classic predator-prey model, where only the fixed point of the predator increases when $r$ is increased, see Eq.~\eqref{eq:CLV}. }
\label{fig:2layers:RSandB}
\end{figure}

\begin{figure}
\begin{subfigure}{.49\textwidth}
  \centering
 		\def\svgwidth{\columnwidth}
    		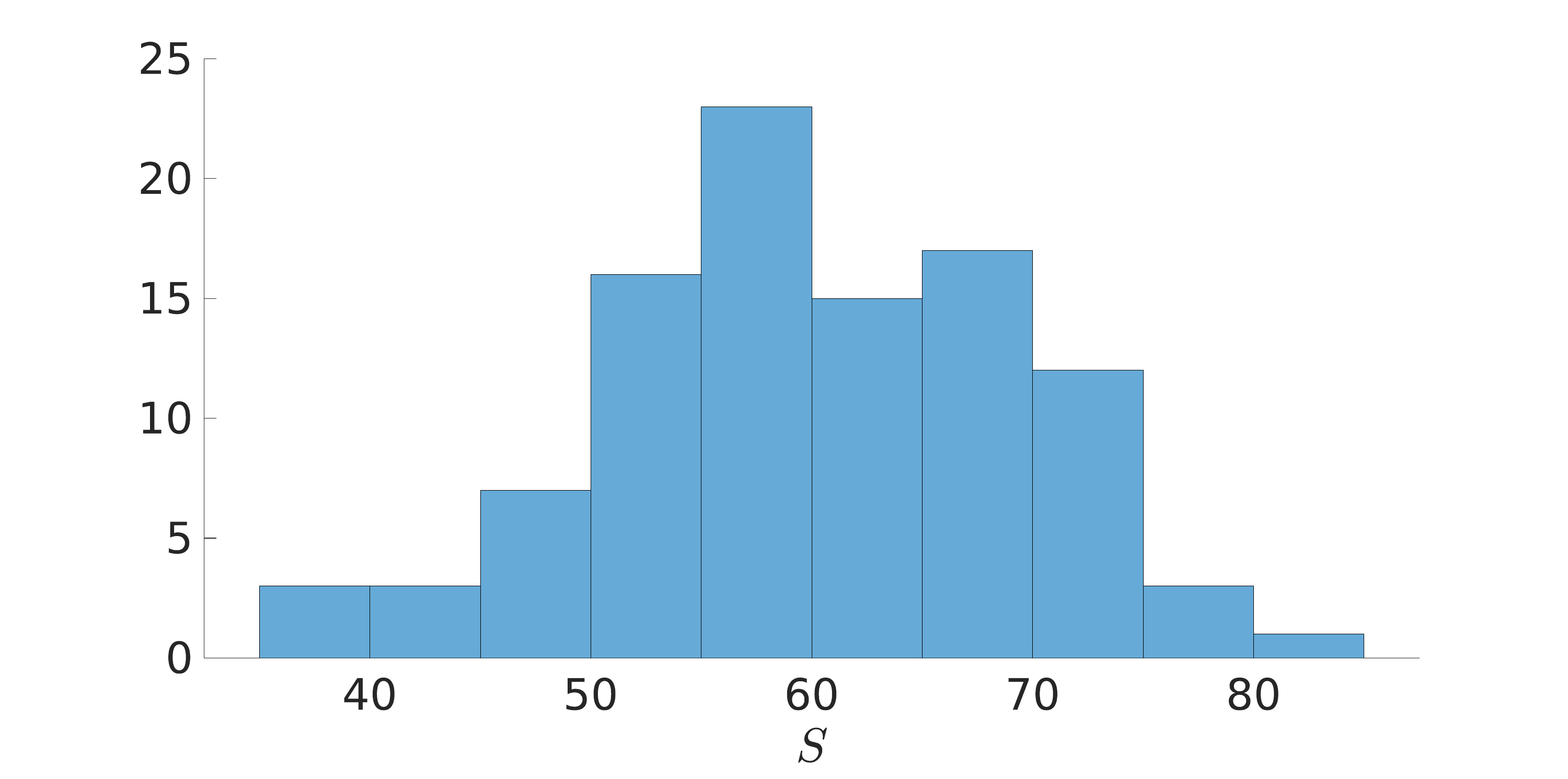
  \caption{}
  \label{fig:2layers:HistogramDelta}
\end{subfigure}
\begin{subfigure}{.49\textwidth}
  \centering
 		\def\svgwidth{\columnwidth}
    		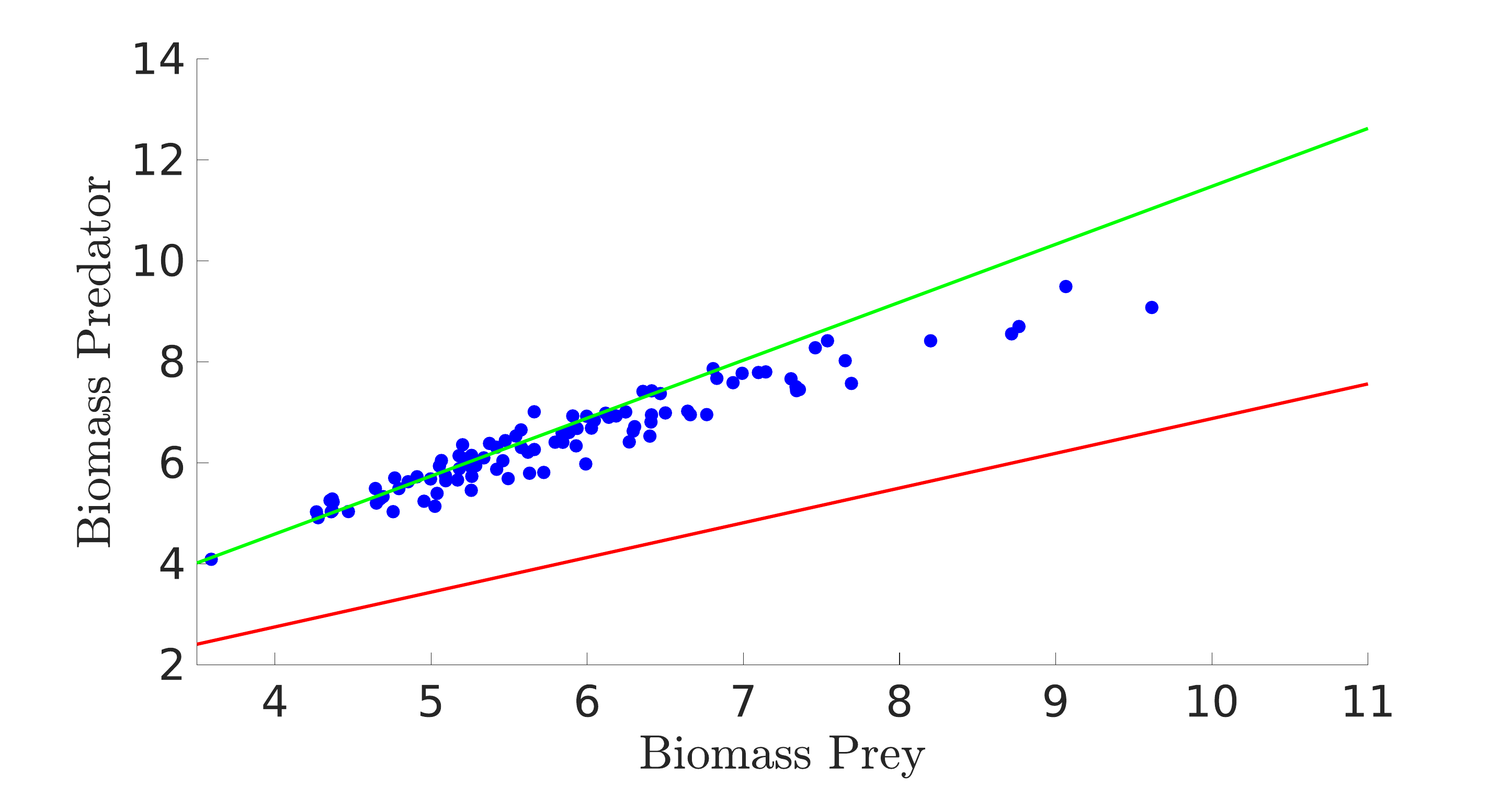
  \caption{}
  \label{fig:2layers:Bpvsrdelta}
\end{subfigure}
\caption{Similar simulation to Fig.~\ref{fig:2layers:HB}, but with  evolving growth rate $r_i$. Figure (a) shows the spread in the total number of species, while Figure (b) shows for all simulations the relation between the predator and prey biomass at the end of the simulation. The red line is the same linear approximation as in Fig.~\ref{fig:2layers:Bpvsr}, while for the green line, the (initial) growth rate $r_1=r$ has been replaced with the average of the growth rates $r_i$ for all species alive at the end of all the 100 simulations.}
\label{fig:2layers:HBdelta}
\end{figure}

We now briefly investigate the characteristics of the dynamics as was done earlier for the fixed~$r$ case. The first results are shown in Fig.~\ref{fig:2layers:RSandB}. Prey species with a higher growth rate $r_i$ have no direct advantage over the other prey species, as there is no explicit competition for resources. However, with a higher growth rate, the size of $x_i$ can become smaller before the derivative $\dot x_i$ becomes negative and the species goes extinct. Hence, over time the average value of $r$ will grow, see Fig.~\ref{fig:2layers:GrowthRAnc} in sec.~\nameref{sec:ClustWithR}. This in turn leads to more predation which explains why the biomass of the predators overtakes the biomass of the prey as shown in Fig.~\ref{fig:2layers:NewModelBvsTdelta}. On average, there are 59.58 species in the system, of which on average 58.79 remain after equilibration, see also Fig.~\ref{fig:2layers:HistogramDelta}. This average is smaller than the previous case with a fixed growth rate $r$, as there is now increased competition between the prey species. When we compare Fig.~\ref{fig:2layers:Bpvsrdelta} with Fig.~\ref{fig:2layers:Bpvsr}, we observe that the very strict correlation is broken up.  However, the correlation is still well approximated by Eq.~(\ref{eq:2layers:biomassratio}) when we replace $r$ with the average of $r_i$ over the living species at the end of the simulation, see the green line in Fig.~\ref{fig:2layers:Bpvsrdelta}. 

Summarizing, we observe that the speciating LV model can generate realistic dynamics on the level of species and ecosystems: the model achieves a dynamic equilibrium in which speciation events are intricately coupled to species size dynamics, yet in which the overall ecosystem size is at all times realistically balanced between predators and prey. Moreover, the variability $\eta$ indeed affects the number of species in the expected manner, and the dynamic equilibrium reached is not dependent on species that without speciation would go extinct. What thus becomes possible to ask are more specific questions about the dynamics of the structure of the ecosystem. One interesting and biologically relevant question is whether a meaningful development of ``clustering'' emerges from the random introduction of random new species. Clustering dynamics is therefore the topic of the next section.

\section{Formation of genealogical and functional clusters}
\label{sec:2layers:drive}
Now that we have established that the speciating LV model can create large stable ecosystems, we study the structure of the community. Is it for example possible that all species are effectively copies of each other? That is to say, all strategies are clustered around an (small number of) optimal strategies as in~\cite{caetano2021evolution,bellavere2023speciation}? Or, in contrast, does speciation results in isolated clusters connected purely on the explicitly modeled mother-daughter interaction as in~\cite{shtilerman2015emergence} (see also the Appendix)? From real-world data, we know that the strength of correlations between two properties of the species varies significantly depending on the genealogical distance between two species~\cite{graham2018phylogenetic}, which we will test in our speciating LV model. 

There are many ways to cluster data; here we will test five different approaches to find that all of them indicate that the random speciation induces, on longer time scales, a clustering of species that are neither just copies of each other, nor are purely based on their ancestral relations. Instead, we shall see that our speciating LV model with two trophic layers induces a clustering of evolving species in which branches of clusters can emerge. Evidence for clustering emerges from the individual clustering methods used, but also from the consistency of the results between different clustering methods.

The first method to cluster applies to the setting with a constant growth rate $r$. In this context, a species is defined by the rows and columns of~$P$, as all the other parameters are constant. Therefore, we can for two species only measure their genealogical distance as defined in sec.~\nameref{sec:setup} and their phenomenological distance, as defined for prey species by
\begin{equation}
\label{eq:DistPrey}
    d(i,\ell)=\sum_k|P_{ik}-P_{\ell k}|^2,
\end{equation}
 and for predators by
\begin{equation}
    d(j,\ell)=\sum_k|P^T_{jk}-P^T_{\ell k}|^2.
\end{equation}
For the prey species, this function $d$ defines a distance in an $S_{\mathrm{pred}}$-dimensional space, hence a clustering algorithm like \textit{kmeans} can be used to define clusters in this $S_{\mathrm{pred}}$-dimensional space. For the predator species, the same holds but then with an $S_{\mathrm{prey}}$-dimensional space. We will call this \emph{kmeans-distance clustering} (KDC) and note that one clustering input parameter is the number of clusters one demands it to find. See Appendix~\ref{sec:AppB} for a discussion on the different settings of this algorithm.

The same type of algorithm cannot be used for clustering on the genealogical tree, as the genealogical distance is defined along the tree. This leads us to test a second, hierarchical clustering algorithm: Matlab's \texttt{cluster}~\cite{matlabstat}. Here, we choose to cluster based on a cut-off in genealogical distance. We will call this clustering \emph{genealogical-distance clustering} (GDC). Note that both kmeans and genealogical clustering approaches only result in clusters within the two trophic layers, i.e. within the prey species and within the predator species. 

To study clusters within the total predator-prey community, we can apply a modularity algorithm~\cite{blondel2008fast} to the full interaction matrix $A$. This approach will group predator and prey species that have strong interactions. We call this clustering approach \emph{interaction-modularity clustering} (IMC).  

Does the choice of clustering algorithm affect the clustering observation? We test for similarity as any clustering algorithm is in essence heuristic: different algorithms or similar algorithms with different parameters, result in different clusters. We will first explore this question for a fixed growth rate $r$. Clustering analysis with the aforementioned methods can be done at a given stage of the evolutionary process, but the high and constantly changing dimension of the parameter space makes it difficult to track the emergence of clusters over time.  To visualize the emergence of functional clusters, we study the case with evolution in the growth rate $r$ of the prey species in sec.~\nameref{sec:ClustWithR}.

\subsection{Clustering with fixed growth rate: Testing KDC, GDC and IMC algorithms}
\label{sec:2layers:KG}
\begin{figure}[t]
\begin{subfigure}{.49\textwidth}
  \centering
 		\def\svgwidth{\columnwidth}
    		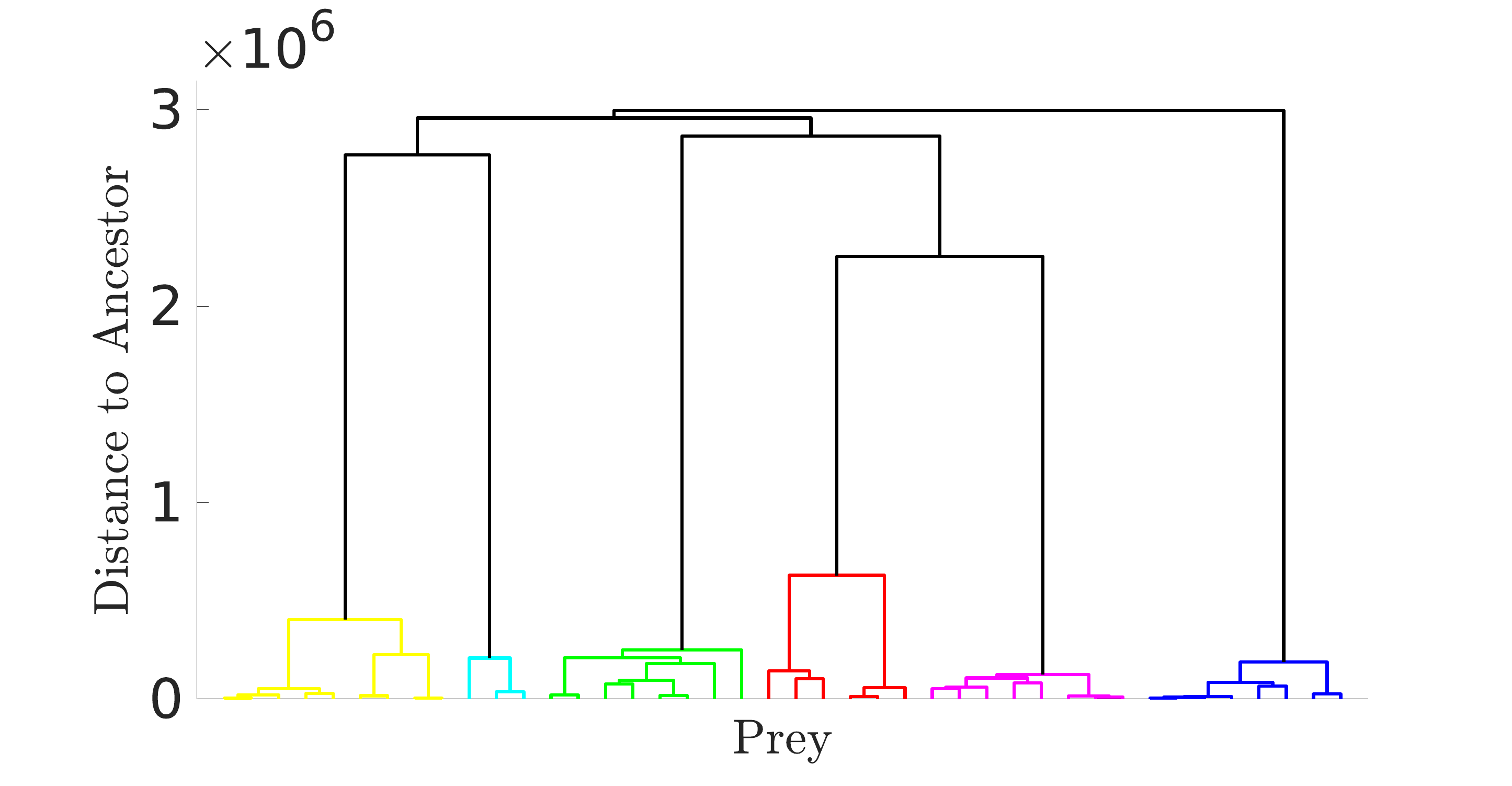
  \caption{}
  \label{fig:2layers:dendrogramNoDeltaP}
\end{subfigure}
\begin{subfigure}{.49\textwidth}
  \centering
 		\def\svgwidth{\columnwidth}
    		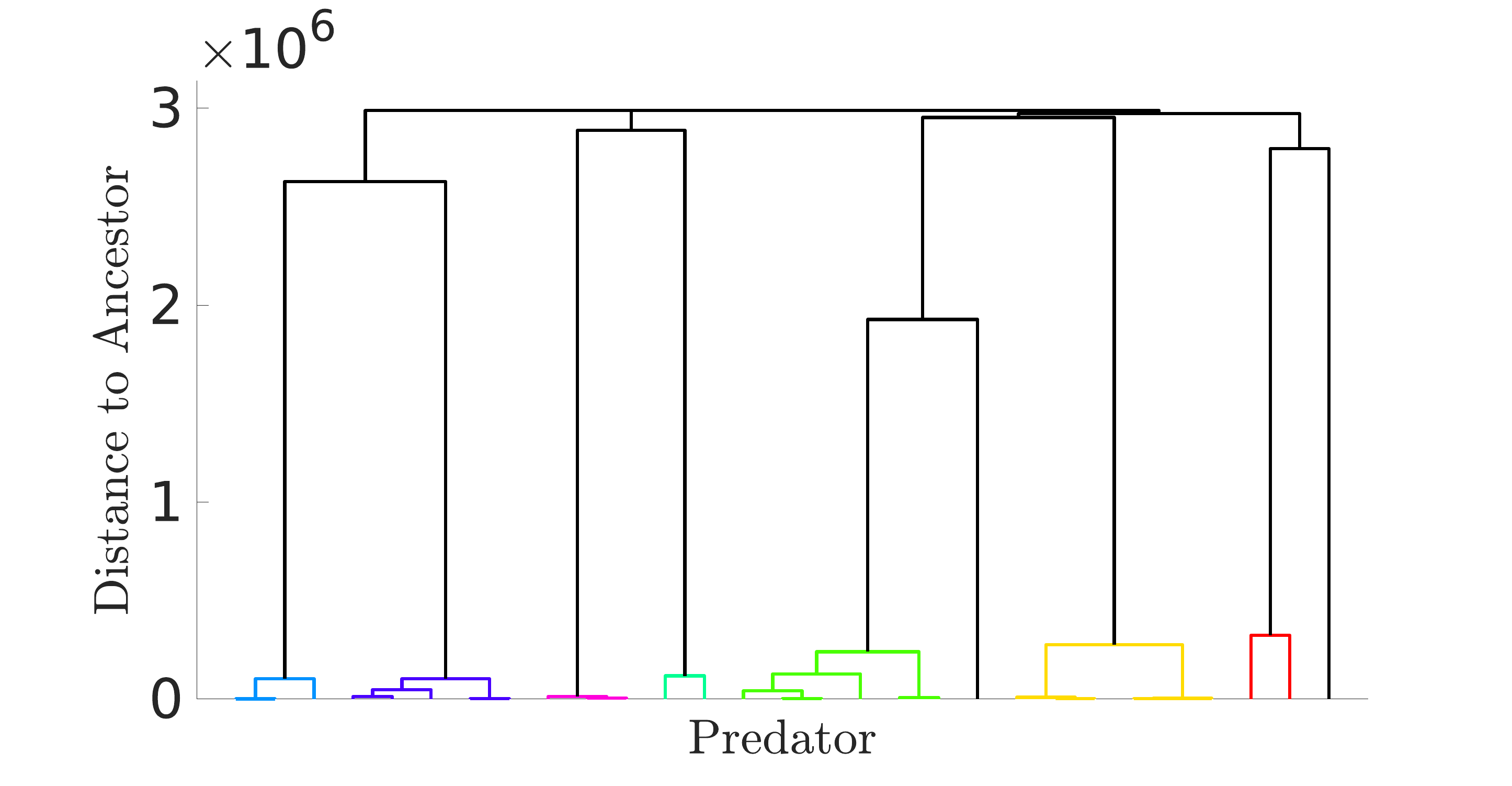
  \caption{}
  \label{fig:2layers:dendrogramNoDeltaR}
\end{subfigure}
\caption{Genealogical trees for the predator and prey species in a simulation with the same settings as Fig.~\ref{fig:2layers}. Clusters of species are colored together when the distance to their common ancestor is less than $10^6$. For this specific iteration of the stochastic model, the clustering is seemingly obvious.}
\label{fig:2layers:dendrogramNoDelta}
\end{figure}

In Fig.~\ref{fig:2layers:dendrogramNoDelta}, we show the genealogical structures of the predator and prey species for the same parameter setting as the simulations in sec.~\nameref{sec:2layers}. Only the species alive at the end of the simulation are shown; not the full tree with extinct species. Notice that there are six respectively nine well-defined clusters found by GDC when we take as genealogical distance cut-off $t=10^6$. These clusters are persistent; a large range of different cut-off values around $t=10^6$ would result in the same clusters. We find the same clusters when we apply KDC to the phylogenetic distance and force it to find six respectively nine clusters, showing the consistency of the two clustering approaches. However, with the KDC, we can further examine the similarity within the clusters with the silhouette criterion~\cite{kaufman2009finding}. Based on this criterion, the optimal number of clusters is five respectively eight: the application of the silhouette criterion merged two clusters into one.

\begin{figure}[t]
\begin{subfigure}{.49\textwidth}
  \centering
 		\def\svgwidth{\columnwidth}
\begingroup%
  \makeatletter%
  \providecommand\color[2][]{%
    \errmessage{(Inkscape) Color is used for the text in Inkscape, but the package 'color.sty' is not loaded}%
    \renewcommand\color[2][]{}%
  }%
  \providecommand\transparent[1]{%
    \errmessage{(Inkscape) Transparency is used (non-zero) for the text in Inkscape, but the package 'transparent.sty' is not loaded}%
    \renewcommand\transparent[1]{}%
  }%
  \providecommand\rotatebox[2]{#2}%
  \newcommand*\fsize{\dimexpr\f@size pt\relax}%
  \newcommand*\lineheight[1]{\fontsize{\fsize}{#1\fsize}\selectfont}%
  \ifx\svgwidth\undefined%
    \setlength{\unitlength}{1387.5bp}%
    \ifx\svgscale\undefined%
      \relax%
    \else%
      \setlength{\unitlength}{\unitlength * \real{\svgscale}}%
    \fi%
  \else%
    \setlength{\unitlength}{\svgwidth}%
  \fi%
  \global\let\svgwidth\undefined%
  \global\let\svgscale\undefined%
  \makeatother%
  \begin{picture}(1,0.51945946)%
    \lineheight{1}%
    \setlength\tabcolsep{0pt}%
    \put(0,0){\includegraphics[width=\unitlength,page=1]{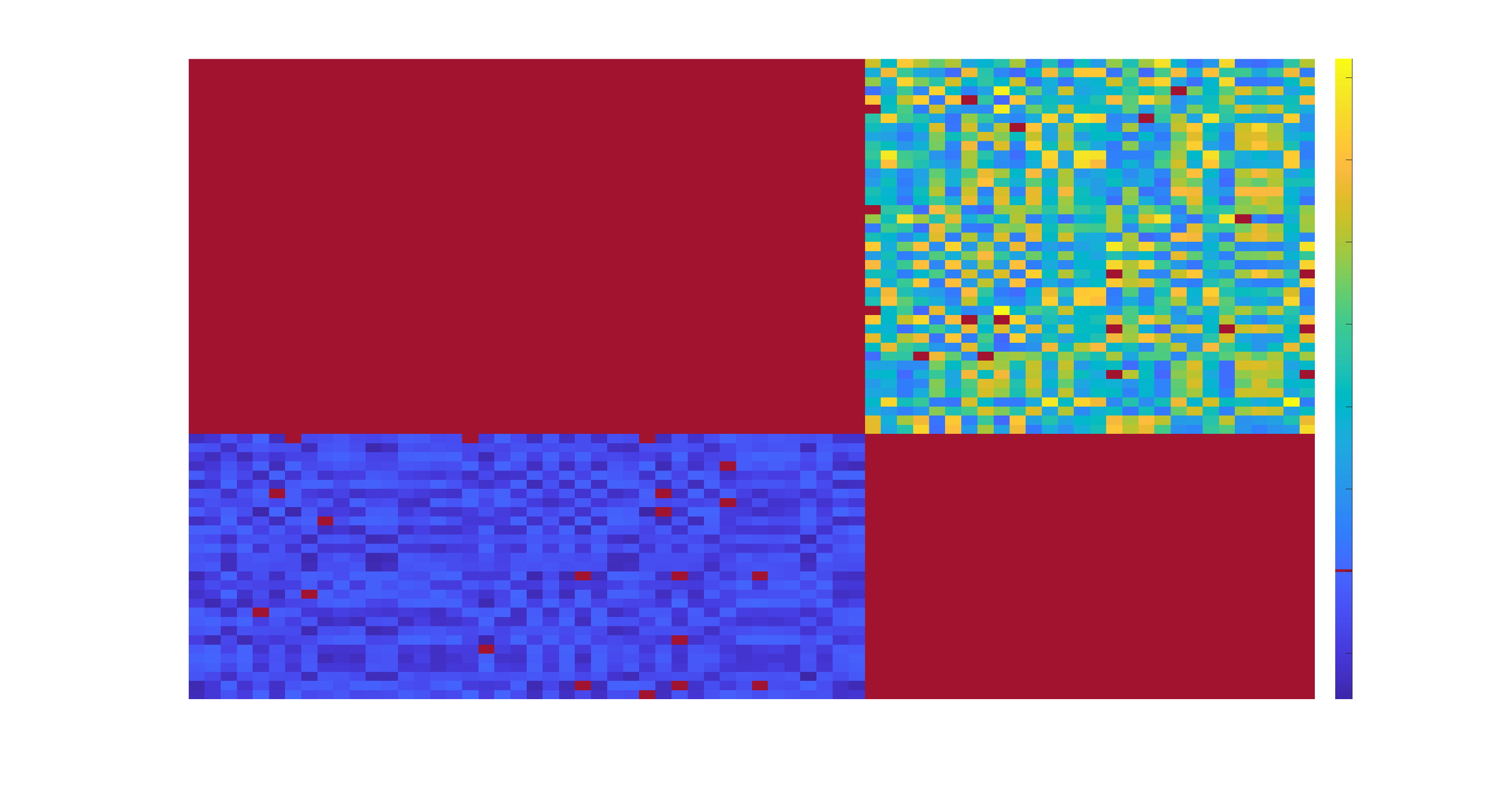}}%
    \put(0.90122524,0.07868097){\makebox(0,0)[lt]{\lineheight{1.25}\smash{\begin{tabular}[t]{l}-0.1\end{tabular}}}}%
    \put(0.90122524,0.13302703){\makebox(0,0)[lt]{\lineheight{1.25}\smash{\begin{tabular}[t]{l}0\end{tabular}}}}%
    \put(0.90122524,0.18737308){\makebox(0,0)[lt]{\lineheight{1.25}\smash{\begin{tabular}[t]{l}0.1\end{tabular}}}}%
    \put(0.90122524,0.24171914){\makebox(0,0)[lt]{\lineheight{1.25}\smash{\begin{tabular}[t]{l}0.2\end{tabular}}}}%
    \put(0.90122524,0.29606514){\makebox(0,0)[lt]{\lineheight{1.25}\smash{\begin{tabular}[t]{l}0.3\end{tabular}}}}%
    \put(0.90122524,0.35041119){\makebox(0,0)[lt]{\lineheight{1.25}\smash{\begin{tabular}[t]{l}0.4\end{tabular}}}}%
    \put(0.90122524,0.40475724){\makebox(0,0)[lt]{\lineheight{1.25}\smash{\begin{tabular}[t]{l}0.5\end{tabular}}}}%
    \put(0.90122524,0.4591033){\makebox(0,0)[lt]{\lineheight{1.25}\smash{\begin{tabular}[t]{l}0.6\end{tabular}}}}%
    \put(0,0){\includegraphics[width=\unitlength,page=2]{ModularityNoDelta.pdf}}%
  \end{picture}%
\endgroup%

  \caption{}
  \label{fig:2layers:ModularityNoDeltaU}
\end{subfigure}
\begin{subfigure}{.49\textwidth}
  \centering
 		\def\svgwidth{\columnwidth}
\begingroup%
  \makeatletter%
  \providecommand\color[2][]{%
    \errmessage{(Inkscape) Color is used for the text in Inkscape, but the package 'color.sty' is not loaded}%
    \renewcommand\color[2][]{}%
  }%
  \providecommand\transparent[1]{%
    \errmessage{(Inkscape) Transparency is used (non-zero) for the text in Inkscape, but the package 'transparent.sty' is not loaded}%
    \renewcommand\transparent[1]{}%
  }%
  \providecommand\rotatebox[2]{#2}%
  \newcommand*\fsize{\dimexpr\f@size pt\relax}%
  \newcommand*\lineheight[1]{\fontsize{\fsize}{#1\fsize}\selectfont}%
  \ifx\svgwidth\undefined%
    \setlength{\unitlength}{1387.5bp}%
    \ifx\svgscale\undefined%
      \relax%
    \else%
      \setlength{\unitlength}{\unitlength * \real{\svgscale}}%
    \fi%
  \else%
    \setlength{\unitlength}{\svgwidth}%
  \fi%
  \global\let\svgwidth\undefined%
  \global\let\svgscale\undefined%
  \makeatother%
  \begin{picture}(1,0.51945946)%
    \lineheight{1}%
    \setlength\tabcolsep{0pt}%
    \put(0,0){\includegraphics[width=\unitlength,page=1]{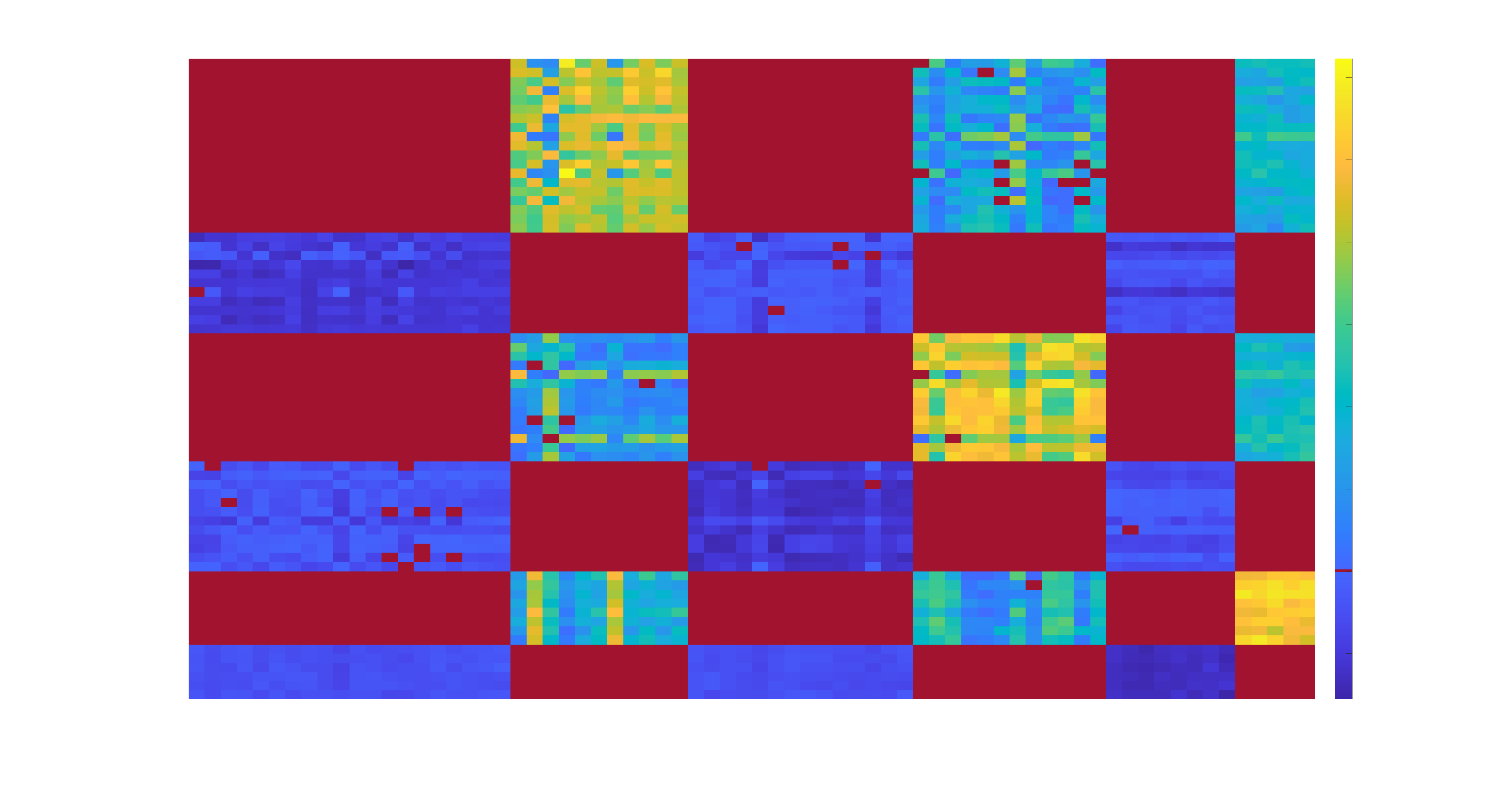}}%
    \put(0.90122524,0.07868097){\makebox(0,0)[lt]{\lineheight{1.25}\smash{\begin{tabular}[t]{l}-0.1\end{tabular}}}}%
    \put(0.90122524,0.13302703){\makebox(0,0)[lt]{\lineheight{1.25}\smash{\begin{tabular}[t]{l}0\end{tabular}}}}%
    \put(0.90122524,0.18737308){\makebox(0,0)[lt]{\lineheight{1.25}\smash{\begin{tabular}[t]{l}0.1\end{tabular}}}}%
    \put(0.90122524,0.24171914){\makebox(0,0)[lt]{\lineheight{1.25}\smash{\begin{tabular}[t]{l}0.2\end{tabular}}}}%
    \put(0.90122524,0.29606514){\makebox(0,0)[lt]{\lineheight{1.25}\smash{\begin{tabular}[t]{l}0.3\end{tabular}}}}%
    \put(0.90122524,0.35041119){\makebox(0,0)[lt]{\lineheight{1.25}\smash{\begin{tabular}[t]{l}0.4\end{tabular}}}}%
    \put(0.90122524,0.40475724){\makebox(0,0)[lt]{\lineheight{1.25}\smash{\begin{tabular}[t]{l}0.5\end{tabular}}}}%
    \put(0.90122524,0.4591033){\makebox(0,0)[lt]{\lineheight{1.25}\smash{\begin{tabular}[t]{l}0.6\end{tabular}}}}%
    \put(0,0){\includegraphics[width=\unitlength,page=2]{ModularityNoDeltaS.pdf}}%
  \end{picture}%
\endgroup%

  \caption{}
  \label{fig:2layers:ModularityNoDeltaS}
\end{subfigure}
\caption{Figure~(a) shows the interaction matrix $A$ with zero entries corresponding to the color red. The block structure with $P$ and $-\beta P^T$ is clearly visible. In Figure~(b), the matrix is re-ordered based on modularity using IMC, resulting in three clusters of predators and prey with strong interactions.}
\label{fig:2layers:ModularityNoDelta}
\end{figure}

The trees in Fig.~\ref{fig:2layers:dendrogramNoDelta} give no information on the relations between predator and prey and how they evolve. To show the interactions, we plot the full interaction matrix $A$ in Fig.~\ref{fig:2layers:ModularityNoDelta}, both sorted and unsorted. The sorting in Fig.~\ref{fig:2layers:ModularityNoDeltaS} is done using IMC that groups predator and prey species with strong interactions. This results in three groups of strongly interacting predators and prey.  

\subsection{Variability in speciation affects clustering}
\label{sec:2layers:eta}
\begin{figure}[tb]
\begin{subfigure}{.49\textwidth}
  \centering
 		\def\svgwidth{\columnwidth}
    		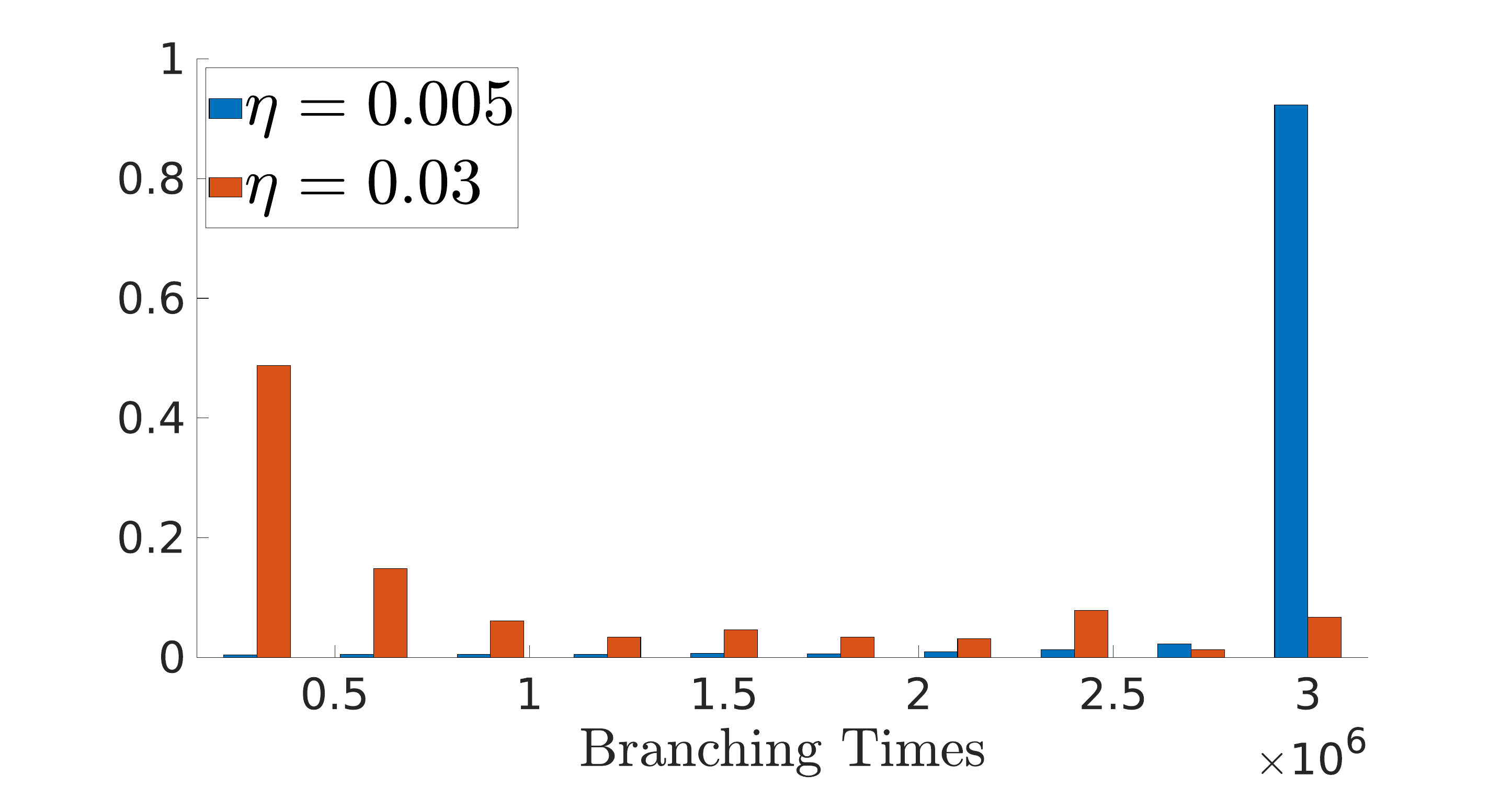
  \caption{}
  \label{fig:2layers:HistAvNorm}
\end{subfigure}
\begin{subfigure}{.49\textwidth}
  \centering
 		\def\svgwidth{\columnwidth}
    		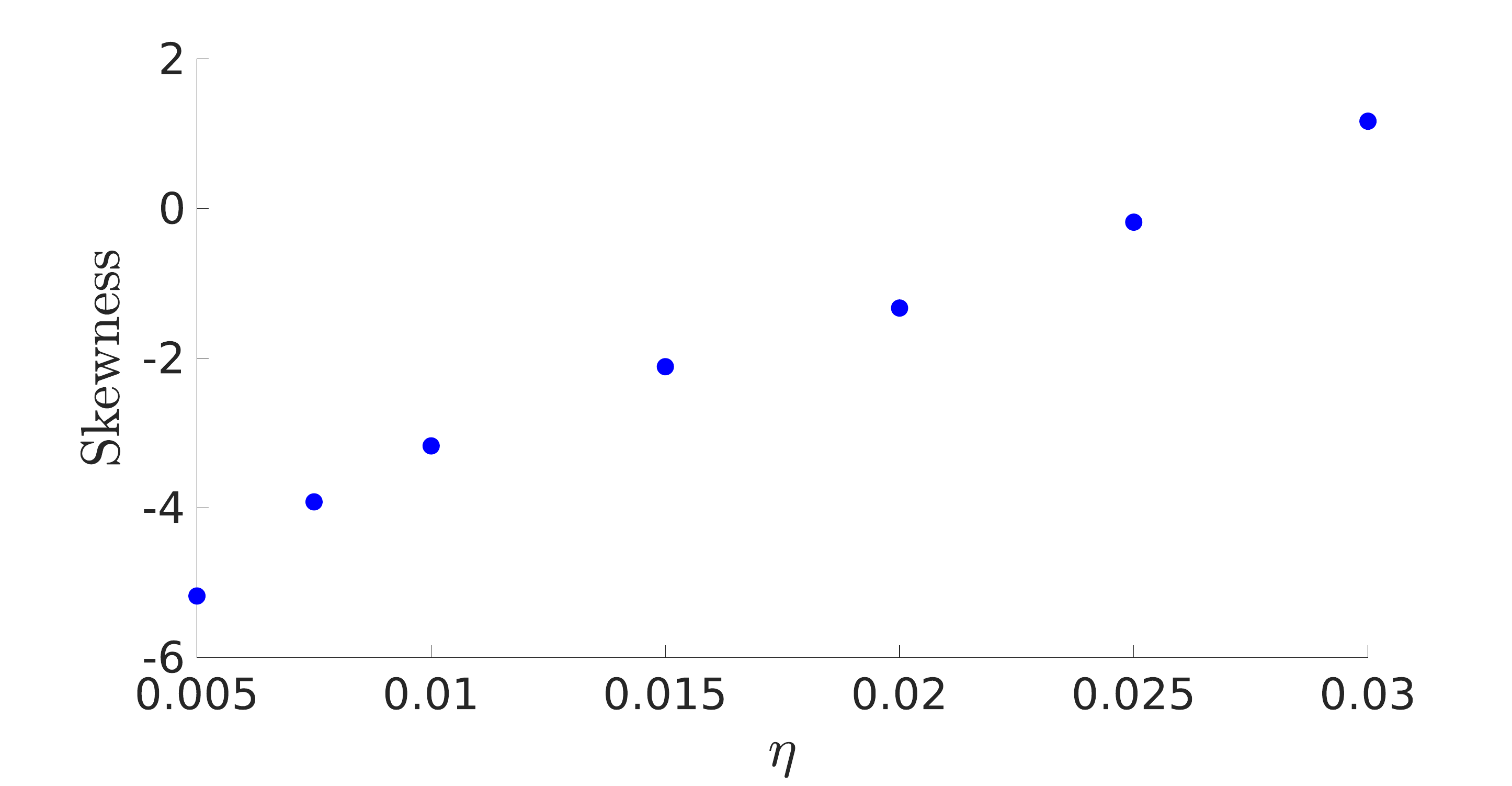
  \caption{}
  \label{fig:2layers:skewnewss}
\end{subfigure}
\caption{Figure~(a) shows two averaged normalized histograms of branching times for the prey species for a large and small variability $\eta$. The time interval $[0,3\cdot 10^6]$ has been subdivided into 10 equal bins. First, the histograms are averaged over 12 iterations and subsequently normalized. The normalization is necessary as the number of branch points differs significantly between large and small $\eta$. Both histograms are asymmetric, which can be quantified using the skewness of the distribution. Figure~(b) shows the skewness for a range of $\eta$ values, indicating that the mass of the distribution indeed switches from long to short branching times as $\eta$ increases.}
\label{fig:treestruc}
\end{figure}

Determining clustering distances with a cut-off is not always as clear as in Fig.~\ref{fig:2layers:dendrogramNoDelta}. Splitting of branches in the range of $[10^6, 1.5\cdot 10^6]$ is not as common as outside this region, but certainly happens. The value of $\eta$ is particularly important to consider for the branching. When $\eta$ is decreased and the number of species hence increases (see Fig.~\ref{fig:2layers:SvsEta}), the number of branches also increases, which in turn makes it harder to distinguish between different clusters. However, this is not the only factor at play. To examine the ability of the model parameters to change the structure in the genealogical tree, we run simulations twelve times for a range of $\eta$ values and make a normalized histogram of the branching times. 
On the one hand, as Fig.~\ref{fig:2layers:HistAvNorm} shows, for large $\eta$ values (indicated in red), the majority of the branching times are close to the end of the simulation, indicating that most of the species are in a small number of clusters and many other branches went extinct. On the other hand, for small~$\eta$ (indicated in blue), many branches survive as indicated by the high peak at long branching times. To evaluate this observation systematically, we compute the skewness, i.e. a measure of the asymmetry, of the histograms as shown in Fig.~\ref{fig:2layers:HistAvNorm}. The result is shown in Fig.~\ref{fig:2layers:skewnewss}. Indeed, the skewness increases with the variability $\eta$, indicating that the mass of the histograms switches from left to right, i.e. from branching times at the beginning of the simulation to branching times at the end. Hence, the strength of the evolution is shown in the assembly of the community, which is supported by data~\cite{oliveira2016species}.  

The results from Fig.~\ref{fig:treestruc} together with the results from Fig.~\ref{fig:2layers:SvsEta} explain our default choice of $\eta=0.02$. To study branching and clustering in a stochastic model we need sufficient statistics for both. We thus aim to be in an intermediate noise level as we need many speciation events, i.e. small $\eta$. However, we also need sufficient statistics on the clustering observations, i.e. high $\eta$. We found that for values of $\eta$ around $0.02$, the distribution of branching times is fairly symmetric and the minimum of the histogram is around the middle of the time interval $[0, 3\cdot 10^6]$, indicating that the cluster formation as in Fig.~\ref{fig:2layers:dendrogramNoDelta} is typical for this $\eta$ value. 

\begin{figure}[t]
\begin{subfigure}{.49\textwidth}
  \centering
 		\def\svgwidth{\columnwidth}
    		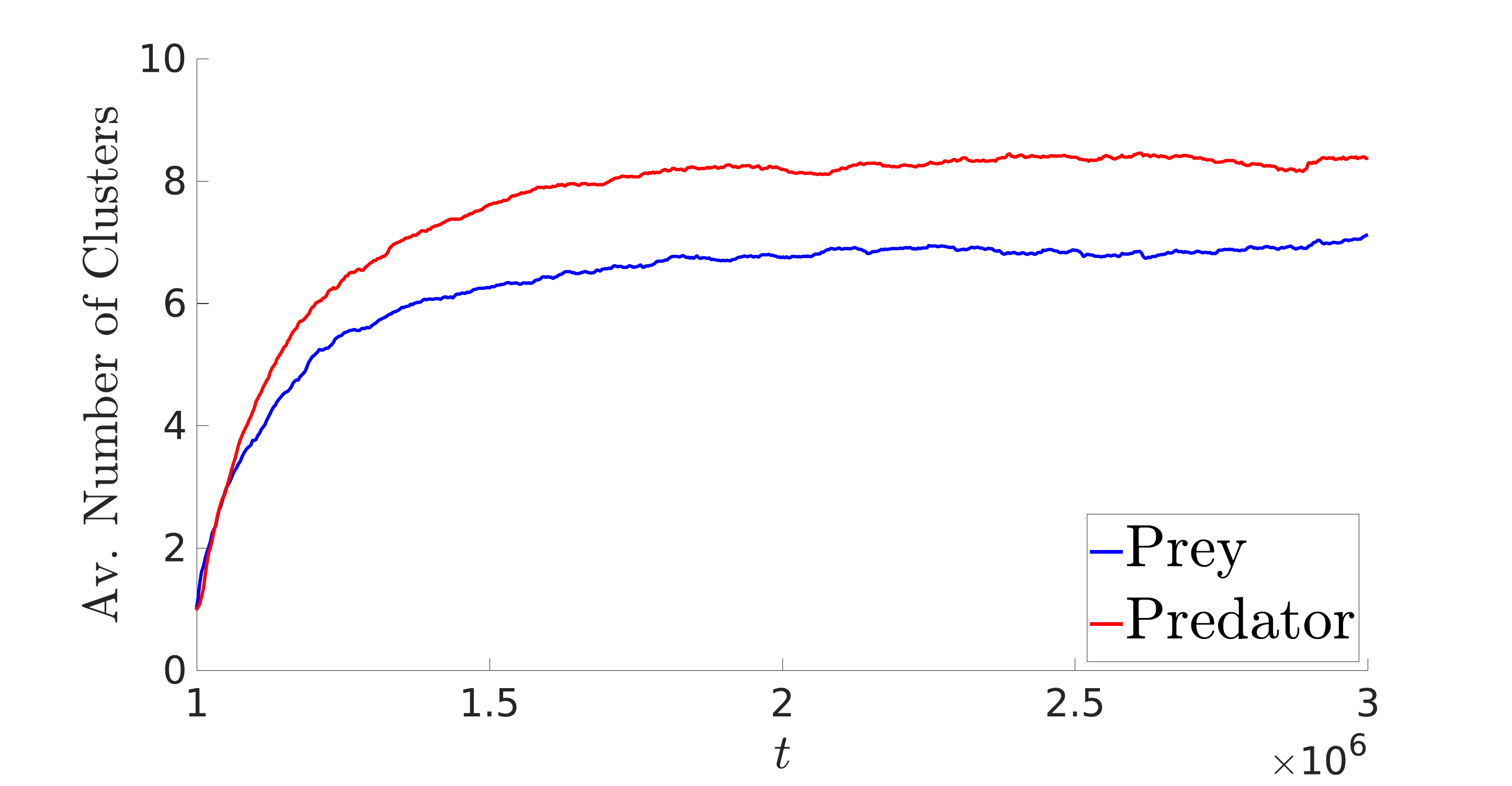
  \caption{}
  \label{fig:2layers:ClustersvsT}
\end{subfigure}
\begin{subfigure}{.49\textwidth}
  \centering
 		\def\svgwidth{\columnwidth}
    		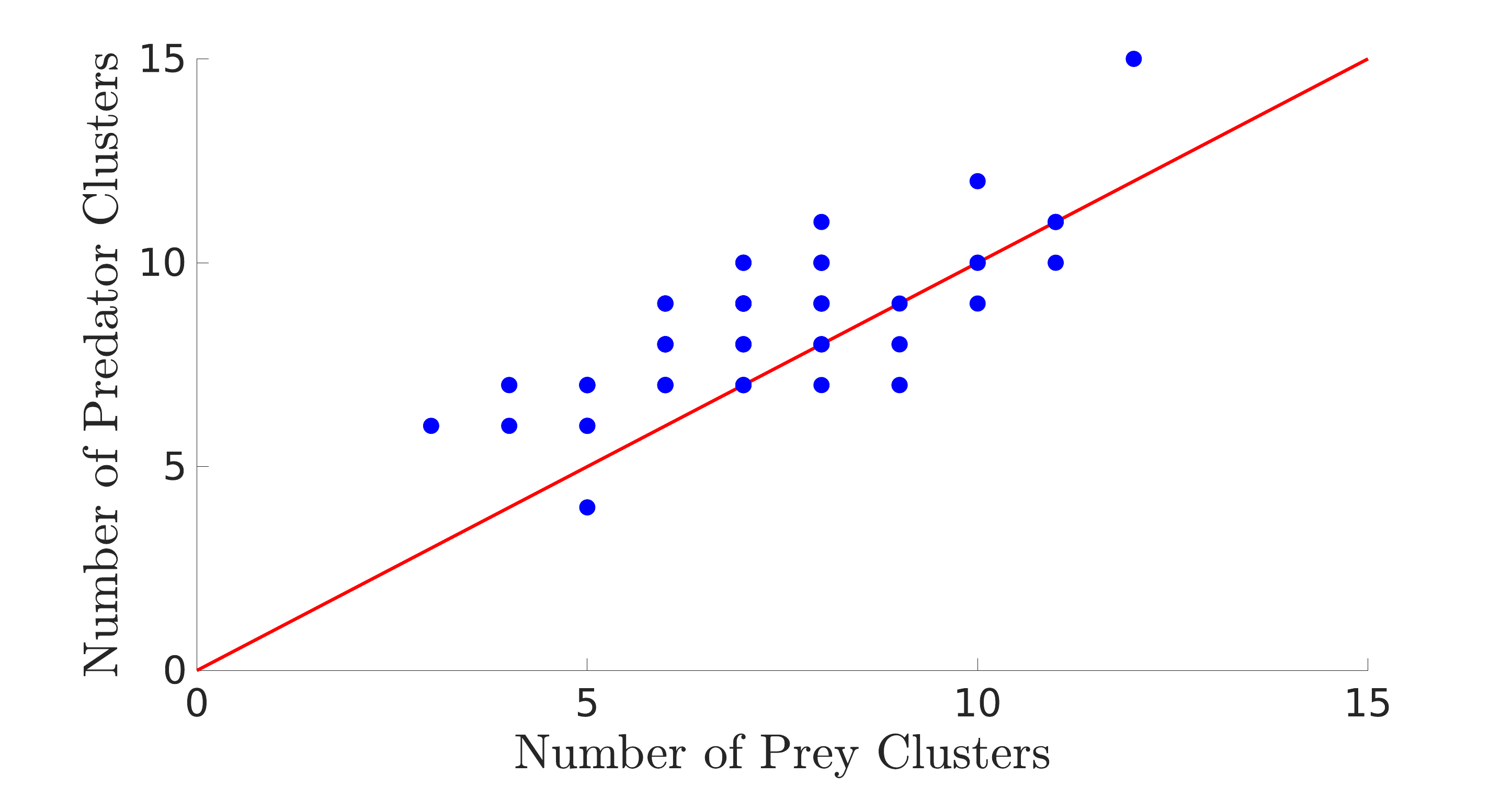
  \caption{}
  \label{fig:2layers:ClustersScatter}
\end{subfigure}
\caption{Figure~(a) shows the evolution of the number of clusters using GDC over time, averaged over 100 runs. The cut-off in the clustering algorithm is fixed at $10^6$, meaning that the number of clusters is always 1 for $t<10^6$. Figure~(b) shows for the final timestep of each simulation the number of predator clusters versus prey clusters and the red line indicates the diagonal. We observe that all pairs lie in a narrow band. }
\label{fig:2layers:dendrogramclust}
\end{figure}

We can also show that using a fixed genealogical cut-off is not an important choice in the clustering. In fact, a given distance cut-off even allows one to track the number of clusters in both predator and prey as a function of time, starting from the minimum cut-off distance. We show this by running the same simulation for 100 iterations and track how the number of clusters evolves over time, using the same distance metric for the clustering on the genealogical tree on all runs, and then compute the cluster statistics. The results on the average number of clusters observed ``in the past'' are shown in Fig.~\ref{fig:2layers:ClustersvsT}. Because we use $t=10^6$ as a minimum distance for the clusters, the number of clusters below this value is always one. Over time, the number of clusters for predator and prey species follow each other quite closely and the number is around six at the end of the simulation for prey, and about eight for predator. Just like the total biomass of predator and prey are correlated, so are the statistics of predator and prey clusters. Fig.\ref{fig:2layers:ClustersScatter} shows that the relation between the number of predator and prey clusters is limited to a narrow band in phase space.

\subsection{Robust clustering with evolution in the growth rate}
\label{sec:ClustWithR}
\begin{figure}
\begin{subfigure}{.49\textwidth}
  \centering
 		\def\svgwidth{\columnwidth}
    		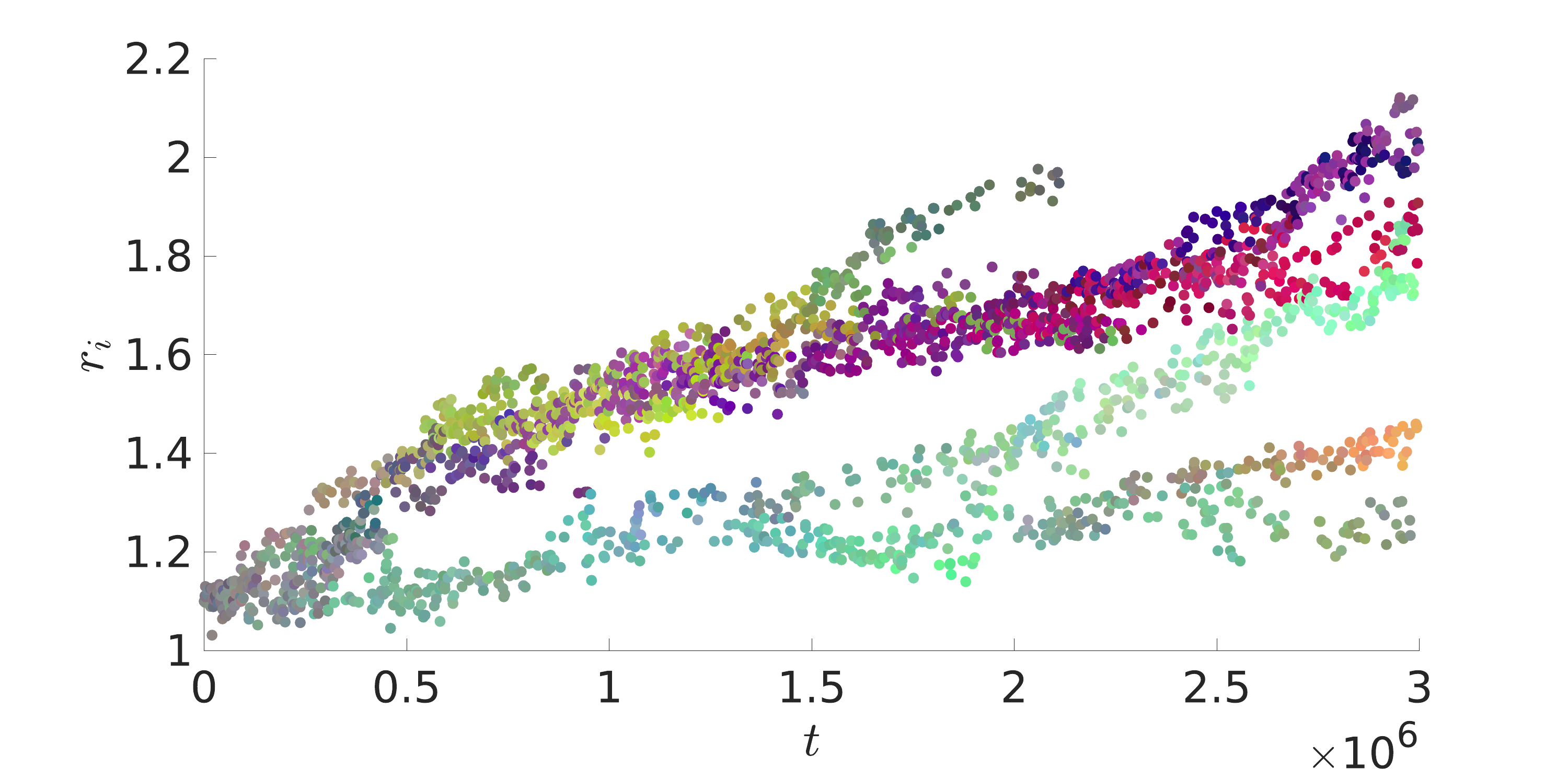
  \caption{}
  \label{fig:2layers:GrowthRAnc}
\end{subfigure}
\begin{subfigure}{.49\textwidth}
  \centering
 		\def\svgwidth{\columnwidth}
    		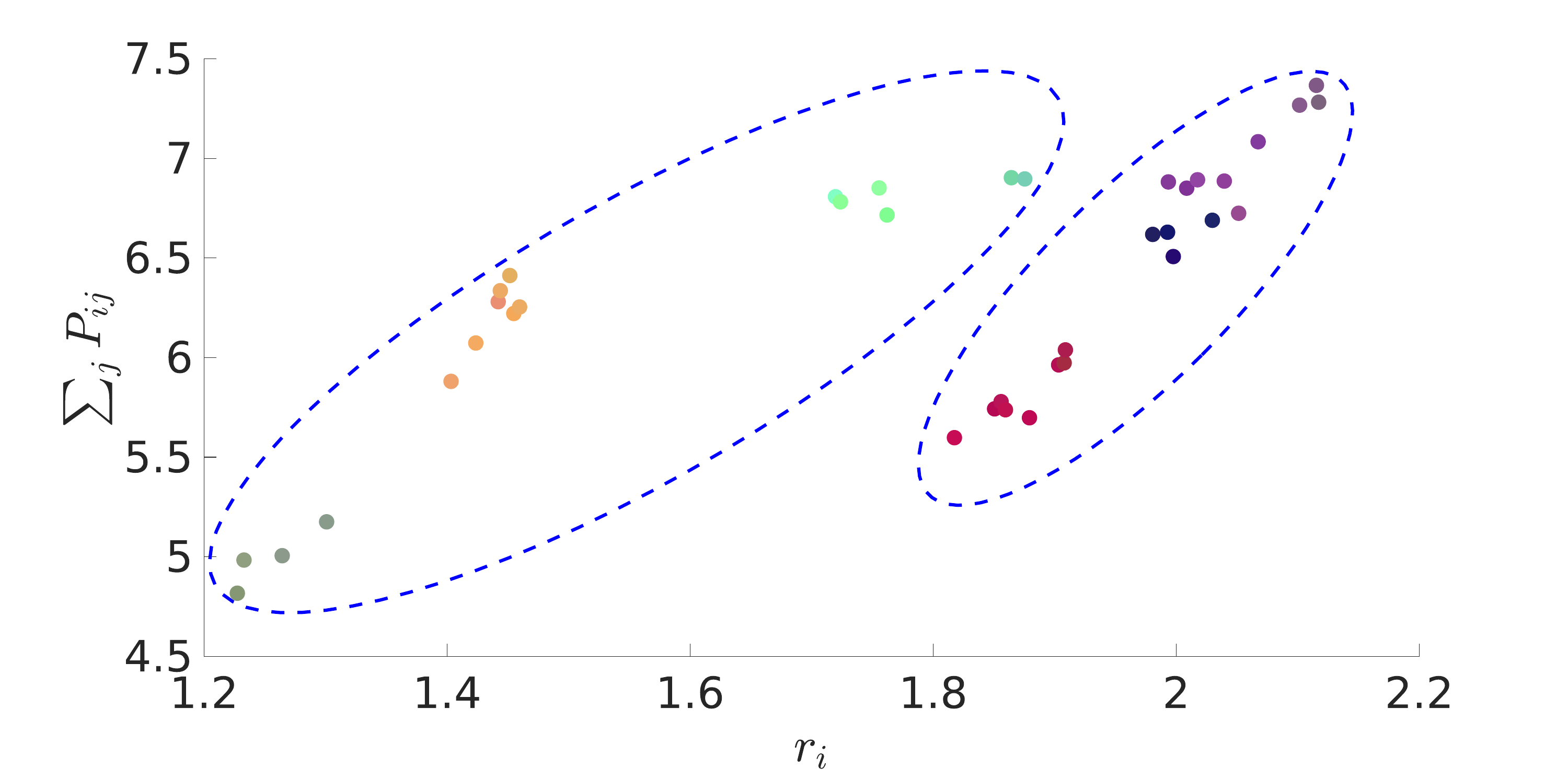
  \caption{}
  \label{fig:2layers:RvsSumP}
\end{subfigure}
\caption{Obtained from the same data as in Fig.~\ref{fig:2layers:RSandB}. Figure~(a) shows for each introduced new prey species the time of speciation versus its value of $r_i$. Clearly, there is a drift upwards in the values of $r_i$. The dots are colored depending on the color of the ancestor. That is, each color is a small perturbation of the RGB color of the ancestor. This highlights the different branches in the evolutionary dynamics. Note however that this figure does not contain any information on the success of each species, i.e. whether it survives and for how long. Figure~(b) shows the value of $r_i$ versus the predation, i.e. the corresponding row sum of $P$, for the 38 alive prey species at the end of the simulation. The same coloring as in Fig.~(a) is used. The two ellipses indicate the lower branch (left) and upper branch (right) of Fig.~(a).}
\label{fig:2layers:R}
\end{figure}

A practical advantage of allowing variation in the growth rate $r$ is that it is a one-dimensional parameter, compared to the high-dimensional rows of $P$. This allows us to track the evolution of this parameter over time and get a fourth, visual type of information on the formation of clusters, without any need for MDC, GDC or IMC, even though these methods can still be applied. Furthermore, we can study correlations between the growth rate and other system parameters. The evolution of $r$ is shown in Fig.~\ref{fig:2layers:GrowthRAnc}, where for each prey $i$, the growth rate $r_i$ is plotted versus the time of spawning. Visually, it is suggested that around $t=5\cdot10^5$, the prey species split into two main branches that later split up into smaller branches.      

We noted in sec.~\nameref{sec:evor} that a higher growth rate~$r_i$ implies that prey $i$ can survive with a relatively low population. On the other hand, this also implies that prey $i$ can survive with a higher value of predation $\sum_jP_{ij}$. Hence, we expect a positive correlation between the growth rates $r_i$ and the row sums of $P$. In Fig.~\ref{fig:2layers:RvsSumP} we show for the prey species alive at the end of the simulation the relation between~$r_i$ and the predation~$\sum_j P_{ij}$. For the species within the two ellipses, there indeed is a strong correlation. The ellipses are drawn based on the evolutionary history, i.e. the left ellipse corresponds to the lower branch in Fig.~\ref{fig:2layers:GrowthRAnc} and the right ellipse to the upper branch. However, for all species together, the correlation is less strong. This is in line with what we expect, as the scale at which one studies correlations in the phylogenetic trees heavily influences the results~\cite{graham2018phylogenetic}.  

\begin{figure}
\begin{subfigure}{.49\textwidth}
  \centering
 		\def\svgwidth{\columnwidth}
    		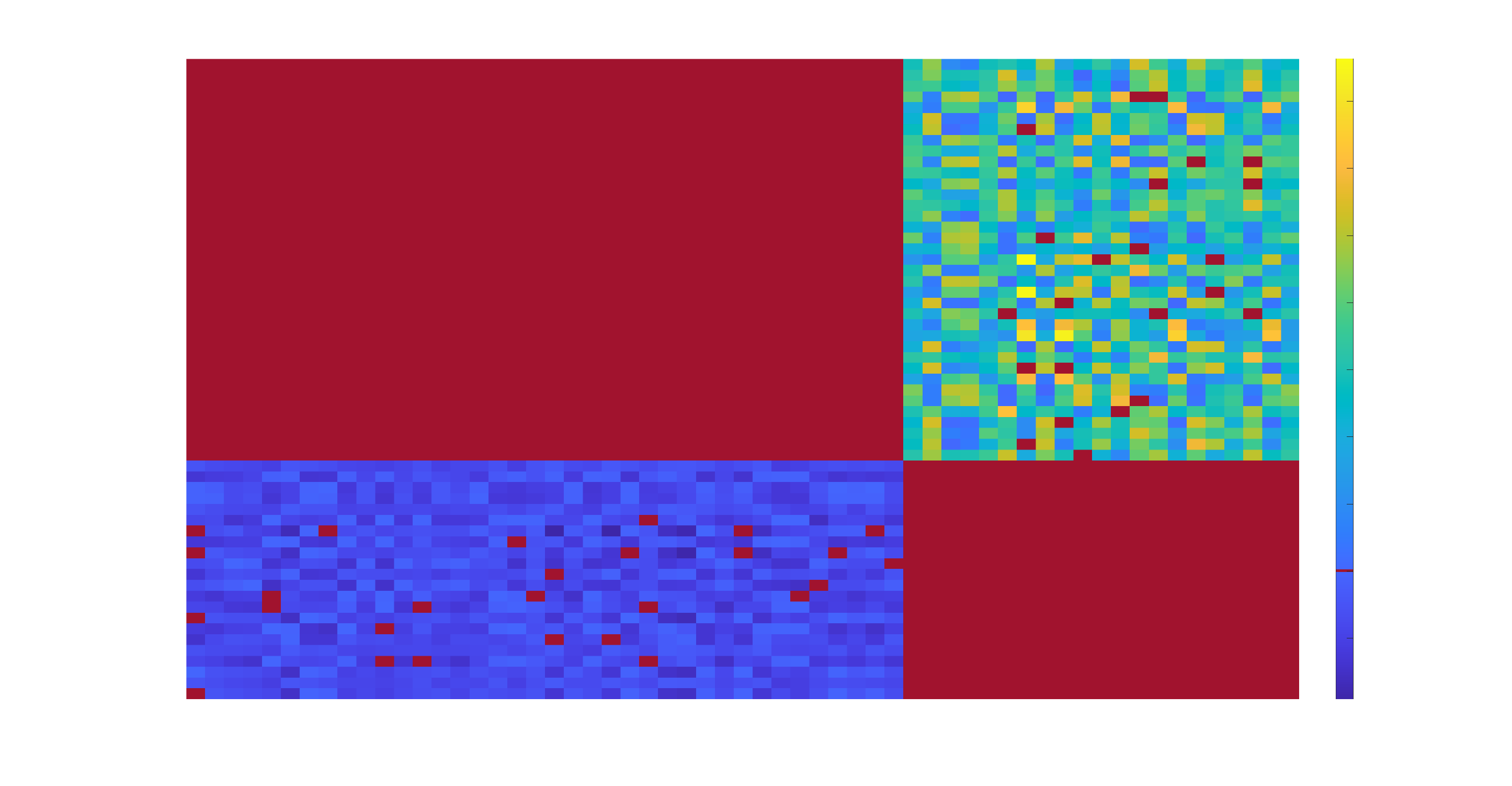
  \caption{}
  \label{fig:2layers:clusterA}
\end{subfigure}
\begin{subfigure}{.49\textwidth}
  \centering
 		\def\svgwidth{\columnwidth}
    		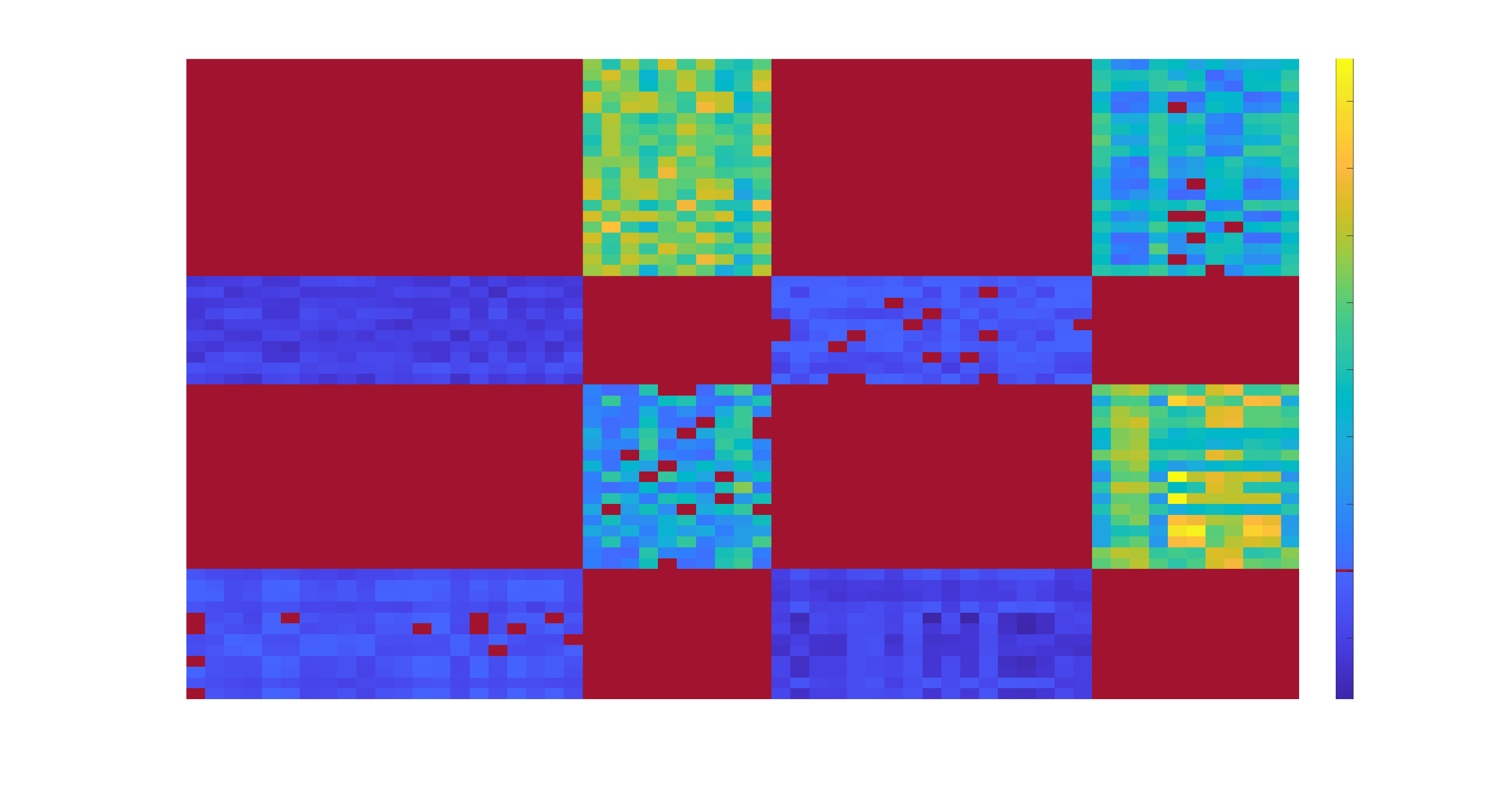
  \caption{}
  \label{fig:2layers:clusterB}
\end{subfigure}
\caption{Figure~(a) shows the interaction matrix $A$ with zero entries corresponding to the color red. The block structure is clearly visible. In Figure~(b) the matrix is re-ordered based on the IMC algorithm. The clustering overlaps completely with the upper and lower branches in Fig.~\ref{fig:2layers:GrowthRAnc}. Hence, this figure indicates that there are two main groups of predators, one that mainly interacts with the upper branch of prey and a group that mainly interacts with the lower branch of prey.}
\label{fig:2layers:cluster}
\end{figure}

\subsubsection{Correspondence of functional and interaction clustering}
Figs.~\ref{fig:2layers:GrowthRAnc} and \ref{fig:2layers:RvsSumP} show that when the green-colored prey species from the lower branch overlap with the red-colored prey species from the upper branch in the $r_i$ versus time plot, their rows of $P$ are not necessarily similar as indicated by the two different ellipses. Therefore, the species in the upper and lower branches are not just different in the growth rate $r_i$, but also differ in their functional interactions with the predators. To further investigate this, we apply the same IMC algorithm as used before in the fixed growth rate context to the interaction matrix $A$. As Fig.~\ref{fig:2layers:cluster} shows, the $2\times2$ (predators versus prey) block structure can be divided into a smaller $4\times4$ structure (two groups of predators versus two groups of prey) such that the interaction within these blocks is significantly stronger than the interaction between the different blocks. This shows that there are indeed two main clusters of interacting prey and predator species. Importantly, the two clusters found by the IMC algorithm exactly overlap with the upper and lower branch of the evolutionary tree. Hence, species with a strong predator-prey interaction share an evolutionary history, which is strong evidence of co-evolution in our model. 

\subsubsection{Correspondence of functional and kmeans clustering}
Fig.~\ref{fig:2layers:RvsSumP} indicates that it must be possible to subdivide the lower and upper branches further into smaller clusters. Specifically, the lower branch should be divisible into the clusters gray, orange and green, and the upper branch into the clusters red, blue and purple. First, we apply GDC to the evolutionary tree of the species present at the end of the simulation. When we take as cut-off that the genealogical distance is less than $5\cdot 10^5$, this results in the six clusters described by the colors above. This is to be expected, as the color coding in Fig.~\ref{fig:2layers:R} depends on the genealogical distance.

\begin{figure}[h]
  \centering
 		\def\svgwidth{\columnwidth}
    		\includegraphics[width=.8\linewidth]{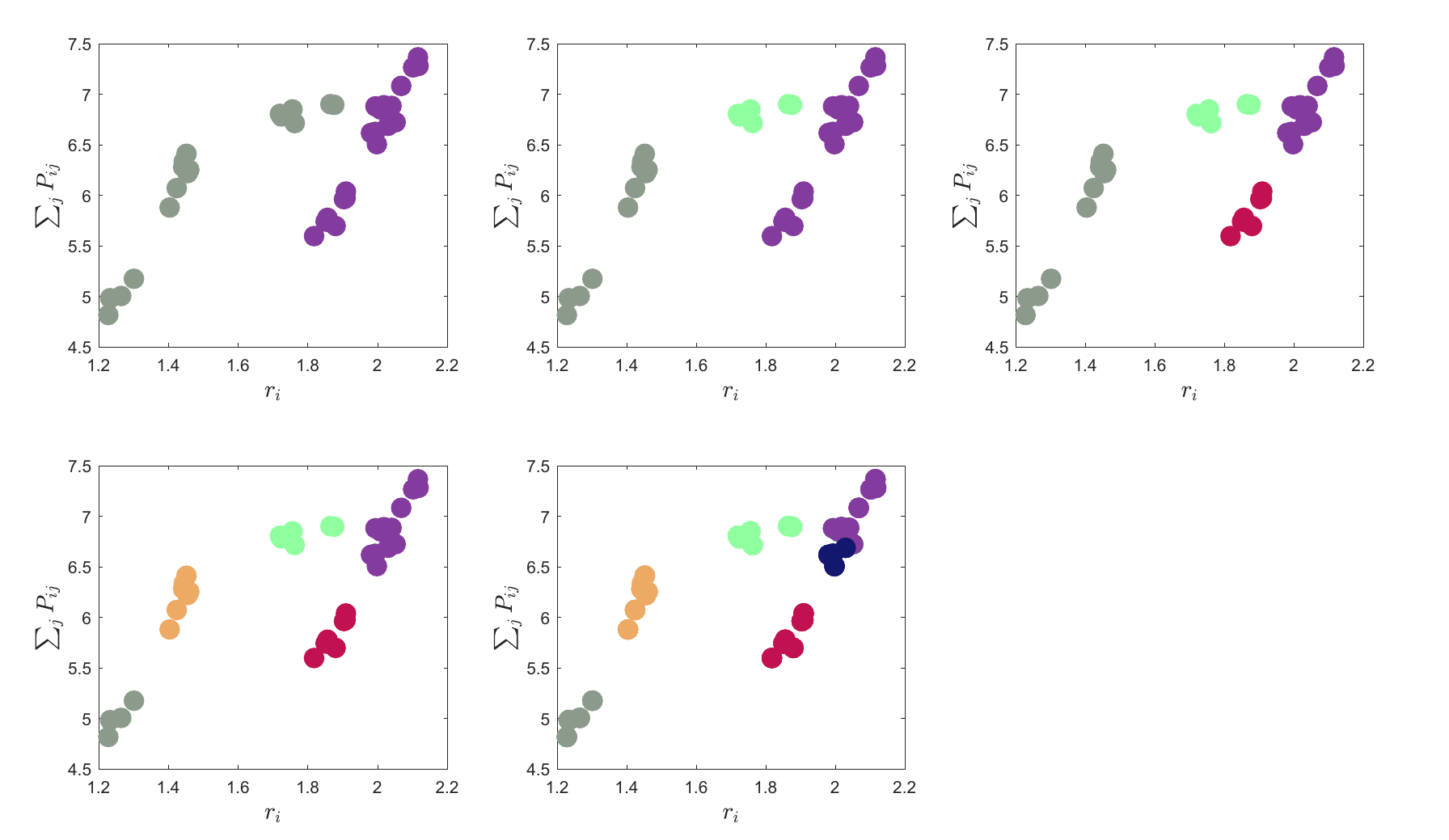}
\caption{Obtained from the same data as in Fig.~\ref{fig:2layers:RSandB}. In the five panels we show for the 38 alive prey species at the end of the simulation, the clustering as found by KMC by forcing the algorithm to find $2, \ldots, 6$ clusters. Similar coloring as in Fig.~\ref{fig:2layers:RvsSumP} is used. The order of the splitting is similar to the order seen in Fig.~\ref{fig:2layers:GrowthRAnc}.}
\label{fig:kmeansprogress}
\end{figure}

For our final approach, we use a more quantitative clustering on the functional differences between the prey species. That is, we update the definition of the phenomenological distance $d$ from Eq.~(\ref{eq:DistPrey}) defining the \emph{distance} between prey species $i$ and $\ell$ as
\begin{equation}
    d(i,\ell)=\alpha|r_i-r_\ell|^2+\sum_k|P_{ik}-P_{\ell k}|^2,
\end{equation}
where the parameter $\alpha$ allows us to balance the influence of $r$ and $P$ on the functional difference. When we apply KMC using the silhouette criterion and $\alpha=3$, we find the same six clusters as with the GDC approach on the genealogical tree. Therefore, at the end of the simulation, the functional differences match with the genealogical history, but we can go one step further. We again apply KMC, but require it to find two, three, four, five and six clusters respectively. As Fig.~\ref{fig:kmeansprogress} shows, the subdivision into smaller clusters follows the branches in Fig.~\ref{fig:2layers:GrowthRAnc}. In other words, the evolutionary history is completely encoded in the structure of $P$ and $r$. 

\section{Discussion \& Outlook}
The fact that structure in ecosystems is not random but clustered is an important observable to reproduce while modeling evolutionary dynamics~\cite{oliveira2016species}. So far, the approaches for resource-consumer models~\cite{bellavere2023speciation,caetano2021evolution}, where consumption is normalized, lead to a small number of clusters determined by the normalization. Alternatively, the competitive LV approach in~\cite{shtilerman2015emergence} leads to more clusters, but these are largely formed by mother-daughter interactions and therefore not reflective of real-world ecosystems~\cite{alexandrou2015evolutionary,mahon2021functional}. Enforcing two trophic layers with predator-prey interactions resulted in stable evolutionary dynamics as shown in sec.~\nameref{sec:2layers}. In the case of large and intermediate variability, clusters of predators and prey arose, and the clusters are robust against variation in the growth rate, but in the limit of small variation, there are too many branches in the evolutionary tree to meaningfully define clusters. 

These clusters lead to the following question: What constitutes a species in our model? One could argue that each cluster is a single species with genetic variations within the species, as is the case in real populations. The question of clustering species is not just limited to theoretical explorations, but divisions between clades, species, subspecies, etc. are still widely discussed, see e.g. the discussion on European herpetofauna~\cite{speybroeck2020species}. In future experiments, we could compare the clustered versus unclustered dynamics and test hypotheses on the resilience of ecosystems versus the genetic diversity. 

The speciating LV model with two trophic layers as discussed here has limitations. In real-world ecosystems, all three types of interaction, competition, predation and mutualism, are possible instead of just predation. We also investigated ecosystem modeling without enforcing predator-prey relations, i.e. allowing competition and mutually beneficial relations without specified trophic layers. In particular, in Appendix~\nameref{sec:AppA}, we study a generalized LV equation with speciation without restrictions on the interaction matrix, but this approach consistently showed a divergent biomass and hence did not lead to stable evolutionary dynamics, rendering cluster analysis moot. 

In the current approach, resources are not explicitly modeled like, for example, in~\cite{bellavere2023speciation}. Hence, the prey species do not compete for resources and each prey species follows its own logistic growth rate, independent of other prey species, which in theory allows for unlimited growth of biomass. Combining predator-prey interactions with resource-consumer models, hence creating three trophic layers, will be an important next step in extending the evolutionary dynamics. 

In our speciating predator-prey LV model, the interaction matrix $A$ is enforced to be anti-symmetric, up to a factor $\beta$. This $\beta$ is fixed for all predator and prey combinations and therefore leaves no room for specialization, as a predator can consume each prey species with similar effort and reward. Letting the $\beta$ be species-dependent and vary with the speciation process allows specialization among the predators. However, in the current setup, this would imply that predators with higher $\sum_j\beta_{ij}$ have a significant advantage over their competitors leading to so-called ``Darwinian Demons'', so a budget for the vector $\beta$ would be necessary in this case, for example by fixing its norm. 

All the species in our model are to some extent artificially similar, as the values of $\delta,K$, and $\beta$ are fixed. The carrying capacity $K$ can be removed by explicitly introducing resources and $\beta$ has been discussed above, but for the mortality $\delta$, it is not {\it{a priori}} clear how to introduce stochastic variability. We have to assume that $\delta>0$, but introducing noise into this parameter automatically leads to a downward drift as a lower mortality rate is advantageous. If a similar approach is taken as in sec.~\nameref{sec:2layers:drive}, there will be species with negative~$\delta$ on the timescale of the simulation (result not shown). Hence, a more detailed description of mortality is needed before it can be modified by noise; this is left as future work.

\section{Conclusions}
We studied random speciation in a predator-prey LV model and found that this stochastic approach of adding new species leads to large complex ecosystems that (a) are stable, (b) have their ecological and evolutionary timescales intertwined and (c) show clustering of species, both functional and genealogical. The observed dynamics are robust to various model parameter variations and cluster definitions. Our results indicate that adding stochastic speciation to a coupled ODE system has both mathematically novel and biologically relevant implications for the better modeling of ecosystems.

\section{Acknowledgments}
The work of CH is funded by the Dutch Institute for Emergent Phenomena (DIEP) at the University of Amsterdam via the program Foundations and Applications of Emergence (FAEME). We thank Marco Saltini for constructive remarks on earlier versions of this paper. 

\section{Data Availability Statement}
All figures in this paper are generated using Matlab scripts written by the authors. The scripts that simulate the speciating two trophic Lotka-Volterra model can be found on \url{https://github.com/chshamster/SpeciatingLV}.

\bibliographystyle{plain}   
\bibliography{ref}

\appendix
\section{Generalized Lotka-Volterra equations with speciation}
\label{sec:AppA}
\setcounter{figure}{0}
\setcounter{table}{0}
\setcounter{equation}{0}
\counterwithin{equation}{section}
\counterwithin{figure}{section}
\counterwithin{table}{section}
In the main text, we enforced the model to have two trophic layers, i.e. Eq.~\eqref{eq:2layers:prey} and \eqref{eq:2layers:pred}. In this appendix, we argue for the necessity of this two-trophic layer condition for the current setup with speciation, by showing that the more general case leads to unphysical behavior. We relax the tropic layer condition, following the approach of \cite{shtilerman2015emergence}, we start with Generalized Lotka-Volterra (GLV) equations
\begin{equation}
\label{eq:app}
    \dot x_i=r_i x_i\left(1-\frac{x_i}{K}\right)+\frac{x_i}{K}\sum_{j=1}^N A_{ij}x_j.
\end{equation}
The biomass $x_i$ of each of the $N$ species present in the system follows a logistic model with growth rate $r_i>0$ and carrying capacity $K$, on top of which interaction with different species is possible.
In the community matrix $A$, each component $A_{ij}$  describes the influence species $j$ has on $i$. In general, $A_{ij}\neq A_{ji}$, so $A$ is not symmetric. Furthermore, we take $A_{ii}=0$, as the self-interaction is modeled by the logistic term. For simplicity, we fix the carrying capacity $K$. The signs of $A_{ij}$ and $A_{ji}$ describe the relationship between the species $i$ and $j$. Note that the signs of $A_{ij}$ are not fixed as, in contrast to the main text, there are no pre-described trophic layers. In contrast to the two trophic layer model, the speciating GLV model therefore has no built-in feedback mechanism to bound the biomass. Each species has a carrying capacity of $K$, but when $A$ has enough positive values, this self-limiting term can in principle be overcome and the biomass in the system diverges. The question is whether the stochastically generated $A_{ij}$ \emph{may} allow for a stable solution at long timescales. As we will see below, we have numerically not been able to find a set of parameters that keeps the biomass finite, but we have no proof that such a divergence must always occur in this model implementation.

\begin{table}[h]
    \centering
    \begin{tabular}{c|c|c}
         parameter& meaning & value \\ \hline
$x_i$ & biomass species $i$ &\\  
       $r_i$  & growth rate & $~\sim \mathcal{N}(1,\sigma^2)$ \\
$K$ & carrying capacity & $10$ \\
$A_{ij}$ & interaction of species $j$ on species $i$ & in $\mathbb{R}$\\
$\mu$&Average of initial distribution $A_{ij}$&-0.3\\
$\sigma$&Standard deviation initial distribution $r$ and $A_{ij}$&0.1\\
$\eta$ & standard deviation of the noise/variability in speciation &0.02\\
$\gamma$ & mother/daughter interaction strength &0.4\\
$p_m(0)$ & Parameter for the exponential waiting times &$10^{-3}$\\
    \end{tabular}
    \caption{Table summarizing the parameters used in the speciating GLV model.}
    \label{tab:app}
\end{table}

\subsection*{Speciation}
\label{subsec:app:speciation}
We initialize the ecosystem with $n$ species with random interaction coefficients:
\begin{equation}
    \begin{split}
    \label{eq:cmatrix3}
    A & =
    \begin{pmatrix}
        0 & \alpha_{12} & \hdots & \alpha_{1n}\\
        \alpha_{21} & 0 & \hdots & \alpha_{2n}\\
        \vdots & \vdots & \ddots & \vdots \\
        \alpha_{n1} & \alpha_{n2} & \hdots & 0
    \end{pmatrix}\,,
     \end{split}
\end{equation}
where all the random variables $\alpha_{ij}$ are drawn independently from a normal distribution $\mathcal{N}(\mu,\sigma^2)$. Typically, we initialize with two species, i.e. $n=2$, and take $\mu=-0.3$ and $\sigma=0.1$. Hence, we typically start with an ecosystem dominated by competition to ensure the stability of the initial configuration. The growth rate $r_i$ is chosen from the distribution $\mathcal{N}(1,\sigma^2)$.

As there are no trophic layers, introducing a new species means that we must take the whole row vector and column vector defining the ancestor $i$ from the interaction matrix $A$ and perturb these. The perturbation of every element is assumed to be distributed as $\mathcal{N}(0,\eta^2)$. Furthermore, we assume the perturbation to be small, so $\eta$ has to be small compared to $\mu$. Typically, $\eta=0.02$, as in the main text, compared to $\mu=-0.3$. The updated matrix, which we denote with $\tilde A$, now becomes
\begin{equation}
    \tilde A=\begin{pmatrix}
        A&\psi_c\\
        \psi_r&0\end{pmatrix},
\end{equation}
where $\psi_c$ is a perturbation of the column $A_{*,i}$ and $\psi_r$ is a perturbation of the row $A_{i,*}$.
In this formulation of speciation, the interaction between daughter and ancestor is given by $\tilde A_{(n+1)i}=A_{ii}+\psi_r(i)=\psi_r(i)\sim\mathcal{N}(0,\eta^2)$, which is weak as the standard deviation $\eta$ is small. That is, mother and daughter are very similar. However, we expect there to be strong competition between the daughter and ancestor species as their niches are similar. Therefore, to model the competition between these two species, we take $\tilde A_{(n+1)i} = -\gamma+\psi_r(i)$ and for $\tilde A_{i(n+1)}$ we take the same approach. For the figures below, we use $\gamma=0.4$. 

\subsection*{Unbounded growth of biomass}
\begin{figure}
\begin{subfigure}{.49\textwidth}
  \centering
 		\def\svgwidth{\columnwidth}
    		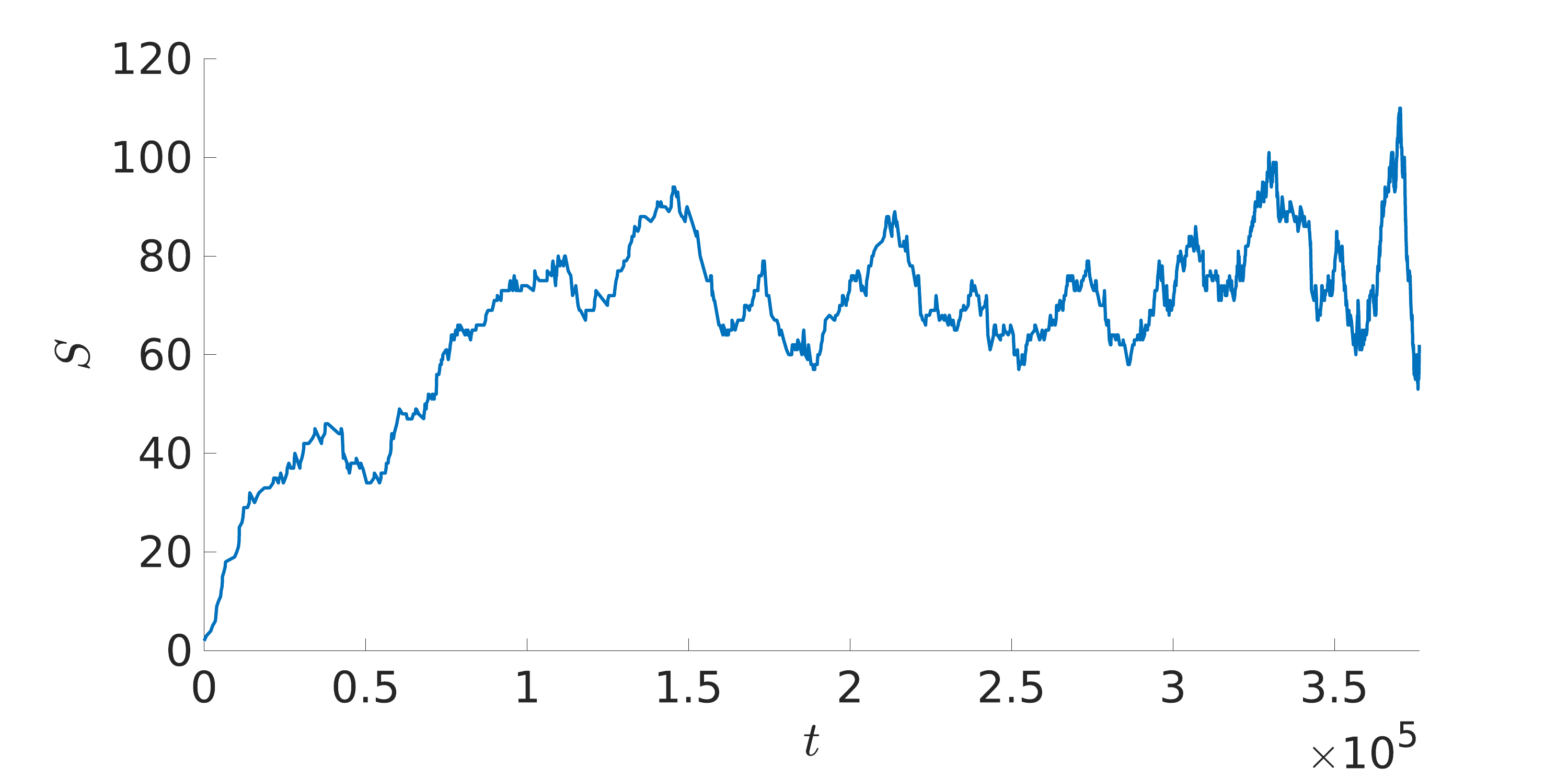
  \caption{}
  \label{fig:app:SvsT}
\end{subfigure}
\begin{subfigure}{.49\textwidth}
  \centering
 		\def\svgwidth{\columnwidth}
    		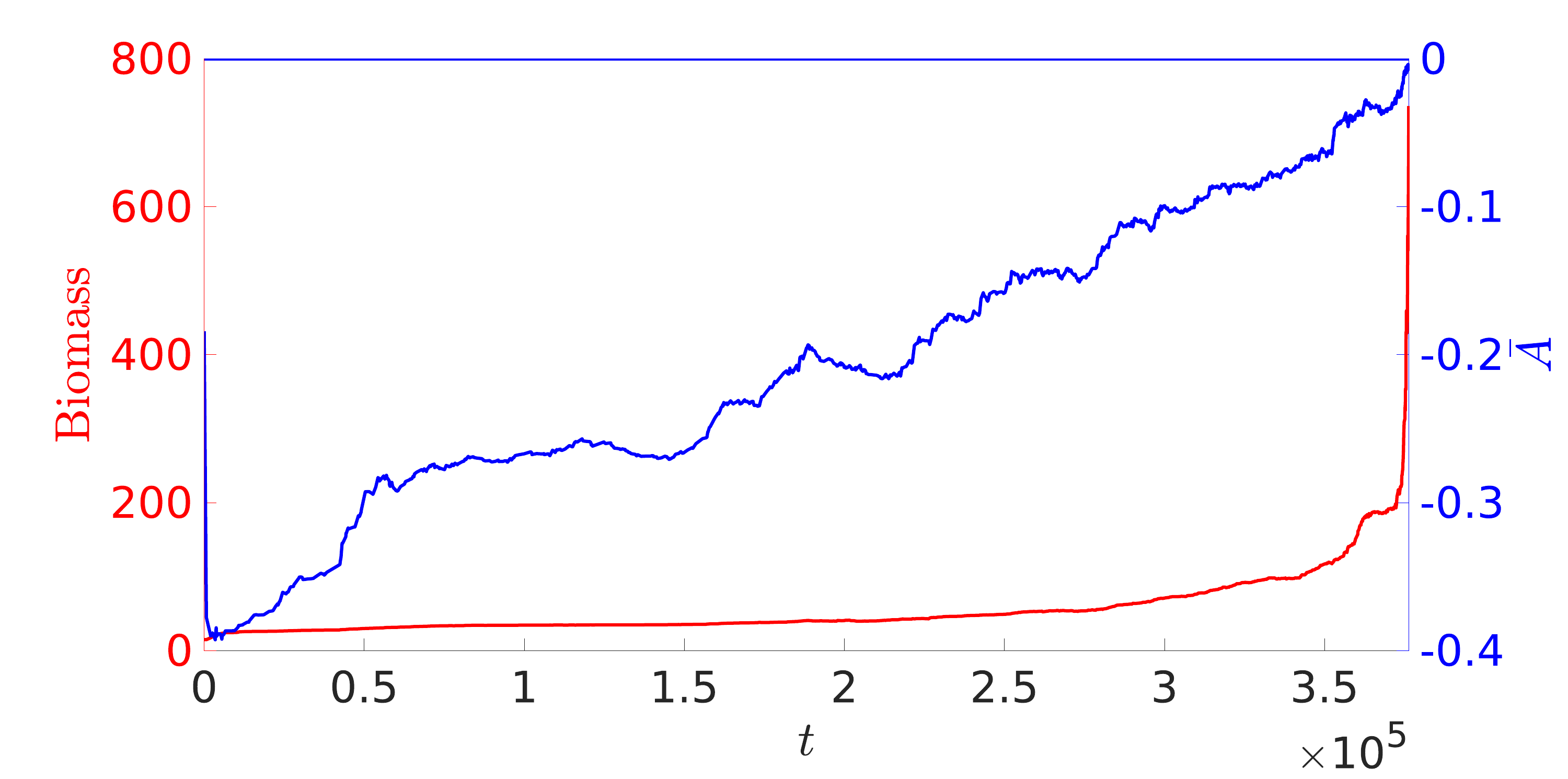
  \caption{}
  \label{fig:app:BAvsT}
\end{subfigure}
  \caption{Figure~(a) shows the number of species of a long-time simulation of the speciating GLV model. Figure~(b) shows the biomass and the growth of the average value of the interaction matrix $A$. Note that the biomass grows unboundedly when $\bar A$ reaches $0$. The parameters are as described in Table~\ref{tab:app}.}
\label{fig:app:BlowUp}
\end{figure}

Biomass divergence in Eq.~\ref{eq:app} occurs when the $A_{ij}$ become positive enough \emph{on average} to induce an unbounded growth for some $x_i$. To investigate if, and when, this happens during speciation, we simulate the system and track the biomass $B$, the species abundance $S$  (as defined in the main text), and the average $\bar A$ of all the interaction matrix coefficients $A_{ij}$. We expect that the biomass initially will remain below $KS$ but that an increase in $\bar A$ should eventually precipitate in a biomass divergence. Indeed, as Fig.~\ref{fig:app:BlowUp} shows, the moment $\bar A$ becomes positive, the biomass grows without bound.

We cannot prove under which conditions for $\bar A$ the biomass divergence occurs. Indeed, there are counterexamples of $2\times2$ matrices with positive averages that lead to stable dynamics. The general case of $N\times N$ matrices is however not easily considered. Stability of biomass also has a temporal dimension: the exact timing of the divergence depends on the value of $\gamma$: we have observed that when the mother-daughter interaction becomes stronger, the divergence of the biomass typically occurs at later times (results not shown). While absence of proof is not proof of absence, we have not been able to find any parameter setting in which the biomass in the speciating GLV model remains finite. The observation of biomass divergence is in fact remarkably robust. Empirical evidence suggests that adjusting the several parameters in the speciating GLV model can only slow down the moment of biomass divergence, but cannot prevent this unphysical behavior.

\subsection{Battling biomass divergence}
\label{sec:app:Preventing}
We will now briefly discuss a possible more biologically motivated or ad-hoc solution to stabilize the unphysical ecosystem dynamics. 
One way to constrain species growth is by coupling the growth rate to a finite (replenishable) resource bath.  In a resource-consumer model like the MacArthur model, considered in a similar context as in this article in~\cite{bellavere2023speciation}, divergence is not possible because biomass increase depletes resources and suppresses growth rates. 

\begin{figure}[h]
\begin{subfigure}{.49\textwidth}
  \centering
 		\def\svgwidth{\columnwidth}
\begingroup%
  \makeatletter%
  \providecommand\color[2][]{%
    \errmessage{(Inkscape) Color is used for the text in Inkscape, but the package 'color.sty' is not loaded}%
    \renewcommand\color[2][]{}%
  }%
  \providecommand\transparent[1]{%
    \errmessage{(Inkscape) Transparency is used (non-zero) for the text in Inkscape, but the package 'transparent.sty' is not loaded}%
    \renewcommand\transparent[1]{}%
  }%
  \providecommand\rotatebox[2]{#2}%
  \newcommand*\fsize{\dimexpr\f@size pt\relax}%
  \newcommand*\lineheight[1]{\fontsize{\fsize}{#1\fsize}\selectfont}%
  \ifx\svgwidth\undefined%
    \setlength{\unitlength}{1387.5bp}%
    \ifx\svgscale\undefined%
      \relax%
    \else%
      \setlength{\unitlength}{\unitlength * \real{\svgscale}}%
    \fi%
  \else%
    \setlength{\unitlength}{\svgwidth}%
  \fi%
  \global\let\svgwidth\undefined%
  \global\let\svgscale\undefined%
  \makeatother%
  \begin{picture}(1,0.51945946)%
    \lineheight{1}%
    \setlength\tabcolsep{0pt}%
    \put(0,0){\includegraphics[width=\unitlength,page=1]{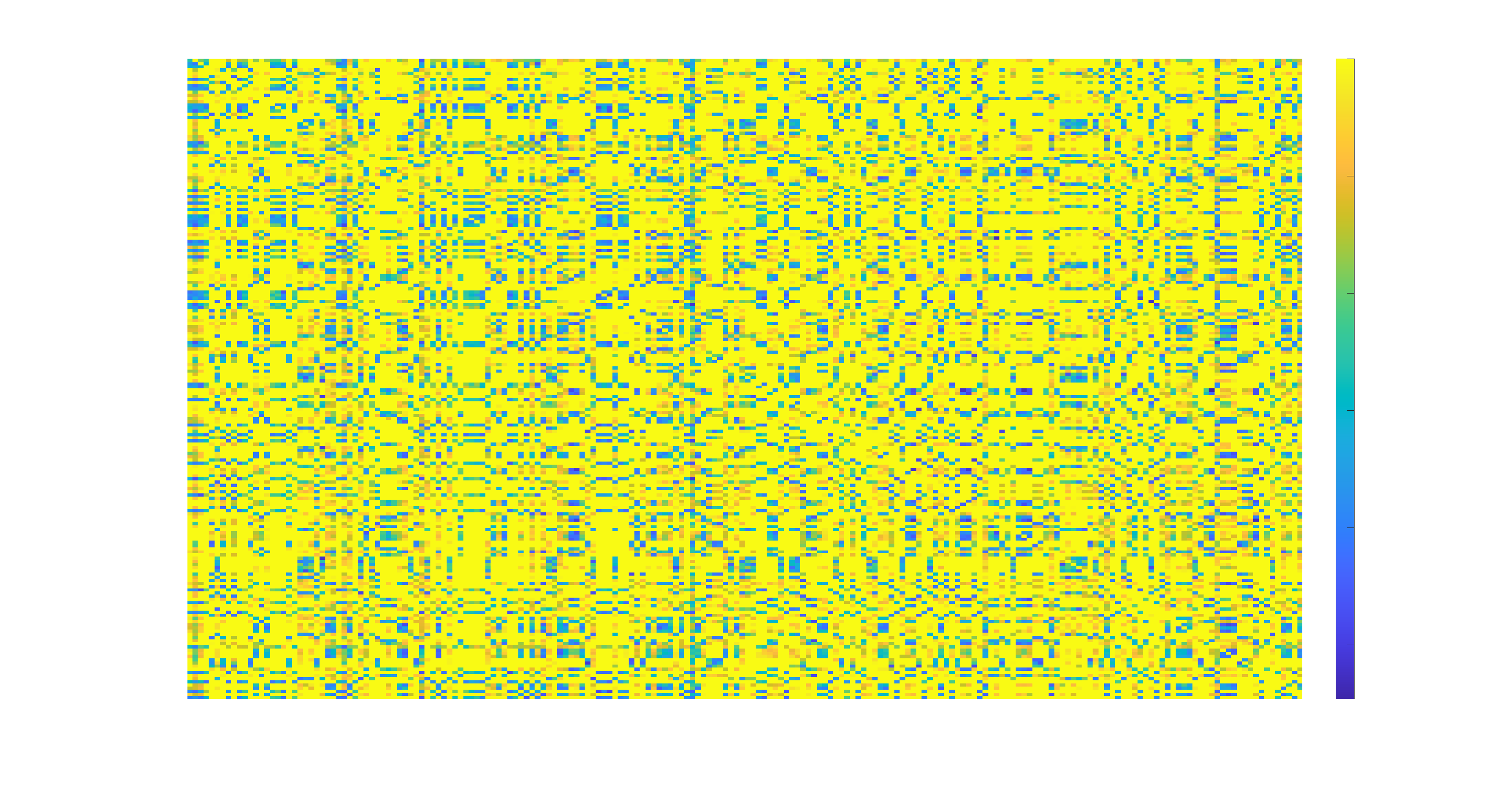}}%
    \put(0.90230632,0.08406708){\makebox(0,0)[lt]{\lineheight{1.25}\smash{\begin{tabular}[t]{l}-0.5\end{tabular}}}}%
    \put(0.90230632,0.161578){\makebox(0,0)[lt]{\lineheight{1.25}\smash{\begin{tabular}[t]{l}-0.4\end{tabular}}}}%
    \put(0.90230632,0.23908892){\makebox(0,0)[lt]{\lineheight{1.25}\smash{\begin{tabular}[t]{l}-0.3\end{tabular}}}}%
    \put(0.90230632,0.31659984){\makebox(0,0)[lt]{\lineheight{1.25}\smash{\begin{tabular}[t]{l}-0.2\end{tabular}}}}%
    \put(0.90230632,0.3941107){\makebox(0,0)[lt]{\lineheight{1.25}\smash{\begin{tabular}[t]{l}-0.1\end{tabular}}}}%
    \put(0.90230632,0.47162162){\makebox(0,0)[lt]{\lineheight{1.25}\smash{\begin{tabular}[t]{l}0\end{tabular}}}}%
    \put(0,0){\includegraphics[width=\unitlength,page=2]{FPclusterA.pdf}}%
  \end{picture}%
\endgroup%

  \caption{}
  \label{fig:app:FPclusterA}
\end{subfigure}
\begin{subfigure}{.49\textwidth}
  \centering
 		\def\svgwidth{\columnwidth}
\begingroup%
  \makeatletter%
  \providecommand\color[2][]{%
    \errmessage{(Inkscape) Color is used for the text in Inkscape, but the package 'color.sty' is not loaded}%
    \renewcommand\color[2][]{}%
  }%
  \providecommand\transparent[1]{%
    \errmessage{(Inkscape) Transparency is used (non-zero) for the text in Inkscape, but the package 'transparent.sty' is not loaded}%
    \renewcommand\transparent[1]{}%
  }%
  \providecommand\rotatebox[2]{#2}%
  \newcommand*\fsize{\dimexpr\f@size pt\relax}%
  \newcommand*\lineheight[1]{\fontsize{\fsize}{#1\fsize}\selectfont}%
  \ifx\svgwidth\undefined%
    \setlength{\unitlength}{1387.5bp}%
    \ifx\svgscale\undefined%
      \relax%
    \else%
      \setlength{\unitlength}{\unitlength * \real{\svgscale}}%
    \fi%
  \else%
    \setlength{\unitlength}{\svgwidth}%
  \fi%
  \global\let\svgwidth\undefined%
  \global\let\svgscale\undefined%
  \makeatother%
  \begin{picture}(1,0.51945946)%
    \lineheight{1}%
    \setlength\tabcolsep{0pt}%
    \put(0,0){\includegraphics[width=\unitlength,page=1]{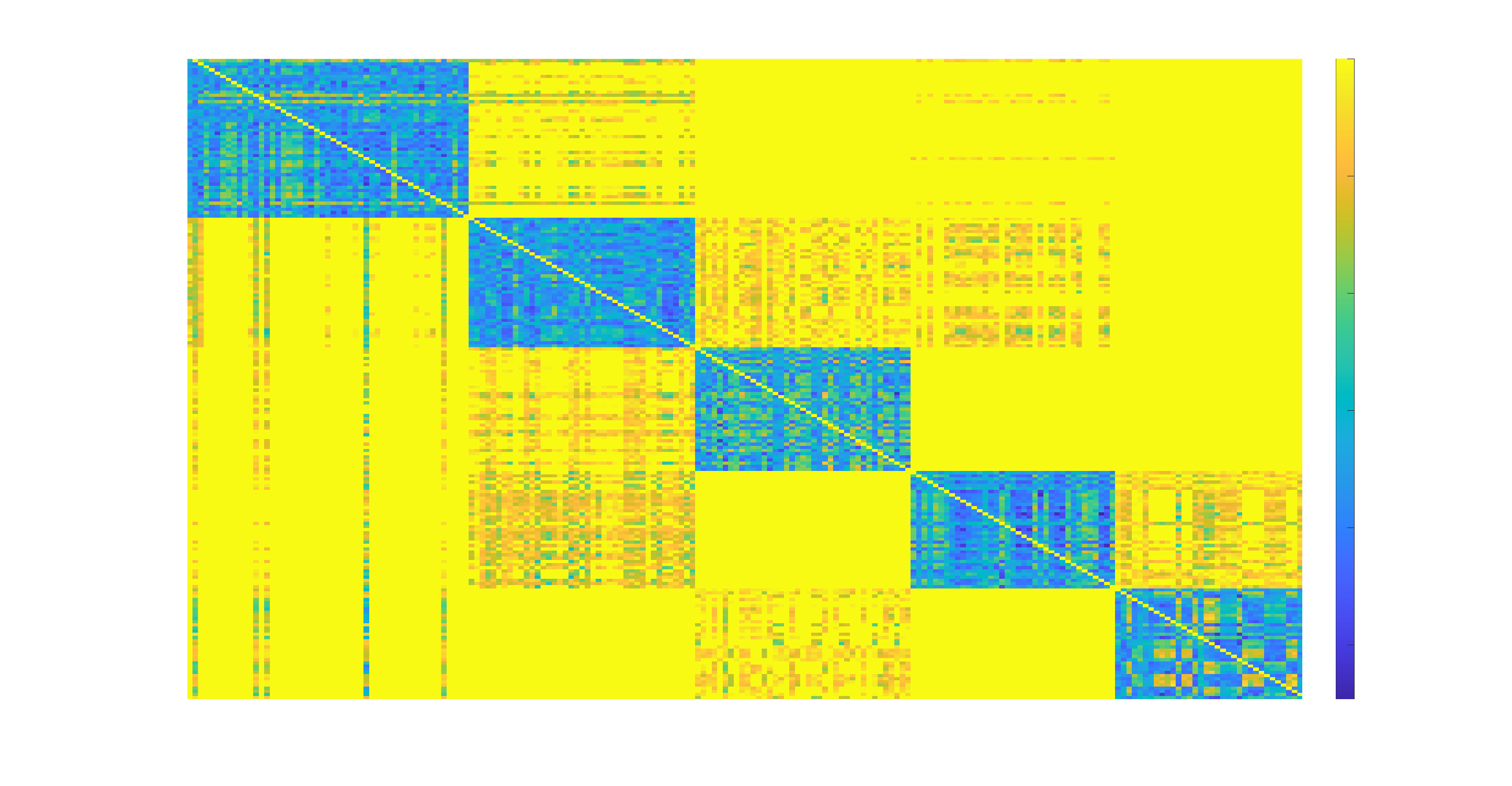}}%
    \put(0.90230632,0.08406708){\makebox(0,0)[lt]{\lineheight{1.25}\smash{\begin{tabular}[t]{l}-0.5\end{tabular}}}}%
    \put(0.90230632,0.161578){\makebox(0,0)[lt]{\lineheight{1.25}\smash{\begin{tabular}[t]{l}-0.4\end{tabular}}}}%
    \put(0.90230632,0.23908892){\makebox(0,0)[lt]{\lineheight{1.25}\smash{\begin{tabular}[t]{l}-0.3\end{tabular}}}}%
    \put(0.90230632,0.31659984){\makebox(0,0)[lt]{\lineheight{1.25}\smash{\begin{tabular}[t]{l}-0.2\end{tabular}}}}%
    \put(0.90230632,0.3941107){\makebox(0,0)[lt]{\lineheight{1.25}\smash{\begin{tabular}[t]{l}-0.1\end{tabular}}}}%
    \put(0.90230632,0.47162162){\makebox(0,0)[lt]{\lineheight{1.25}\smash{\begin{tabular}[t]{l}0\end{tabular}}}}%
    \put(0,0){\includegraphics[width=\unitlength,page=2]{FPclusterB.pdf}}%
  \end{picture}%
\endgroup%

  \caption{}
  \label{fig:app:FPclusterB}
\end{subfigure}
\begin{subfigure}{.49\textwidth}
  \centering
 		\def\svgwidth{\columnwidth}
    		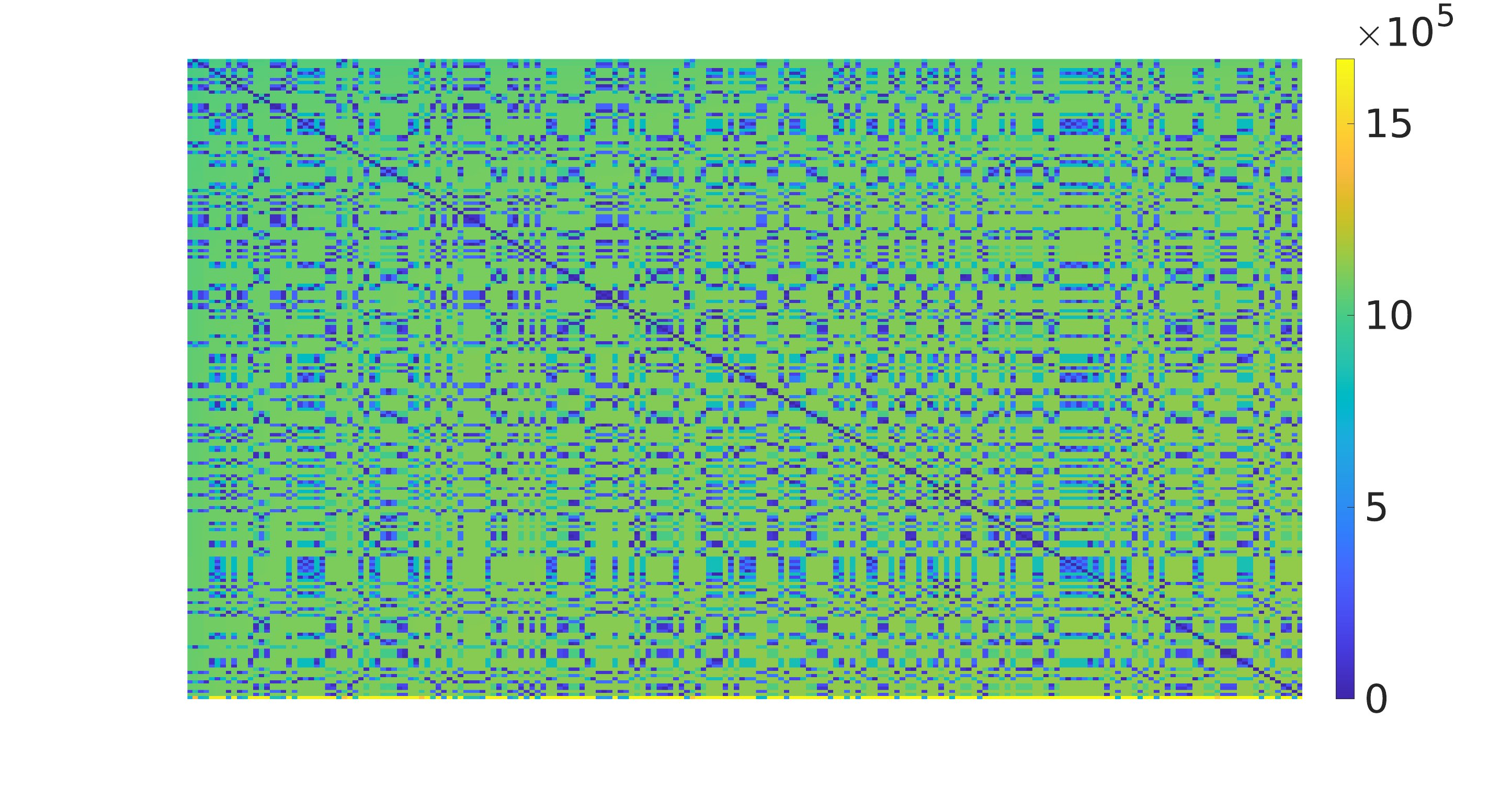
  \caption{}
  \label{fig:app:FPclusterC}
\end{subfigure}
\begin{subfigure}{.49\textwidth}
  \centering
 		\def\svgwidth{\columnwidth}
    		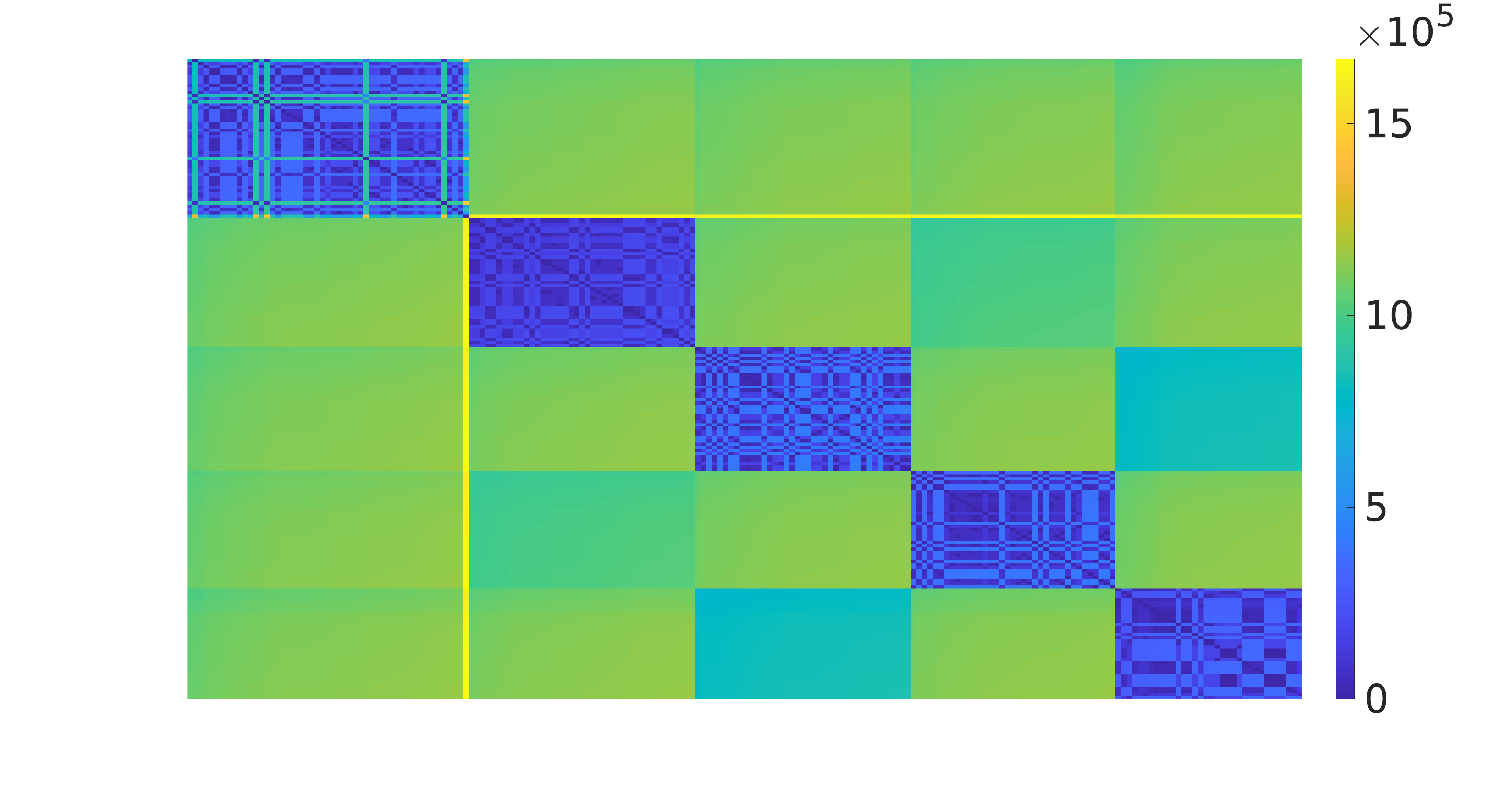
  \caption{}
  \label{fig:app:FPclusterD}
  \end{subfigure}
    \caption{Figure~(a) shows the interaction matrix enforced to be nonpositive as defined in sec.~\nameref{sec:app:Preventing}. When applying a clustering algorithm, we find in Figure~(b) that five clear clusters exist. Figures~(c) and (d) show the genealogical distance between different species, where (d) is clustered as (b). This indicates that interaction strength correlates with genealogical distance and interaction between the different clusters is effectively negligible.}
    \label{fig:app:clusters}
\end{figure}

Another way to prevent biomass divergence is to enforce that the coefficients in the interaction matrix are nonpositive. This can be realized by setting all positive values in $A$ to zero after a speciation step. However, the drift upwards in the coefficients of $A$ due to speciation is still observed (result not shown). This will result in a situation with many species, but the competition is effectively only with recent ancestors, as in every speciation step a $-\gamma$ niche interaction term is introduced. To show this effect, we follow~\cite{shtilerman2015emergence} and cluster the living species at the end of a simulation according to their interaction strengths, using the same ICM algorithm~\cite{blondel2008fast} as in the main text. Figs.~\ref{fig:app:FPclusterA} and \ref{fig:app:FPclusterB} show the unclustered and clustered interaction matrix $A$. These figures indeed show that the surviving 203 species are clustered in five groups of species that strongly interact. The question now is whether this clustering correlates to genealogical distance. The matrix of all genealogical distances is shown in Fig.~\ref{fig:app:FPclusterC}, but when we sort it according to the clustering in Fig.~\ref{fig:app:FPclusterB}, we see in Fig.~\ref{fig:app:FPclusterD} that the genealogical distance within these clusters is short. In terms of the model, this implies that the interaction is dominated by the introduction of the mother-daughter interaction $\gamma$. As pointed out in~\cite{shtilerman2015emergence}, this outcome supports Darwin's competition-relatedness hypothesis. However, studies in real-world ecosystems suggest that competition is much more based on functional similarities instead of genealogical similarities~\cite{alexandrou2015evolutionary,mahon2021functional}. Hence, the absence of interaction between the different clusters here appears to be artificial and not reflective of real ecosystems. In contrast, this interaction between clusters is present in the two trophic layer model as discussed in the main text, see in particular Fig.~\ref{fig:2layers:cluster}.

\section{Sensitivity analysis of model parameters}
\label{sec:AppB}

In this appendix, we discuss some of the system parameters from the main text and explain why certain choices were made, and why some parameter choices do not significantly affect the modeling outcomes. 

\begin{figure}[t]
\begin{subfigure}{.49\textwidth}
  \centering
 		\def\svgwidth{\columnwidth}
    		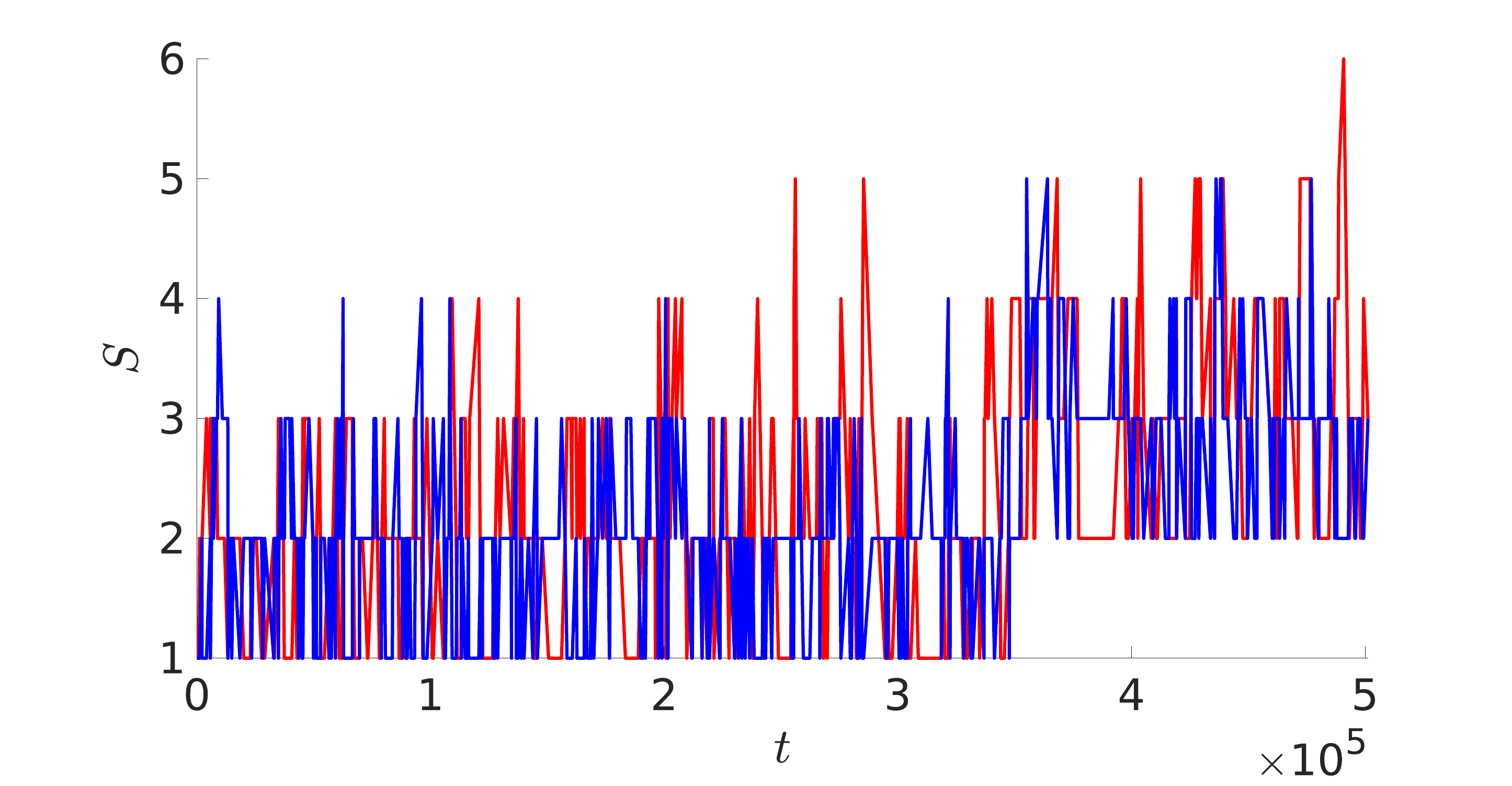
  \caption{}
  \label{fig:app:SvsKinf}
\end{subfigure}
\begin{subfigure}{.49\textwidth}
  \centering
 		\def\svgwidth{\columnwidth}
    		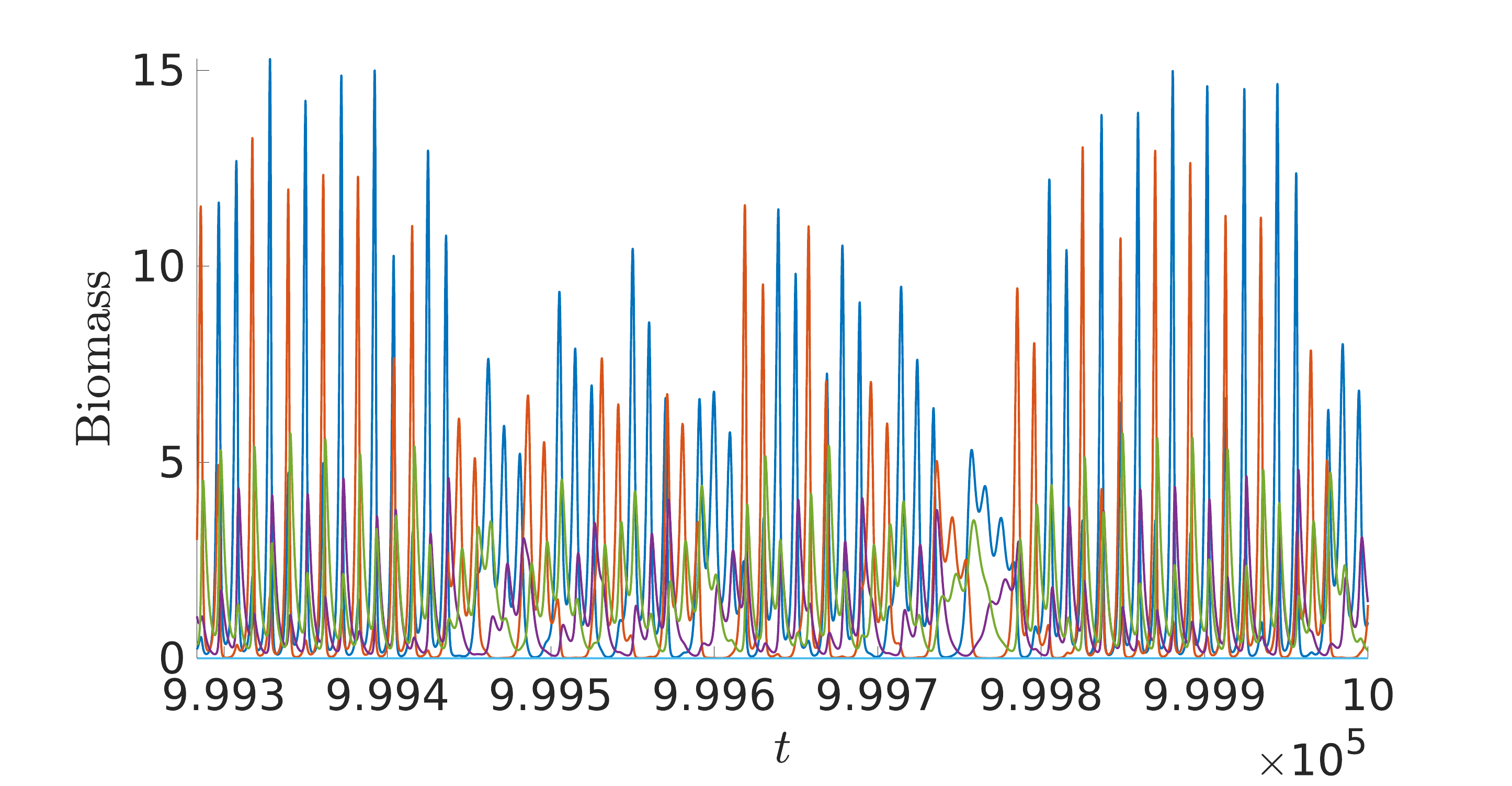
  \caption{}
  \label{fig:app:SvsKinfChaos}
\end{subfigure}
\caption{Single run of the model for $K\rightarrow\infty$, i.e. no carrying capacity. As is shown, the total number of predators (red) and prey (blue) $S$ remains low. Out of the six species that remain at $t=5\cdot 10^5$, four survive after speciation is stopped. In Figure~(b), we show the dynamics of these four remaining species after a long integration time to remove any transients, but the dynamics remains chaotic, as it scored a 0.9987 using the 0-1 chaos test from \cite{gottwald2009implementation}.}
\end{figure}

\paragraph{Carrying capacity $K$ ---} We chose to work with a carrying capacity $K=10$. Before we discuss this specific choice, let us study the case of $K\rightarrow\infty$, i.e. no carrying capacity. This choice makes the model linear, and all else being equal, this results in oscillatory dynamics instead of an asymptotic fixed point. This in turn implies that the total biomass fluctuates strongly. An example run is shown in Fig.~\ref{fig:app:SvsKinf}. We observe that in this setting, the number of species is small. Furthermore, when speciation is stopped at the end of the simulation, the resulting system with two predator and two prey species is chaotic, see Fig.~\ref{fig:app:SvsKinfChaos}.  We conclude that a carrying capacity is needed to obtain systems with a high number of species. 

\begin{figure}[t]
\begin{subfigure}{.49\textwidth}
  \centering
 		\def\svgwidth{\columnwidth}
    		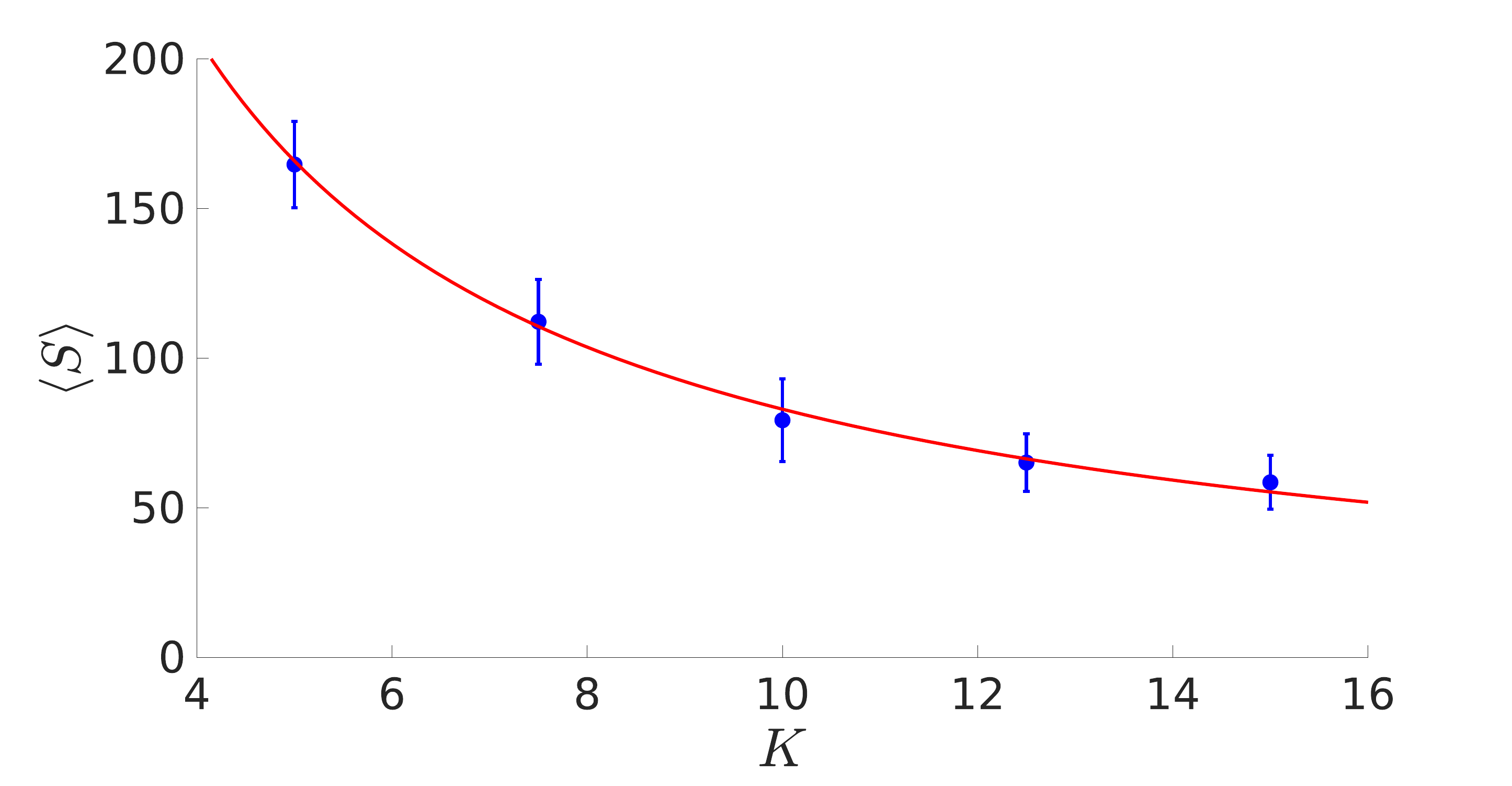
  \caption{}
  \label{fig:app:SvsK}
\end{subfigure}
\begin{subfigure}{.49\textwidth}
  \centering
 		\def\svgwidth{\columnwidth}
    		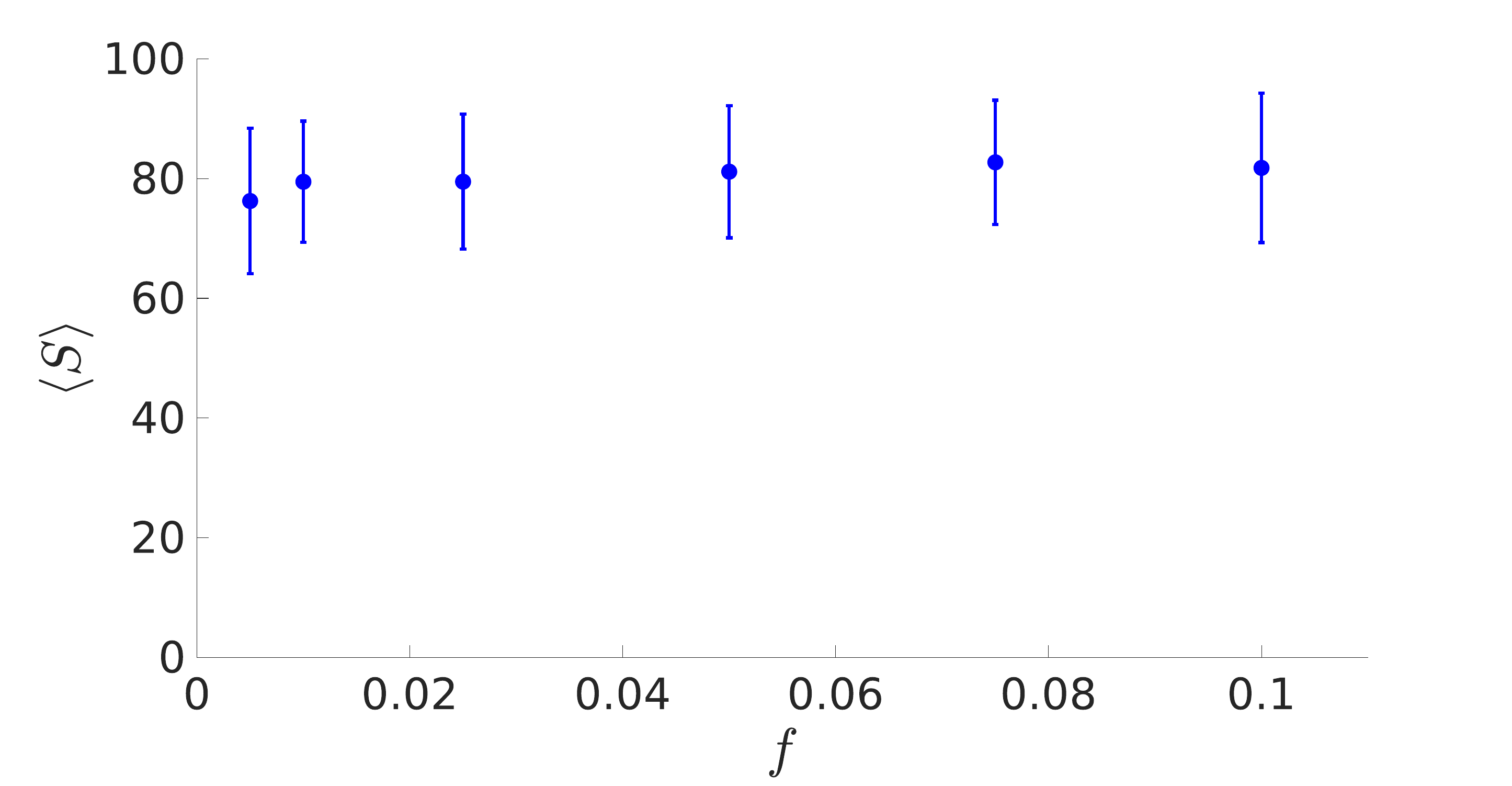
  \caption{}
  \label{fig:app:Svsf}
\end{subfigure}
\caption{Figure (a) shows $\langle S\rangle$ as a function of the carrying capacity $K$, the red line is a fit to the power law $\langle S\rangle=aK^{-1}$, with $a \approx 829$. Figure (b) shows $\langle S\rangle$ as a function of $f$, the fraction of biomass that is transferred to the daughter species. This figure shows that $f$ does not meaningfully influence the average number of species. }
\end{figure}

For the fixed point \eqref{eq:CLV} to exist, we need $K>\delta/(\beta P)$. The latter equals four for the values in Table~\ref{tab:2layers}. In Fig.~\ref{fig:app:SvsK}, we plot the average number of species $\langle S\rangle$ for a range of $K$-values. Clearly, $\langle S\rangle\sim K^{-1}$, as indicated by the red line which is a fit to $\langle S\rangle =aK^{-1}$ with $a \approx 829$. A fit with a free exponent yields similar results; we choose not to fit an offset because of the limiting behavior of $K \to \infty$ discussed above. The interpretation of the $1/K$ scaling relationship is beyond the scope of this work, but we can see that our model choice of $K=10$ can be considered a trade-off between making sure the model is not very sensitive to $K$, while keeping enough species for the clustering analysis.   

\paragraph{Ratio Ancestor-Daughter species $f$ ---} In the main text, we set the fraction $f$ of biomass that is transferred from the ancestor to the daughter species to 0.05. Changing this parameter does not essentially influence the dynamics, as is shown in Fig.~\ref{fig:app:Svsf}. Only when $f$ is chosen close to or below the extinction threshold, $\langle S \rangle$ trivially decreases. 

\begin{figure}[t]
\begin{subfigure}{.49\textwidth}
  \centering
 		\def\svgwidth{\columnwidth}
    		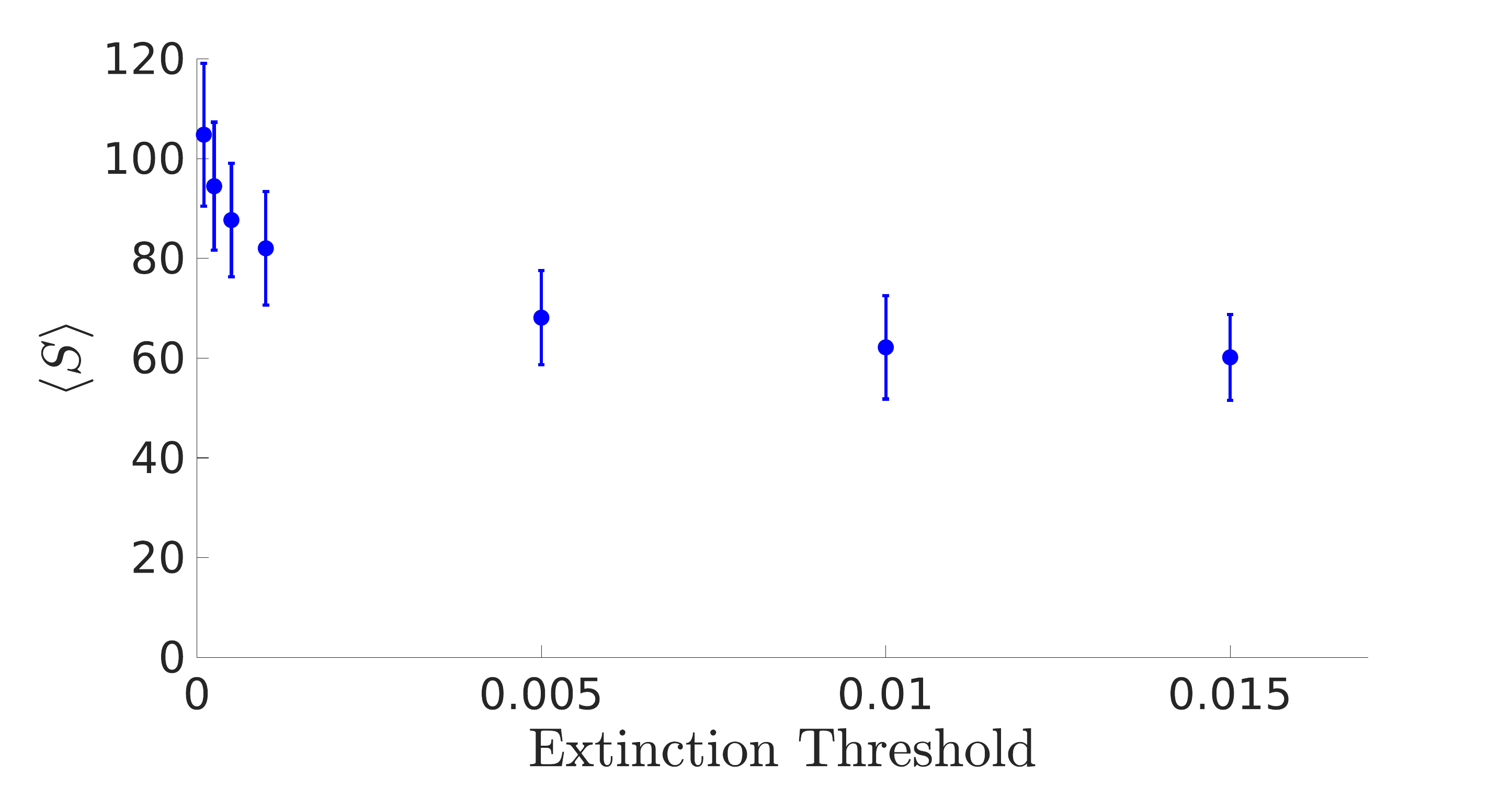
  \caption{}
  \label{fig:app:SvsET}
\end{subfigure}
\begin{subfigure}{.49\textwidth}
  \centering
 		\def\svgwidth{\columnwidth}
    		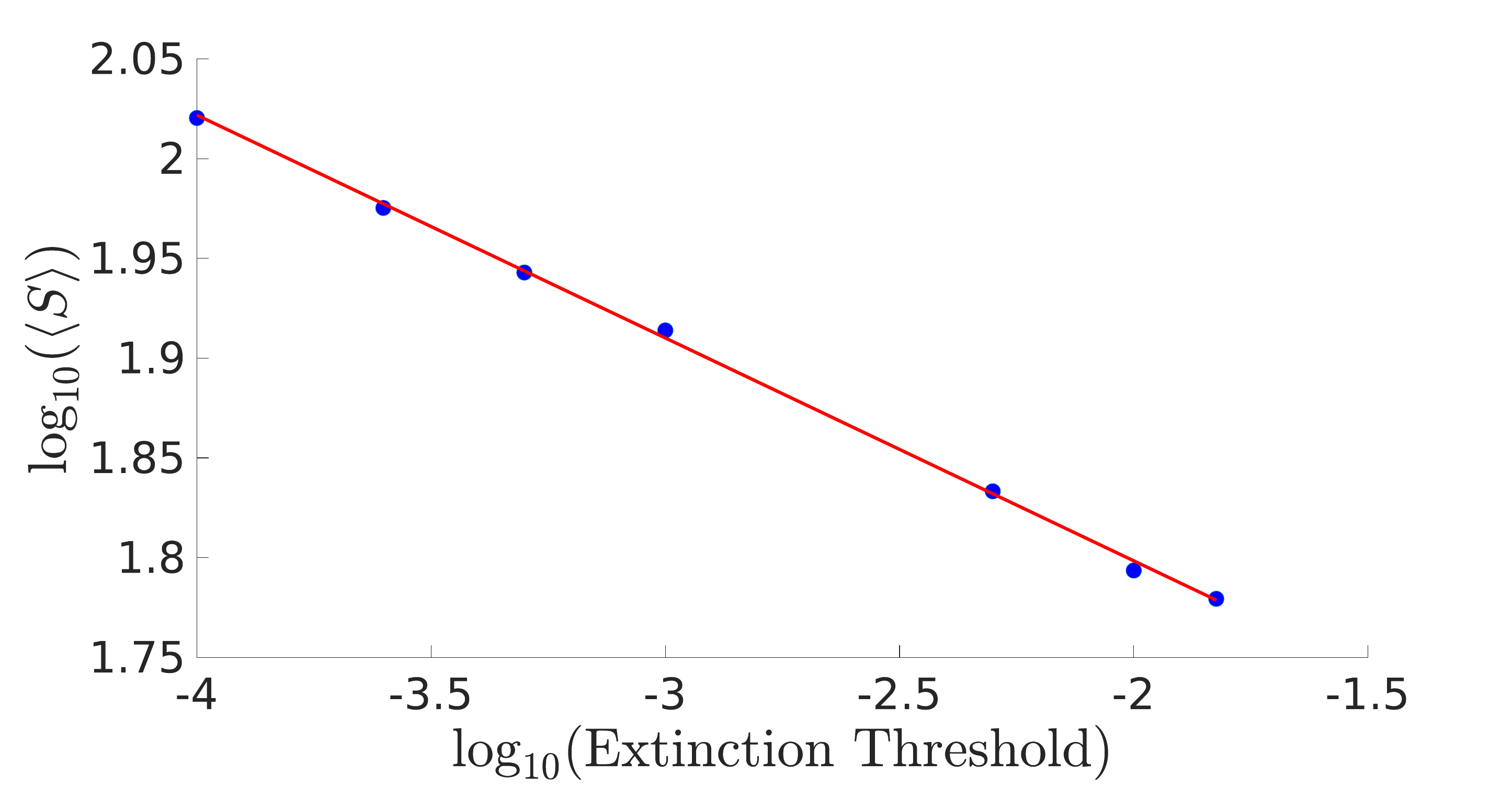
  \caption{}
  \label{fig:app:SvsETlog}
\end{subfigure}
\caption{Figure (a) shows the average number of species $\langle S\rangle$ as a function of the extinction threshold. From this figure, the decay rate is difficult to probe, but in the log-log plot in Figure (b), there is a clear indication of a power law. However, the decay is small as the red line has a slope of $-0.11$. }
\end{figure}

\paragraph{Existence Threshold ---} The average number of species $\langle S \rangle$ as a function of the extinction threshold has two trivial limits. When the threshold is larger than the total biomass in the system, the number of species is 0. When there is no extinction threshold, no species go extinct as species decay exponentially to 0. Given these extreme limits, we plot the dependence on the threshold over two orders of magnitude and indeed observe the expected negative relation, but the decay is slow, the fit in Fig.~\ref{fig:app:SvsETlog} indicates a weak power law decay with exponent -0.11. Hence, the sensitivity of $\langle S\rangle$ on the extinction threshold is small. 

\subsection{Parameters in clustering algorithms}
For the genealogical-distance clustering, the dependence of the number of clusters on the cut-off in the genealogical distance can be understood visually from Fig.~\ref{fig:2layers:dendrogramNoDelta}. A smaller distance implies more clusters. For Matlab's \texttt{kmeans} algorithm, however, it is not immediately clear how the different options influence the outcome. Given the matrix $P$ as defined in \eqref{eq:DefA} and the vector $r$ of growth rates, we write in Matlab
\begin{lstlisting}[style=Matlab-editor]
Clusters=kmeans([3*r,P],6,'Distance','sqEuclidean','Replicates',500);
silh=silhouette([3*r,P],Clusters,'sqEuclidean');
mean(silh)
\end{lstlisting}
The score for the clustering is given by \texttt{mean(silh)}, where a higher score is better. The number of replicates here must be high to ensure repeatability. For a low number of replicates, the clustering is different each turn. Matlab has a range of choices for the distance function, \texttt{sqEuclidean, cityblock, cosine} and \texttt{correlation}. When these different options are applied to the data from Fig.~\ref{fig:kmeansprogress} for 
cluster numbers 5, 6 and 7, all options return the same clustering with 6 clusters as optimal.

\end{document}